\documentclass[twocolumn,twoside]{IEEEtran}
\usepackage{mathpple}
\usepackage{times}
\usepackage{amsmath}  
\usepackage{amssymb}  
\usepackage{mathrsfs} 
\usepackage{theorem}  
\usepackage{cite}     
\usepackage{comment}  
\usepackage{upref}
\usepackage{amsfonts}
\usepackage{verbatim}
\usepackage[dvipsnames,usenames]{color}
\usepackage{enumerate}
\usepackage{cuted}
\usepackage[shortlabels]{enumitem}

\usepackage{algorithm}
\usepackage{times,color}
\usepackage{graphicx}
\usepackage{float}
\usepackage{color}
\usepackage{mathtools}
\usepackage{graphicx}
\usepackage{listings}
\usepackage{tikz}
\usepackage{multicol}
\usepackage{psfrag}

\usepackage{pgfplots}
\usepackage{hhline}
\usepackage{xcolor,colortbl}
\usepackage{algorithm}
\usepackage[noend]{algpseudocode}

\usepackage{wrapfig}
\usepackage{subfigure}
\usepackage[top=0.58in, bottom=0.58in, left=0.6in, right=0.6in]{geometry}
\usepackage{kantlipsum,cuted}
\usepackage[all]{xy}
\usepackage{algorithmicx}
\algrenewcommand\alglinenumber[1]{\scriptsize #1:}
\algrenewcommand\algorithmicindent{1em}%
\usepackage{latexsym}
\usepackage{multirow}

\usepackage{tabu}
\usepackage{lipsum}
\usepackage{blkarray, bigstrut}
\usepackage{comment}

\newcommand{\RN}[1]{%
	\textup{\uppercase\expandafter{\romannumeral#1}}%
}

\newtheorem{innercustomgeneric}{\customgenericname}
\providecommand{\customgenericname}{}
\newcommand{\newcustomtheorem}[2]{%
	\newenvironment{#1}[1]
	{%
		\renewcommand\customgenericname{#2}%
		\renewcommand\theinnercustomgeneric{##1}%
		\innercustomgeneric
	}
	{\endinnercustomgeneric}
}
\newcustomtheorem{customproposition}{Proposition}
\newcustomtheorem{customtheorem}{Theorem}
\newcustomtheorem{customlemma}{Lemma}
\newcustomtheorem{customclaim}{Claim}
\newcustomtheorem{customcorollary}{Corollary}

\makeatletter
\def\BState{\State\hskip-\ALG@thistlm}

\makeatother

\algnewcommand{\Initialize}[1]{%
	\State \textbf{Initialize:}
	\Statex \hspace*{\algorithmicindent}\parbox[t]{.8\linewidth}{\raggedright #1}
}

\newcommand{\be}[1]{\begin{equation}\label{#1}}
\newcommand{\ee}{\end{equation}}
\newcommand{\eq}[1]{(\ref{#1})}

\newcommand{\bc}{\begin{center}}
\newcommand{\ec}{\end{center}}

\newcommand{\cA}{{\cal A}}
\newcommand{\cB}{{\cal B}}
\newcommand{\cC}{{\cal C}}
\newcommand{\cD}{{\cal D}}
\newcommand{\cE}{{\cal E}}
\newcommand{\cF}{{\cal F}}
\newcommand{\cG}{{\cal G}}

\newcommand{\cP}{{\cal P}}

\newcommand{\cR}{{\cal R}}
\newcommand{\cS}{{\cal S}}
\newcommand{\cT}{{\cal T}}
\newcommand{\cU}{{\cal U}}

\newcommand{\bfc}{{\boldsymbol c}}

\newcommand{\bfg}{{\boldsymbol g}}
\newcommand{\bfh}{{\boldsymbol h}}

\newcommand{\bfu}{{\boldsymbol u}}
\newcommand{\bfv}{{\boldsymbol v}}

\newcommand{\bfx}{{\boldsymbol x}}
\newcommand{\bfy}{{\boldsymbol y}}

\renewcommand{\le}{\leqslant}
\renewcommand{\leq}{\leqslant}
\renewcommand{\ge}{\geqslant}
\renewcommand{\geq}{\geqslant}

\newcommand{\Tref}[1]{Theo\-rem\,\ref{#1}}

\newcommand{\Lref}[1]{Lem\-ma\,\ref{#1}}
\newcommand{\Cref}[1]{Co\-rol\-la\-ry\,\ref{#1}}

\theoremstyle{plain} \theorembodyfont{\normalfont\slshape}

\newtheorem{thm}{Theorem$\!$}
\newenvironment{theorem}{\begin{thm}\hspace*{-1ex}{\bf.}}{\end{thm}}

\newtheorem{prop}[thm]{Proposition$\!$}

\newtheorem{lem}[thm]{Lemma$\!$}
\newenvironment{lemma}{\begin{lem}\hspace*{-1ex}{\bf.}}{\end{lem}}

\newtheorem{cor}[thm]{Corollary$\!$}
\newenvironment{corollary}{\begin{cor}\hspace*{-1ex}{\bf.}}{\end{cor}}

\newtheorem{cons}[thm]{Construction$\!$}
\newenvironment{construction}{\begin{cons}\hspace*{-1ex}{\bf.}}{\end{cons}}

\newtheorem{defi}[thm]{Definition$\!$}
\newenvironment{definition}{\begin{defi}\hspace*{-1ex}{\bf.}}{\end{defi}}

\newtheorem{cl}{Claim}
\newenvironment{claim}{\begin{cl}\hspace*{-1ex}{\bf .}}{\end{cl}}
\theorembodyfont{\normalfont}

\newtheorem{exam}{Example$\!$}
\newenvironment{example}{\begin{exam}\hspace*{-1ex}{\bf .}}{\end{exam}}

\newtheorem{remrk}{Remark$\!$}

\newtheorem{proper}[thm]{Property$\!$}
\newenvironment{property}{\begin{proper}\hspace*{-1ex}{\bf.}}{\end{proper}}

\newcommand{\qbin}[3]{\begin{bmatrix}{#1}\\ {#2}\end{bmatrix}_{#3}}

\definecolor{Codecolor}{named}{White}  

\newcommand{\Copen}{\mbox{\{\kern-5.50pt\{}}
\newcommand{\Cclose}{\mbox{\}\kern-5.50pt\}}}
\newcommand{\Cslash}{\mbox{$\backslash\kern-6.02pt\backslash$}}

\newcommand{\e}[1]{\textcolor{red}{#1}}

\begin{document}
		
		\title{Array Codes for Functional PIR and Batch Codes}
		
		\author{\large Mohammad~Nassar,~\IEEEmembership{Student Member,~IEEE},  and Eitan~Yaakobi,~\IEEEmembership{Senior Member,~IEEE} 
		  \thanks{This work was partially funded by the Israel Science Foundation (grant \#1817/18) and by the Technion Hiroshi Fujiwara Cyber Security Research Center and the Israel National Cyber Directorate. This article was presented in part at the IEEE International Symposium on Information Theory (ISIT), Los Angeles, CA, June 2020 (reference~\cite{NY20}).}
			\thanks{M. Nassar and E. Yaakobi are with the Department of Computer Science, Technion --- Israel Institute of Technology, Haifa 3200003, Israel (e-mail: \texttt{\{mohamadtn,yaakobi\}@cs.technion.ac.il}).}}

		\maketitle		
\begin{abstract}
A \emph{functional PIR array code} is a coding scheme which encodes some $s$ information bits into a $t\times m$ array such that every linear combination of the $s$ information bits has $k$ mutually disjoint \emph{recovering sets}. Every recovering set consists of some of the array's columns while it is allowed to read at most $\ell$ encoded bits from every column in order to receive the requested linear combination of the information bits. \emph{Functional batch array codes} impose a stronger property where every multiset request of $k$ linear combinations has $k$ mutually disjoint recovering sets. \emph{Locality functional array codes} demand that the size of every recovering set is restrained to be at most $r$. Given the values of $s, k, t, \ell,r$, the goal of this paper is to study the optimal value of the number of columns $m$ such that these codes exist. Several lower bounds are presented as well as explicit constructions for several of these parameters. 
\end{abstract}

\section{Introduction}\label{sec:intro}
\renewcommand{\baselinestretch}{1}\normalsize
\emph{Private information retrieval (PIR) codes} and {batch codes} are families of codes which have several applications such as PIR protocols~\cite{BIKR02,CKGS98,DvirGopi16_1,FVY15,G04,Y10}, erasure codes in distributed storage systems~\cite{PHO13,RPDV14,tamo2014family}, one-step majority-logic decoding~\cite{LC04,M63}, load balancing in storage, cryptographic protocols~\cite{IKOS04}, switch codes~\cite{BCSY17,CGTZ15,WKCB17}, and more. They have been recently generalized to \emph{functional PIR} and \emph{functional batch} codes~\cite{ZYE19}. In this work we study these families of codes when they are used as array codes. 

The setup of storing information in array codes works as follows. Assume $s$ bits are encoded to be stored in a $t\times m$ array, where each column corresponds to a \emph{server} that stores the encoded bits. The encoded bits should satisfy several properties which depend upon whether the resulting code is a PIR, batch, functional PIR, or functional batch codes. Given a design parameter $k$ of the code, it is required in PIR codes that every information bit has $k$ mutually disjoint \emph{recovering sets}. Here, a recovering set is a set of columns, i.e., servers, in which given the encoded bits in the columns of the recovering set it is possible to recover the information bit. In case it is possible to read only a portion of the encoded bits in every column, we denote this parameter by $\ell$. An array code with these parameters and properties is defined as an \emph{$(s,k,m,t,\ell)$ PIR array code}. Furthermore, it will be called an \emph{$(s,k,m,t,\ell)$ batch array code} if every \emph{multiset} request of $k$ information bits has $k$ mutually disjoint recovering sets. In case the requests are not only of information bits but any linear combination of them, we receive an \emph{$(s,k,m,t,\ell)$ functional PIR array code}, if the same linear combination is requested $k$ times or \emph{$(s,k,m,t,\ell)$ functional batch array code} for a multiset request of $k$ linear combinations. Yet another family of codes that will be studied in this paper will be referred by \emph{locality functional array codes}. Here we assume that $\ell=t$ and an \emph{$(s,k,m,t,r)$ locality functional array code} guarantees that every linear combination $\bfv$ of the information bits has $k$ mutually disjoint recovering sets, where each is of size of at most $r$.

The main figure of merit when studying these families of codes is to optimize the number of columns, i.e., servers, given the values of $s,k,t,\ell$. Thus, the smallest $m$ such that an $(s,k,m,t,\ell)$ PIR, batch, functional PIR, functional batch code exists, is denoted by $P_{t,\ell}(s,k), B_{t,\ell}(s,k), FP_{t,\ell}(s,k), FB_{t,\ell}(s,k)$, respectively. Studying the value of $P_{t,\ell}(s,k)$ has been initiated in~\cite{FVY15} and since then several more results have appeared; see e.g.~\cite{BE17,BE19,CKYZ19,ZWWG19}. Note that the first work~\cite{IKOS04} which studied batch codes defined them in their array codes setup and only later on they were studied in their one-dimensional case, also known as \emph{primitive batch codes}; see e.g.~\cite{AY17,LS15,RSDG16,VY16, ZS16}. Functional PIR and batch codes have been recently studied in~\cite{ZYE19} but only for vectors, that is, $t=\ell=1$. Thus, this paper initiates the study of functional PIR and batch codes in the array setup. 


The motivation to study functional PIR and batch codes originates from the observation that in many cases and protocols, such as PIR, the user is not necessarily interested in one of the information bits, bur rather, some linear combination of them. Furthermore, functional batch codes are closely related to the family of \emph{random I/O (RIO) codes}, introduced by Sharon and Alrod~\cite{SA13}, which are used to improve the random input/output performance of flash memories. A variant of RIO codes, called \emph{parallel RIO codes}, was introduced in~\cite{YM16}, and linear codes of this family of codes have been studied in~\cite{YKL17}. It was then shown in~\cite{ZYE19} that in fact linear parallel RIO codes are equivalent to functional batch codes.

The rest of the paper is organized as follows. In Section~\ref{sec:defs}, we formally define the codes studied in the paper, discuss some of the previous related work, and list several basic properties. In Section~\ref{sec:bounds}, we show lower bounds on the number of servers for functional PIR and batch array codes. Section~\ref{sec:cons} lists several code constructions which are based on the Gadget Lemma, covering codes, and several more results for $k=1,2$. Section~\ref{sec:cons_spec} presents three constructions of array codes and in Section~\ref{sec:analysis} the rates of these codes are studied. Section~\ref{sec:Locality} studies locality functional array codes. Lastly, Section~\ref{sec:conc} concludes the paper. 

\section{Definitions and Preliminaries}\label{sec:defs}
This work is focused on five families of codes, namely \emph{private information retrieval} (\emph{PIR}) codes that were defined recently in~\cite{FVY15}, \emph{batch codes} that were first studied by Ishai et al. in~\cite{IKOS04}, their extension to \emph{functional PIR codes} and \emph{functional batch codes} that was investigated in~\cite{ZYE19}, \emph{and locality functional codes}. In these five families of codes, $s$ information bits are encoded to $m$ bits. While for PIR codes it is required that every information bit has $k$ mutually disjoint recovering sets, batch codes impose this property for every multiset request of $k$ bits. Similarly, for functional PIR codes it is required that every linear combination of the information bits has $k$ mutually disjoint recovering sets, and functional batch codes impose this property for every multiset request of $k$ linear combination of the bits. Lastly, similar to functional PIR codes, for locality functional codes it is required that the size of every recovering set is limited to be at most $r$. While this description of the codes corresponds to the case of one-dimensional codewords, the goal of this work is to study their extension as \emph{array codes}, which is defined as follows. The set $[n]$ denotes the set of integers $\{1,2,\ldots,n\}$ and $\Sigma=\mathbb{F}_2$.  

We start with the formal definition of the first four families of codes that will be studied in the paper, while we defer the definition of locality functional array codes to Section~\ref{sec:Locality}.
\begin{definition}\label{ArrayCodesDef}
\begin{enumerate}

\item An $(s,k,m,t,\ell)$ \textbf{PIR array code} over $\Sigma$ is defined by an encoding map $\cE:\Sigma^s \rightarrow (\Sigma^t)^m$ that encodes $s$ information bits $x_1,\dots,x_s$ into a $t\times m$ array and a decoding function $\cD$ that satisfies the following property. For any $i \in [s]$ there is a partition of the columns into $k$ recovering sets $S_1,\ldots,S_k \subseteq [m]$ such that $x_i$ can be recovered by reading at most $\ell$ bits from each column in $S_j,j\in[k]$.

\item An $(s,k,m,t,\ell)$ \textbf{batch array code} over $\Sigma$ is defined by an encoding map $\cE:\Sigma^s \rightarrow (\Sigma^t)^m$ that encodes $s$ information bits $x_1,\dots,x_s$ into a $t\times m$ array and a decoding function $\cD$ that satisfies the following property. For any multiset request of $k$ bits $i_1,\ldots,i_k\in[s]$ there is a partition of the columns into $k$ recovering sets $S_1,\ldots,S_k \subseteq [m]$ such that $x_{i_j}, j\in[k]$ can be recovered by reading at most $\ell$ bits from each column in $S_j$.

\item An $(s,k,m,t,\ell)$ \textbf{functional PIR array code} over $\Sigma$ is defined by an encoding map $\cE:\Sigma^s \rightarrow (\Sigma^t)^m$ that encodes $s$ information bits $x_1,\dots,x_s$ into a $t\times m$ array and a decoding function $\cD$ that satisfies the following property. For any request of a linear combination $\bfv$ of the information bits, there is a partition of the columns into $k$ recovering sets $S_1,\ldots,S_k \subseteq [m]$ such that $\bfv$ can be recovered by reading at most $\ell$ bits from each column in $S_j,j\in[k]$.

\item An $(s,k,m,t,\ell)$ \textbf{functional batch array code} over $\Sigma$ is defined by an encoding map $\cE:\Sigma^s \rightarrow (\Sigma^t)^m$ that encodes $s$ information bits $x_1,\dots,x_s$ into a $t\times m$ array and a decoding function $\cD$ that satisfies the following property. For any multiset request of $k$ linear combinations $\bfv_1,\ldots, \bfv_k$ of the information bits, there is a partition of the columns into $k$ recovering sets $S_1,\ldots,S_k \subseteq [m]$ such that $\bfv_{j}, j\in[k]$ can be recovered by reading at most $\ell$ bits from each column in $S_j$.
\end{enumerate}
\end{definition}

We refer to each column as a \emph{bucket} and to each entry in a bucket as a \emph{cell}. Furthermore, it is said that a cell stores a \emph{singleton} if one of the information bits is stored in the cell. In the rest of the paper we will refer to every linear combination of the information bits as a binary vector of length $s$, which indicates the information bits in this linear combination. Our goal is to fix the values of $s,k,t$ and $\ell$ and then seek to optimize the value of $m$. In particular, we will have that $t$ and $\ell$ are fixed, where $t\geq \ell$, and then study the growth of $m$ as a function of $s$ and $k$. Hence, we denote by $P_{t,\ell}(s,k), B_{t,\ell}(s,k), FP_{t,\ell}(s,k), FB_{t,\ell}(s,k)$ the smallest $m$ such that an $(s,k,m,t,\ell)$ PIR, batch, functional PIR, functional batch code exists, respectively. In case $\ell=t=1$ we will simply remove them from these notations.

The following upper and lower bounds on the number of buckets for PIR array codes have been shown in~\cite{BE19,CKYZ19,ZWWG19} and are stated in the following theorem.
\begin{theorem}\label{theorem:PIRLB}
\begin{enumerate}
	\item $P_{t,t}(s,k) \geq \frac{2\cdot k \cdot s}{s+t}$,~\cite[Th. 3]{BE19}.\label{theorem:part1}
	
	\item For any integer $t \geq 2$ and any integer $s>t$, $P_{t,t}(s,k) \geq \frac{k\cdot s \cdot(2s - 2t + 1)}{(2s - 2t+1)t+(s-t)^2}$,~\cite[Th. 4]{BE19}.\label{theorem:part2}
	
	\item For any integer $t \geq 2$ and any integer $s>2t$, $P_{t,t}(s,k) \geq \frac{2k\cdot s \cdot(s + 1)}{(s-t)^2 + 3st - t^2 + 2t}$,~\cite[Th. 16]{ZWWG19}.\label{theorem:part4}
	
	\item For any integer $t \geq 2$ and any integer $t < s \leq 2t$, $P_{t,t}(s,k) \leq \frac{k\cdot s \cdot(2s - 2t +1)}{(2s - 2t+1)t+(s-t)^2}$,~\cite[Th. 6]{BE19}.\label{theorem:part3}
	
	\item For any integers $p,t$ with $p\leq t+1$, $P_{t,t}(pt,k) \leq m$, where $k = {t \choose t-p+1}{s \choose t}$ and $m = {t \choose t-p+1}{s \choose t}+{s-p \choose t-p+1}{s-1 \choose p-1}$,~\cite[Th. 10]{CKYZ19}.\label{theorem:part5}
	
	\end{enumerate}
\end{theorem}
Note that for any two integers $t \geq 2$ and $s>t$, the bound in Theorem~\ref{theorem:PIRLB}\eqref{theorem:part2} improves upon the bound in Theorem~\ref{theorem:PIRLB}\eqref{theorem:part1}. This is verified by showing that $\frac{k\cdot s \cdot(2s - 2t + 1)}{(2s - 2t+1)t+(s-t)^2} - \frac{2\cdot k \cdot s}{s+t} \geq 0$ by basic algebraic manipulations. However the lower bound in Theorem~\ref{theorem:PIRLB}\eqref{theorem:part1} holds for all values of $s$, while the one in Theorem~\ref{theorem:PIRLB}\eqref{theorem:part2} only for $s>t$. Also, in~\cite{ZWWG19} it was shown that for any two integers $t \geq 2$ and $s>2t$, the bound in Theorem~\ref{theorem:PIRLB}\eqref{theorem:part4} is stronger than the bound in Theorem~\ref{theorem:PIRLB}\eqref{theorem:part2}.

The result in Theorem~\ref{theorem:PIRLB}\eqref{theorem:part3} is achieved by Construction 1 in~\cite{BE19}. The authors of~\cite{BE19} presented another construction which is not reported here due to its length. For the exact details please refer to~\cite[Construction 4 and Th.8]{BE19}. This construction was then improved in~\cite{ZWWG19} and in~\cite{CKYZ19}. Several more constructions of PIR array codes have also been presented in~\cite{CKYZ19,ZWWG19}.

The following theorem summarizes some of the known basic previous results, as well as several new ones. The proofs are rather simple and are thus omitted.
\begin{theorem}\label{theorem:Basic}
 For every $s,k,t,\ell,a$ positive integers:
\begin{enumerate}
    \item $P_{t,\ell}(s,1) = B_{t,\ell}(s,1) = \lceil s/t\rceil$.
    \item $FP_{t,\ell}(s,k_1+k_2) \leq FP_{t,\ell}(s,k_1) + FP_{t,\ell}(s,k_2)$ (also for $P$, $B$, and $FB$).
    \item $FP_{t,\ell}(s,a \cdot k) \leq a \cdot FP_{t,\ell}(s,k)$ (also for $P$, $B$, and $FB$). \label{theorem:partak}
    \item $FP_{t,\ell}(s_1 + s_2,k) \leq FP_{t,\ell}(s_1,k) + FP_{t,\ell}(s_2,k)$ (also for $P$, $B$, and $FB$). \label{theorem:parts1s2}
    \item $FP_{t,\ell}(a \cdot s,k) \leq a \cdot FP_{t,\ell}(s,k)$ (also for $P$, $B$, and $FB$). \label{theorem:partas}
    \item $FP_{t,\ell}(s,k) \leq a \cdot FP_{a\cdot t,\ell}(s,k)$ (also for $P$, $B$, and $FB$). \label{theorem:partat} 
\end{enumerate}
\end{theorem}

One of the simplest ways to construct array PIR and batch codes uses the Gadget Lemma, which was first proved in~\cite{IKOS04}.

\begin{lemma}(\textbf{The Gadget Lemma})\label{lem:gadget}
Let $\cC$ be an $(s,k,m,1,1)$ batch code, then for any positive integer $t$ there exists an $(ts,k,m,t,1)$ batch array code $\cC'$ (denoted also by $t\cdot \cC$).
\end{lemma}
It is easily verified that the Gadget Lemma holds also for PIR codes and therefore $P_{t,\ell}(s,k) \leq P_{t,1}(s,k) \leq P(\lceil s/t\rceil,k)$ and $B_{t,\ell}(s,k) \leq B_{t,1}(s,k) \leq B(\lceil s/t\rceil,k)$. However, unfortunately, the Gadget Lemma does not hold in general for functional PIR and batch codes. Even a weaker variation of the Gadget Lemma, where $\ell=t$, does not hold in general for functional PIR and batch codes either.  Assume by contradiction that if there is an $(s,k,m,1,1)$ functional PIR code $\cC$, then for any positive integer $t$ there exists a $(ts,k,m,t,t)$ functional PIR array code. Then, this will imply that $FP_{t,t}(ts,k) \leq FP(s,k)$. However, it is known that $FP(2,2) = 3$ by the simple parity code. Thus, under this assumption it would hold that $FP_{2,2}(4,2) \leq FP(2,2) = 3$. But, according to a lower bound on functional PIR array codes, which will be shown in Theorem~\ref{theorem:LBFP4}, it holds that $FP_{2,2}(4,2) \geq \frac{2\cdot 2 \cdot 15}{15 + 3} > 3$, which is a contradiction.

\section{Lower Bounds on Array Codes}\label{sec:bounds}
In this section we present several lower bounds on functional PIR and batch array codes. Let ${a \brace b}$ be the Stirling number of the second kind, which calculates the number of partitions of a set of $a$ elements into $b$ nonempty subsets. It is well known that
$ {a\brace b}=\frac{1}{b!}\sum_{i=0}^b (-1)^{b-i}{b\choose i} i^a. $

\begin{theorem}\label{theorem:LBFB}
For all $s,k,t$ and $\ell$ positive integers $FB_{t,\ell}(s,k) \geq m^* $, where $m^*$ is the smallest positive integer such that
$$ \sum_{i=k}^{m^*} {m^* \choose i} \cdot {i \brace k}  \cdot \left(\sum_{j=1}^{\ell} {t \choose j} \right)^{i} \geq  {2^s + k - 2 \choose k}.$$
\end{theorem}
\begin{IEEEproof}
Let $\cC$ be an optimal $(s,k,m^*,t,\ell)$ functional batch array code. Since there are $s$ information bits, there are $(2^s - 1)$ possible linear combination requests and there are ${2^s + k - 2 \choose k}$ possible multiset requests of length $k$. For each multiset request of $k$ linear combinations $\bfv_1,\ldots, \bfv_k$ of the information bits, there is a partition of the buckets of the code $\cC$ into $k$ recovering sets $S_1,\ldots,S_k \subseteq [m^*]$ such that $\bfv_{j}, j\in[k]$ can be recovered by reading at most $\ell$ bits from each column in $S_j$.


In each bucket there are $t$ cells where at most $\ell$ cells from them can be read. Thus, there are $\sum_{j=1}^{\ell} {t \choose j}$ nonzero linear combinations that can be obtained from one bucket. For any positive integer $n$, there are $(\sum_{j=1}^{\ell} {t \choose j})^n$ nonzero linear combinations that can be obtained from $n$ buckets while using all the $n$ buckets.

In order to satisfy a multiset request, the buckets must be divided into $k$ disjoint recovering sets such that each set can satisfy one requested linear combination. There are 
$$\sum_{i = k}^{m^*} {m^* \choose i} \cdot {i \brace k} $$
possibilities to divide at most $m^*$ buckets into $k$ nonempty disjoint sets. Each subset of the buckets of size at least $k$ can be divided into $k$ nonempty sets. Thus, we take the sum over all the subsets of the buckets of size at least $k$, where for each such subset we count the number of possibilities to divide it into $k$ nonempty subsets using Stirling number of the second kind. From each subset of size $p$ where $k\leq p\leq m^*$, there exist $(\sum_{j=1}^{\ell} {t \choose j})^{p}$ linear combinations. Therefore, for a given partition of $i, k\leq i \leq m^*$ buckets into $k$ subsets such that the sizes of the subsets are $p_1,p_2,\ldots,p_k$ where $\sum_{j=1}^{k} p_j = i$, the number of different $k$-sets of linear combinations such that each linear combination taken from one subset is $$\prod_{p\in\{p_1,p_2,\cdots,p_k\}} \left(\sum_{j=1}^{\ell} {t \choose j} \right)^{p} = \left(\sum_{j=1}^{\ell} {t \choose j} \right)^{i}.$$

In order to satisfy each multiset request by a set of $k$ linear combinations such that each linear combination satisfies one requested linear combination. It must hold that the number of different $k$-sets of linear combinations such that each linear combination taken from one subset of the buckets, for all partitions of the $m^*$ buckets into $k$ nonempty disjoint subsets, is larger than the number of multiset requests. Thus, 
\begin{equation}\label{eq:LBFB}
\sum_{i=k}^{m^*} {m^* \choose i} \cdot {i \brace k}  \cdot \left(\sum_{j=1}^{\ell} {t \choose j} \right)^{i} \geq  {2^s + k - 2 \choose k}.
\end{equation}
\end{IEEEproof}

A similar lower bound can be obtained for functional PIR array codes. While in functional batch array codes there exist ${2^s + k - 2 \choose k}$ possible multiset requests, in functional PIR array codes there exist $2^s - 1$ possible requests.
\begin{corollary}\label{cor:LBFP3}
For all $s,k,t$ and $\ell$ positive integers $FP_{t,\ell}(s,k) \geq m^* $, where $m^*$ is the smallest positive integer such that
\begin{equation}\label{eq:LBFP3}
\sum_{i=k}^{m^*} {m^* \choose i} \cdot {i \brace k}  \cdot \left(\sum_{j=1}^{\ell} {t \choose j} \right)^{i} \geq  2^s - 1.
\end{equation}
\end{corollary}

Another combinatorial bound for functional PIR array codes is shown in the following theorem.
\begin{theorem}\label{theorem:LBFP1}
For all $s,k,t$ and $\ell$ positive integers $FP_{t,\ell}(s,k) \geq m^*$, where $m^*$ is the smallest positive integer such that
$$ \sum_{i=1}^{m^*-k+1} {m^* \choose i} \cdot \left(\sum_{j=1}^{\ell} {t \choose j} \right)^{i} \geq k \cdot (2^s -1).$$
\end{theorem}
\begin{IEEEproof}
Let $\cC$ be an optimal $(s,k,m^*,t,\ell)$ functional PIR array code. Since there are $s$ information bits, there are $(2^s - 1)$ possible requests. The code $\cC$ must satisfy each request $k$ times by $k$ linear combinations from $k$ disjoint recovering sets. In other words, for each request there are $k$ nonempty disjoint recovering sets, such that each set has a linear combination equal to the request. Each recovering set must be of size at most $m^* - k + 1$, in order to have other $k-1$ nonempty recovering sets.

In each bucket there are $t$ cells where at most $\ell$ cells from them can be read. Thus, there are $\sum_{i=1}^{\ell} {t \choose i}$ nonzero linear combinations that can be obtained from one bucket and $(\sum_{j=1}^{\ell} {t \choose j})^n$ from $n$ buckets, for any positive integer $n$, while using all the $n$ buckets. We are interested in counting the different linear combinations that can be obtained from at most $m^*-k+1$ buckets. Thus, there are $$\sum_{i=1}^{m^*-k+1} {m^* \choose i} \cdot \left(\sum_{j=1}^{\ell} {t \choose j} \right)^{i}$$ such linear combinations.
It must hold that the number of different linear combinations that can be got from at most $m^* - k + 1$ buckets is larger than $k$ times the number of the possible requests. Thus,
\begin{equation}\label{eq:LBFP1}
\sum_{i=1}^{m^*-k+1} {m^* \choose i} \cdot \left(\sum_{j=1}^{\ell} {t \choose j} \right)^{i} \geq k \cdot (2^s -1).
\end{equation}
\end{IEEEproof}

The following corollary is derived from Theorem~\ref{theorem:LBFP1}. 
\begin{corollary}\label{theorem:LBFP2}
$FP_{t,\ell}(s,k) \geq  \left\lceil \frac{\log_2(k(2^s-1)+1)}{\log_2(\sum_{i=0}^{\ell} {t \choose i})}\right\rceil$, for all $s,k,t$ and $\ell$ positive integers.
\end{corollary}
\begin{IEEEproof}
The proof of \Tref{theorem:LBFP1} can be modified by using a weaker constraint, that the size of each subset is at most $m$. Thus, it must hold that $\sum_{i=1}^{m} {m \choose i} \cdot \left(\sum_{j=1}^{\ell} {t \choose j} \right)^{i} \geq k \cdot (2^s -1)$. From the equality $\sum_{i=0}^{m} {m \choose i} \cdot x^i = (x+1)^m$, we get that,
\begin{align*}
 \sum_{i=1}^{m} {m \choose i} \cdot \left(\sum_{j=1}^{\ell} {t \choose j} \right)^{i} &= \left(1+\sum_{j=1}^{\ell} {t \choose j}\right)^m -1 &\\
 & = \left(\sum_{j=0}^{\ell} {t \choose j}\right)^m -1 \geq k \cdot (2^s -1).
 \end{align*}
 Therefore, a lower bound over the minimal number of buckets, is $FP_{t,\ell}(s,k) \geq \left\lceil \frac{\log_2(k(2^s-1)+1)}{\log_2(\sum_{j=0}^{\ell} {t \choose j})}\right\rceil$.
\end{IEEEproof}


Lastly in this section we show a different lower bound for functional PIR array codes, which is motivated by the corresponding lower bound for PIR array codes from~\cite[Th. 3]{BE19}.
\begin{theorem}\label{theorem:LBFP4}
For any $s,k,t$ and $\ell$ positive integers, $FP_{t,\ell}(s,k) \geq \frac{2\cdot k \cdot (2^s-1)}{(2^s-1)+(\sum_{i=1}^{\ell} {t \choose i})}$.
\end{theorem}
\begin{IEEEproof}
Suppose there exists an $(s,k,m,t,\ell)$ functional PIR array code. 
There are $2^s-1$ possible linear combination requests which are denoted by $\bfu_i$ for $1\leq i\leq 2^s-1$. For $i\in[2^s-1]$, we define by $\alpha_i$ to be the number of recovering sets of size $1$ of the $i$-th linear combination request $\bfu_i$. 

Since it is possible to read at most $\ell$ bits from each bucket, every bucket can satisfy at most $\sum_{i=1}^{\ell} {t \choose i}$ linear combinations. Thus, the number of recovering sets of size $1$ is $m \cdot \sum_{i=1}^{\ell} {t \choose i}$, and 
$\sum_{j=1}^{2^s-1} \alpha_j \leq m \cdot \sum_{i=1}^{\ell} {t \choose i}$. Hence, there exists $q \in[2^s-1]$ such that $\alpha_q \leq \frac{m\cdot \sum_{i=1}^{\ell} {t \choose i}}{2^s-1}$, so out of its $k$ disjoint recovering sets of $\bfu_q$, at most $\alpha_q$ of them are of size $1$, and the size of each of the remaining $k-\alpha_q$ subsets is at least $2$. Hence,
$$m \geq \alpha_q + 2(k-\alpha_q) = 2k - \alpha_q  \geq 2k - \frac{m\cdot\sum_{i=1}^{\ell} {t \choose i}}{2^s-1},$$ and therefore $m(1+\frac{\sum_{i=1}^{\ell} {t \choose i}}{(2^s-1)}) \geq 2k$, which implies that 
$FP_{t,\ell}(s,k) \geq \frac{2k(2^s-1)}{(2^s-1)+\sum_{i=1}^{\ell} {t \choose i}}.$
\end{IEEEproof}

\section{General Constructions of Array Codes}\label{sec:cons}
In this section we present several constructions of array codes for functional PIR and batch codes.

\subsection{Basic Constructions}
Even though the Gadget Lemma cannot be extended in general for functional PIR and batch codes, here we show a variation of it that will hold.
For any positive integer $i$, $\mathbf 0^{i}$ denotes the zero vector of length $i$, and for any two vectors $\bfv$ and $\bfu$, the vector $\bfv\bfu$ is defined to be the concatenation of $\bfu$ after $\bfv$.
\begin{lemma}\label{lemma:FBGadget}
For any positive integer $p$, if there exists an $(s,p\cdot k,m,t,\ell)$ functional batch array code, then there exists an $(p\cdot s,k,m,p\cdot t,\ell)$ functional batch array code. Therefore, 
$$FP_{p \cdot t,\ell}(s,k) \leq FB_{p \cdot t,\ell}(p \cdot s, k) \leq FB_{t,\ell}(s,p \cdot k),$$
and in particular, $FP_{t,1}(s,k) \leq FB_{t,1}(s,k) \leq FB(\lceil \frac{s}{t}\rceil,t\cdot k)$.
\end{lemma}
\begin{IEEEproof}
Let $\cC$ be an $(s,p\cdot k,m,t,\ell)$ functional batch array code with encoding function $\cE$ and decoding function $\cD$. We construct an $(p\cdot s,k,m,p \cdot t,\ell)$ functional batch array code $\cC'$ by using the code $\cC$. Let $\cS = \{x_{i,j}: 1 \leq i \leq p, 1 \leq j \leq s\}$ be the set of $p\cdot s$ information bits. The $p\cdot s$ information bits can be partitioned into $p$ parts, each of size $s$, such that part $i,i\in[p]$ is $\cS_i = \{x_{i,j}: 1\leq j \leq s\}$. 
The code $\cC'$ will be represented by a $pt \times m$ array $A$, that contains $p$ subarrays $A_1,A_2,\ldots,A_p$ each of dimension $t \times m$. In the encoding function of the code $\cC'$, the $i$-th subarray $A_i$ stores the encoded bits of the set $\cS_i $ by applying the encoding function $\cE$ of the code $\cC$ over the information bits in the set $\cS_i$. 


Let $R = \{\bfv_1,\bfv_2,\ldots,\bfv_k\}$ be a multiset request of size $k$ of the $p \cdot s$ information bits, where $\bfv_i,i\in[k]$ is a binary vector of length $ps$ that represents the $i$-th request. 
For each $i\in[k]$, denote $\bfv_i = (\bfv_i^1, \bfv_i^2,\ldots,\bfv_i^p)$ where $\bfv_i^j,j\in[p]$ is a vector of length $s$ that represents the linear combination of the bits in $\cS_j$. 
Let $R^{*} = \{\bfv_i^j: 1\leq i \leq k, 1\leq j \leq p\}$ be a multiset request of size $pk$, that has $pk$ vectors of length $s$ each. By using the decoding function $\cD$ of the code $\cC$ with the request $R^{*}$ we get $pk$ recovering sets. For each $i\in[k]$ and $j\in[p]$, let $B_i^j = \{(h_{i,1},\bfu_{i,1}), (h_{i,2},\bfu_{i,2}), \ldots, (h_{i,a_i},\bfu_{i,a_i})\}$ be a recovering set for $\bfv_i^j$ of size $a_i$, where for each $g\in[a_i]$, $(h_{i,g},\bfu_{i,g})$ is a pair of a bucket $h_{i,g}$ with a vector $\bfu_{i,g}$ of length $t$ that indicates the cells which are read from the bucket $h_{i,g}$. For each $B_i^j$ and $f\in[p]$, let $B_{i,f}^j = \{( h_{i,1},\mathbf 0^{t(f-1)} \bfu_{i,1}\mathbf 0^{t(p-f)}),\ldots,( h_{i,a_i},\mathbf 0^{t(f-1)} \bfu_{i,a_i}\mathbf 0^{t(p-f)})\}$ be a recovering set for $\bfv_i^j$, that reads the cells of subarray $A_f$. 
For each $i\in[k]$, to satisfy the request $\bfv_i$, the union $\cup_{f=1}^{p} B_{i,f}^{f}$ is taken, since for each $f\in[p]$ the subset $B_{i,f}^{f}$ can satisfy the request $\bfv_i^f$.

For each $f_1,f_2\in[p]$, $i_1,i_2\in[k]$ and $j_1,j_2\in[p]$, $B_{i_1,f_1}^{j_1}$ and $B_{i_2,f_2}^{j_2}$ have disjoint subsets of buckets if $i_1\neq i_2$ or $j_1\neq j_2$, because $B_{i_1}^{j_1}$ and $B_{i_2}^{j_2}$ have disjoint subsets of buckets if $i_1\neq i_2$ or $j_1\neq j_2$. Thus, for any $i\neq j\in[k]$, $\cup_{f=1}^{p} B_{i,f}^{f}$ and $\cup_{f=1}^{p} B_{j,f}^{f}$ have disjoint subsets of buckets.



It remains to show that we read at most $\ell$ cells from each bucket. For any $\bfv_i,i\in[k]$ it is clear that if the recovering set $B_{i,f_1}^{j}$ was used then $f_1 = j$, which implies that the recovering sets $B_{i,f_2}^{j}$ for each $f_2 \neq f_1$ was not used. Thus, the recovering sets that were
used to satisfy $\bfv_i$ have disjoint subsets of buckets. Thus, each bucket can appear in at most one of these recovering sets, and it is known that each one of these subsets uses at most $\ell$ cells from each bucket from the properties of the code $\cC$. 

The last claim in the lemma holds by setting $p=t$ and $t=1$. 
\end{IEEEproof}

Another general construction is stated in the next theorem.
\begin{theorem}\label{theorem:t+1Tot}
For any positive integers, $s,k,t,t_0$, and $\ell$, $FB_{t,\ell}(s,k) \leq m +m_0$, where $m = FB_{t+t_0,\ell}(s,k)$ and $m_0 = FB_{t,\ell}(m\cdot t_0,k)$.
\end{theorem}
\begin{IEEEproof}
Let $\cC_1,\cC_2$ be an $(s,k,m,t+t_0,\ell),(m \cdot t_0,k,m_0,t,\ell)$ functional batch array code, respectively. We construct an $(s,k,m+m_0,t,\ell)$ functional batch array code $\cC$ by using the codes $\cC_1,\cC_2$. First, the $s$ information bits are encoded using the encoder function of the code $\cC_1$ to get a $(t+t_0)\times m$ array $A$. Then, the $t_0 \cdot m$ bits in the last $t_0$ rows of $A$ are encoded into a $t\times m_0$ array $B$ using the encoder function of the code $\cC_2$.
The code $\cC$ will be represented by a $t\times (m+m_0)$ array, where
the first $m$ buckets (columns) will be the first $t$ rows of the array $A$ and the last $m_0$ buckets will be the array $B$.

Let $R = \{\bfv_1,\ldots,\bfv_k\}$ be a multiset request of size $k$, where $\bfv_i,i\in[k]$ is a binary vector of length $s$ that represents the $i$-th request. Denote by $\{E_1,\ldots,E_k\}$ the $k$ recovering sets that are obtained by using the decoding function of the code $\cC_1$ with the request $R$. For each $i\in[k]$, assume that $|E_i| = p_i$ and denote $E_i = \{(h_{i,1},\bfu_{i,1}), \ldots, (h_{i,p_i},\bfu_{i,p_i})\}$ where for each $j\in[p_i]$, $(h_{i,j},\bfu_{i,j})$ is a pair of a bucket $h_{i,j}$ with a vector $\bfu_{i,j}$ of length $t+t_0$ that indicates the cells which are read from the bucket $h_{i,j}$.
For each $i\in[k]$ and $j\in[p_i]$, let $\bfu'_{i,j}$ be the vector with the last $t_0$ entries of $\bfu_{i,j}$ and let $R'_{i,j}$ be the sum of the bits in the cells that indicated by $\bfu'_{i,j}$.

Let $R' = \{\sum_{j=1}^{p_1} R'_{1,j},\dots,\sum_{j=1}^{p_k} R'_{k,j}\}$ be a multiset request of size $k$.
Denote by $\{F_1,\ldots,F_k\}$ the $k$ recovering sets that are obtained by using the decoding function of the code $\cC_2$ with the multiset request $R'$.
To satisfy $\bfv_i$, the code $\cC$ can use the recovering set $F_i \cup E'_i$, where $E'_i = \{(h_{i,1},\bfu''_{i,1}), \ldots, (h_{i,k},\bfu''_{i,k})\}$ where for each $j\in[k]$, $\bfu''_{i,j}$ is the vector with the first $t$ entries of $\bfu_{i,j}$.

It remains to show that at most $\ell$ cells are read from each bucket. Each $\bfv_i,i\in[k]$ has a recovering set $F_i\cup E'_i$, where the recovering set $F_i$ of $\cC_2$ uses at most $\ell$ cells from each bucket from the property of the code $\cC_2$. Also, the recovering set $E_i$ of $\cC_1$ uses at most $\ell$ cells from each bucket from the property of the code $\cC_1$. Thus, $E'_i$ also uses at most $\ell$ cells.
\end{IEEEproof}

Note that a similar statement can hold for functional PIR array code, where for any positive integers $s,k,t,t_0$, and $\ell$, $FP_{t,\ell}(s,k) \leq m +m_0$, where $m = FP_{t+t_0,\ell}(s,k)$ and $m_0 = FB_{t,\ell}(m\cdot t_0,k)$.

\subsection{Constructions based upon Covering Codes}
In this section it is shown how covering codes are used to construct array codes. Denote by $d_H(\bfx,\bfy)$ the Hamming distance between two vectors $\bfx,\bfy$, and denote by $w_H(\bfx)$ the Hamming weight of $\bfx$. Also define $\langle \bfx, \bfy\rangle$ as the inner product of the two vectors $\bfx,\bfy$.
Next we remind the definition of covering codes~\cite{CHLL97}. 
\begin{definition}
Let $n\geq1$, $R\geq0$ be integers. A code $\cC \subseteq \mathbb{F}_q^n$ is called an \textbf{$R$-covering code} if for every word $\bfy\in \mathbb{F}_q^n$ there is a codeword $\bfx\in \cC$ such that $d_{H}(\bfx,\bfy)\leq R$.
The notation $[n,k,R]_q$ denotes a linear code over $\mathbb{F}_q$ of length $n$, dimension $k$, and covering radius $R$. The value $g[n,R]_q$ denotes the smallest dimension of a linear code over $\mathbb{F}_q$ with length $n$ and covering radius $R$. The value $h[s,R]_q$ is the smallest length of a linear code over $\mathbb{F}_q$ with covering radius $R$ and redundancy $s$. In case $q=2$ we will remove it from these notations.

\end{definition}

The following property is well known for linear covering codes; see e.g.~\cite[Th. 2.1.9]{CHLL97}.
\begin{property}\label{property:covering}
For an $[n,k,R]$ linear covering code with some parity check matrix $H$, every syndrome vector $s \in \Sigma^{n-k}$ can be represented as the sum of at most $R$ columns of $H$.
\end{property}

The connection between linear codes and functional batch array codes is established in the next theorem. 
\begin{theorem}\label{theorem:CoveringToBucket}
Let $\cC$ be a $[t,t-s,\ell]$ linear covering code. Then, there exists an $(s,1,1,t,\ell)$ functional batch array code. In particular, $FB_{t,\ell}(t-g[t,\ell],1) = 1$.
\end{theorem}
\begin{IEEEproof}
Let $\bfx = (x_1,\ldots,x_s)$ the vector of dimension $1\times s$ with the $s$ information bits, and let $H$ be a parity check matrix of the code $\cC$, with dimension $s\times t$. We construct an $(s,1,1,t,\ell)$ functional batch array code $\cC'$ by taking each entry of the vector $\bfc = (\bfx H)^{\intercal}$ as a cell in the code. The dimension of $\bfc$ is $t\times 1$, and thus, we get one bucket with $t$ cells where each cell has a linear combination of the $s$ information bits.


Let $\bfu\in\Sigma^s$ be a request which represents the linear combination $\langle \bfu, \bfx\rangle$ of the $s$ information bits. 
From Property~\ref{property:covering}, we know that there exists a vector $\bfy\in\Sigma^t$ such that $\bfy\cdot H^{\intercal} = \bfu$, where $w = w_H(\bfy) \leq \ell$. Let {$\cA = \{i : i\in[t], y_{i}~=~1\}$}, where $y_i$ is the entry number $i$ of $\bfy$. Thus, $\langle \bfu, \bfx\rangle = \bfu \cdot \bfx^{\intercal} = \bfy \cdot H^{\intercal}\cdot \bfx^{\intercal} = \bfy \cdot \bfc = \sum_{i\in\cA} c_i$, where $c_i$ is the entry number $i$ of $\bfc$. Therefore, to satisfy the request $\langle \bfu, \bfx\rangle$ we should read $\left|\cA\right| = w \leq \ell$ cells from the code $\cC'$.  


Recall that $g[t,\ell]$ is the smallest dimension of a linear code with length $t$ and covering radius $\ell$. Thus, there exists a $[t,g[t,\ell],\ell]$ linear covering code. We get that there exists a $(t-g[t,\ell],1,1,t,\ell)$ functional batch array code, which implies that $FB_{t,\ell}(t-g[t,\ell],1) = 1$.
\end{IEEEproof}

Theorem~\ref{theorem:CoveringToBucket} holds also for functional PIR array code and thus the following results are derived.
\begin{corollary}\label{cor:coveringgtell}
Let $s,k,t$ and $\ell$ be positive integers. Then,
\begin{enumerate}
\item\label{cor:coveringk} $FP_{t,\ell}(s,k) \leq FB_{t,\ell}(s,k) \leq k\cdot \left\lceil \frac{s}{t - g[t,\ell]} \right\rceil$.
\item\label{cor13:part2} $FP_{t+t_0,\ell}(s,k) \leq FP_{t,t}(s,k)$, where $t_0 = g[t+t_0,\ell]$. Also works for FB.
\item\label{cor:coveringt+t_0} $FP_{t,\ell}(s,k) \leq FB_{t,\ell}(s,k) \leq k\cdot \left(\left\lceil \frac{s}{\alpha} \right\rceil + 1\right)$, where $\left\lceil \frac{s}{\alpha} \right\rceil \leq t-g[t,\ell]$, and $\alpha = (t+1)-g[(t+1), \ell]$.

\end{enumerate}
\end{corollary}

The third claim of Corollary~\ref{cor:coveringgtell} is derived from Theorem~\ref{theorem:CoveringToBucket} and Theorem~\ref{theorem:t+1Tot}.

\subsection{The Cases of $k = 1,2$}
Even though the cases of $k=1,2$ are the most trivial ones when the codewords are vectors, they are apparently not easily solved for array codes. In this section we summarize some of our findings on these important and interesting cases.
\begin{theorem}\label{theorem:ArrayCodek1}
For each $s,t,\ell$ positive integers:
\begin{enumerate}
    
    \item\label{theorem:Boundk1} $FP_{t,\ell}(s,1) \geq \left\lceil \frac{s}{\log_2\left(\sum_{i=0}^{\ell} {t \choose i}\right)} \right\rceil $.
        
    \item\label{theorem:ArrayCodek1tt} $FP_{t, t}(s,1) =  \left\lceil \frac{s}{t} \right\rceil$.
    
    \item\label{theorem:ArrayCodek1logt1} $FP_{t,1}(\lfloor \log_2(t+1)\rfloor,1) = 1$ and 
    $\left\lceil \frac{s}{\log_2(t+1)}\right\rceil \leq FP_{t,1}(s,1) \leq  \left\lceil \frac{s}{\lfloor \log_2(t+1)\rfloor}\right\rceil$.
    
    \item\label{thk1:part4} $FP_{t, \alpha \cdot t}(s,1) \leq  \left\lceil \frac{s}{t-g[t, \alpha \cdot t]} \right\rceil$, where $0 < \alpha < 1$.
        
    \item\label{th11:5} $FP_{t, t/2}(s,1) =  \frac{s}{t} + 1$, where $t$ is even, $\frac{s}{t}$ is integer, and $\frac{s}{t} \leq t-1$.

\end{enumerate}
\end{theorem}
\begin{IEEEproof}
\begin{enumerate}

	 \item From corollary~\ref{theorem:LBFP2}.
	
	\item The lower bound over $FP_{t,t}(s,1)$ is obtained by using the lower bound from the first claim of this theorem, $FP_{t,t}(s,1) \geq \left\lceil \frac{s}{\log_2\left(\sum_{i=0}^{t} {t \choose i}\right)} \right\rceil =  \left\lceil \frac{s}{t} \right\rceil$.
	The upper bound can be verified by showing that there exists an $(s,1,\left\lceil\frac{s}{t} \right\rceil,t,t)$ functional PIR array code. There are $t$ cells in each buckets. Then, in order to write all the $s$ information bits there is a need to $\lceil \frac{s}{t}\rceil$ buckets. Each request is a linear combination of the $s$ information bits. Thus, each request can be satisfied by reading the information bits which included in the request. It was shown that $FP_{t,t}(s,1) \geq \left\lceil \frac{s}{t}\right\rceil$ and there exists an $(s,1,m,t,t)$ functional PIR array code. Therefore, $FP_{t,t}(s,1) = \left\lceil \frac{s}{t}\right\rceil$.
			
	\item  A $(\lfloor \log_2(t+1)\rfloor,1,1,t,1)$ functional PIR array code $\cC$ can be obtained by writing all the $2^{\lfloor \log_2(t+1)\rfloor - 1} \leq t$ linear combinations of the information bits in at most $t$ cells of one bucket. Each request is a linear combination of the information bits, and hence, for each request there exists a cell in the bucket that satisfies it. Thus, the appropriate cell can satisfy the request. The minimum number of buckets is $1$. Thus, $FP_{t,1}(\lfloor \log_2(t+1)\rfloor,1) = 1$. The lower bound over $FP_{t,1}(s,1)$ is derived from the first claim of this theorem. Thus $FP_{t,1}(s,1) \geq \left\lceil \frac{s}{\log_2(\sum_{i=0}^{1}{t \choose i})}\right\rceil = \left\lceil \frac{s}{\log_2(t+1)}\right\rceil$. The upper bound is shown by using Theorem~\ref{theorem:Basic}\eqref{theorem:partas},
	\begin{align*}
	FP_{t,1}(s,1) &= FP_{t,1}\left(\frac{s}{\lfloor \log_2(t+1)\rfloor} \cdot \lfloor \log_2(t+1)\rfloor,1\right) &\\
	& \leq FP_{t,1}\left(\left\lceil\frac{s}{\lfloor \log_2(t+1)\rfloor}\right\rceil \cdot \lfloor \log_2(t+1)\rfloor,1\right) &\\
	&\leq \left\lceil\frac{s}{\lfloor \log_2(t+1)\rfloor}\right\rceil \cdot FP_{t,1}\left(\lfloor \log_2(t+1)\rfloor,1\right) &\\ 
	&\leq \left\lceil\frac{s}{\lfloor \log_2(t+1)\rfloor} \right\rceil.&
	\end{align*}
	
	\item From Corollary~\ref{cor:coveringgtell}\eqref{cor:coveringk}.
	
	\item The lower bound over $FP_{t,t/2}(s,1)$ can be found using the lower bound from the first claim of this theorem,
	\begin{align*}
	FP_{t,t/2}(s,1) & \geq \left\lceil \frac{s}{\log_2(\sum_{i=0}^{t/2} {t \choose i})} \right\rceil &\\
	&\geq  \left\lceil \frac{s}{\log_2(\sum_{i=0}^{t} {t \choose i}) } \right\rceil +1 &\\
	& =  \left\lceil \frac{s}{t}\right\rceil +1.
	\end{align*}
	
	
	
	
	For the upper bound, from Corollary~\ref{cor:coveringgtell}\eqref{cor:coveringt+t_0} we get that $FP_{t,t/2}(s,1) \leq \left\lceil \frac{s}{(t+1)-g[(t+1), t/2]} \right\rceil + 1$. 
	Since $g[t+1,t/2] = 1$, then $FP_{t,t/2}(s,1) \leq s/t + 1$. Lastly we need to show that $\left\lceil \frac{s}{(t+1)-g[(t+1), t/2]} \right\rceil \leq t-g[t,\ell]$ in order to use Corollary~\ref{cor:coveringgtell}\eqref{cor:coveringt+t_0}. Since $s/t \leq t-1$, it is derived that $\left\lceil \frac{s}{(t+1)-g[(t+1), t/2]} \right\rceil = \frac{s}{t} \leq t-1 = t - g[t+1,t/2] = g[t,t/2]$. Thus, $FP_{t,t/2} = \frac{s}{t} + 1$.
	
	
	
\end{enumerate}
\end{IEEEproof}

\begin{example}
In this example we demonstrate the construction of a $(12,1,4,4,2)$ functional PIR array code according to Theorem~\ref{theorem:ArrayCodek1}\eqref{th11:5}. The construction is given in Table~\ref{tableEx1}. It can be verified that $FP_{4,2}(12,1) = 4$.
Note that in this example and in the rest of the paper the notation $x_{i_1}x_{i_2}\cdots x_{i_h}$ is a shorthand to the summation $x_{i_1}+x_{i_2}+\cdots + x_{i_h}$.
\begin{table}
\caption{$(12,1,4,4,2)$ functional PIR array code}\label{tableEx1}
\begin{center}
\begin{tabular}{ |c|c|c|c| }
 \hline
 1 & 2 & 3 & 4\\
 \hline
 \hline
 $x_1$ & $x_5$ & $x_9$ & $x_1x_2x_3x_4$ \\ 
 \hline
 $x_2$ & $x_6$ & $x_{10}$ & $x_5x_6x_7x_8$ \\
 \hline
 $x_3$ & $x_7$  & $x_{11}$ &  $x_9x_{10}x_{11}x_{12}$ \\
 \hline
 $x_4$ & $x_8$  & $x_{12}$ &  $x_1x_2\cdots x_{12}$ \\ 
 \hline
\end{tabular}
\end{center}
\end{table}
\end{example}

An improvement for the case of $\ell=1$ is proved in the following theorem. 
\begin{theorem}\label{theorem:FPIRell1}
For any positive integers $s_1,s_2,$ and $t$, $$FP_{t,1}(s_1+s_2,1) \leq \left\lceil\frac{s_1}{\left\lfloor \log_2(t+1) \right\rfloor}\right\rceil + 1,$$ where 
$2^{s_2} -1 \hspace{-0.5ex}\leq \left(\left\lceil\frac{s_1}{\left\lfloor \log_2(t+1) \right\rfloor}\right\rceil+1\right) (t - (2^{\lfloor \log_2(t+1) \rfloor}-1)).$

\end{theorem}

\begin{IEEEproof}
A construction of an $(s_1+s_2,1,m,t,1)$ functional PIR array code for $m = \left\lceil\frac{s_1}{\left\lfloor \log_2(t+1) \right\rfloor}\right\rceil + 1$ is presented. The first $s_1$ information bits are divided into $m-1$ parts, where $h_i,i\in[m-1]$ is the size of part $i$, and $h_i \leq \lfloor \log_2(t+1) \rfloor$. Then, all the linear combinations of part $i\in[m-1]$ are written in the $i$-th bucket, 
so in each of the first $m-1$ buckets there are at least $t - (2^{\lfloor \log_2(t+1) \rfloor}-1)$ empty cells.
In the last bucket, the parity of each of the first $2^{\lfloor \log_2(t+1) \rfloor}-1$ rows is stored. 
Since $2^{s_2} - 1 \leq m \cdot (t - (2^{\lfloor \log_2(t+1) \rfloor}-1))$, each of the $2^{s_2} - 1$ linear combinations of the $s_2$ bits can be written in the empty cells of the $m$ buckets.


Let $\bfv = (\bfv_1,\ldots,\bfv_m)$ be a request such that for any $i\in[m-1]$ the length of $\bfv_i$ is $h_i$, the length of $\bfv_m$ is $s_2$, and for simplicity assume that they are all nonzero. 
The linear combination $\bfv_m$ is satisfied by the cell where it is stored and assume it is in the $j$-th bucket, where $j<m$.
Assume that the cell in the $j$-th bucket where the linear combination $\bfv_j$ is stored is in row $r$.
We read from each bucket $b\in[m-1]$, where $b\neq j$ the cell with the linear combination represented by $\bfv_b + \bfu_b$, where $\bfu_b$ is the vector that represents the cell in bucket $b$ in row $r$, but if $\bfv_b + \bfu_b = \mathbf0$ do not read from bucket $b$. Also, we read the cell in row $r$ from the last bucket. Then, the obtained linear combination is the combination that is represented by $(\bfv_1,\ldots,\bfv_{m-1})$, because $\sum_{1\leq b \leq m, b\neq j} \bfu_b = \bfv_j$ and for each $b\in[m-1]$ where $b\neq j$ we read the linear combination that is represented by $\bfv_b+\bfu_b$ from bucket $b$.

\end{IEEEproof}

For any $t,s_1,s_2$ where $s = s_1+s_2$ and $s_2 \geq \lfloor \log_2(t+1) \rfloor$, the upper bound in Theorem~\ref{theorem:FPIRell1} improves upon the one in Theorem~\ref{theorem:ArrayCodek1}\eqref{theorem:ArrayCodek1logt1} since  $ \left\lceil \frac{s}{\lfloor \log_2(t+1)\rfloor}\right\rceil \geq \left\lceil\frac{s_1}{\left\lfloor \log_2(t+1) \right\rfloor}\right\rceil + 1$.



\begin{example}
In this example the construction of a $(15,1,7,4,1)$ functional PIR array code is demonstrated based on Theorem~\ref{theorem:FPIRell1}. It can be verified that the parameters $t=4, s_1 = 12$ and $s_2 = 3$ satisfy the constraints of Theorem~\ref{theorem:FPIRell1}. 
The construction is given in Table~\ref{tb1}.
The first $s_1 = 12$ information bits are partitioned into $6$ parts, each part of size $2$. All the nonzero linear combinations of part $i,i\in[6]$ are written in the $i$-th bucket with one cell remains empty. The sum of each of the first $3$ rows is written. Now, there are still $7$ empty cells, which are used to store all the nonzero linear combinations of the last $s_2=3$ bits in the empty cells. 
It can be concluded that $FP_{4,1}(15,1)\leq 7$, and from Theorem~\ref{theorem:ArrayCodek1}\eqref{theorem:ArrayCodek1logt1} we get that $FP_{4,1}(15,1) \geq 7$. Thus, $FP_{4,1}(15,1) = 7$. 
\begin{table}
\begin{center}
\caption{$(15,1,7,4,1)$ functional PIR array code}\label{tb1}
\begin{tabular}{ |c|c|c|c|c|c|c| } 
 \hline
 1 & 2 & 3 & 4 & 5 & 6 & 7\\
 \hline
 \hline
 $x_1$ & $x_3$ & $x_5$ & $x_7$ & $x_9$ & $x_{11}$ & $x_1x_3x_5x_7x_9x_{11}$ \\ 
 \hline
 $x_2$ & $x_4$ & $x_6$ & $x_8$ & $x_{10}$ & $x_{12}$ & $x_2x_4x_6x_8x_{10}x_{12}$ \hspace{-2ex} \\
 \hline
 $x_1x_2$ & $x_3x_4$ & $x_5x_6$ & $x_7x_8$ & $x_9x_{10}$ & $x_{11}x_{12}$ & $x_1\cdots x_{12}$ \\
 \hline
 $x_{13}$ & $x_{14}$ & $x_{15}$ & $x_{13}x_{14}$ & $x_{13}x_{15}$ & $x_{14}x_{15}$ & $x_{13}x_{14}x_{15}$ \\ 
 \hline
\end{tabular}
\end{center}
\end{table}
\end{example}

Lastly, we report on several results for $k=2$.

\begin{table}
\begin{center}
\caption{$(8,2,7,2,2)$ functional PIR array code}\label{tb82722}
\begin{tabular}{ |c|c|c|c|c|c|c|} 
 \hline
 1 & 2 & 3 & 4 & 5 & 6 & 7\\
 \hline
 \hline
 $x_1$ & $x_2$ & $x_1x_2$ & $x_5$ & $x_6$ & $x_5x_6$ & $x_1x_2x_5x_6$ \\ 
 \hline
 $x_3$ & $x_4$ & $x_3x_4$ & $x_7$ & $x_8$ & $x_7x_8$ & $x_3x_4x_7x_8$ \\ 
 \hline
\end{tabular}
\end{center}
\end{table}

\begin{theorem}\label{theorem:FB2282} 
$6 \leq FB_{2,2}(8,2) \leq 7$.
\end{theorem}
\begin{IEEEproof}
The lower bound is obtained from Theorem~\ref{theorem:LBFB}. The upper bound is verified using the construction which appears in Table~\ref{tb82722}, i.e., the construction gives an $(8,2,7,2,2)$ functional batch array code.
There are 8 information bits, 7 buckets, each one with 2 cells, and we show that this code can satisfy each multiset request of size 2. Let $\cS_1 = \{x_1,x_2,x_3,x_4\}$ be a set of the first $4$ information bits and $\cS_2 = \{x_5,x_6,x_7,x_8\}$ be a set of the last $4$ information bits. Let $R = \{\bfv_1,\bfv_2\}$ be a multiset request of size $2$, where $\bfv_1$ and $\bfv_2$ are vectors of size $8$. For each $i\in[2]$, $\bfv_i = (\bfv_i^1,\bfv_i^2)$ where $\bfv_i^j,j\in[2]$ is a vector of length $4$ that represents a linear combination of the bits in $\cS_j$.
The possible linear combinations of $\cS_1$ are divided into four different types in the following way. 
\begin{enumerate}
	\item The first type $\cT_1$ includes the vectors that can be satisfied by using only one bucket from the buckets $1-3$.
	
	\item The second type $\cT_2$ includes any vector $\bfu$ that satisfies the following constraint. The vectors $\bfu+(1,1,0,0)$ and $\bfu + (0,0,1,1)$ can be satisfied by one bucket from buckets $1-3$. (The vector (1,1,0,0) represents the linear combination $x_1+x_2$.)
	
	
	\item The third type $\cT_3$ includes any vector $\bfu$ that satisfies the following constraint. The vectors $\bfu+(1,1,1,1)$ and $\bfu + (1,1,0,0)$ can be satisfied by one bucket from the buckets $1-3$.
	
	\item The fourth type $\cT_4$ includes any vector $\bfu$ that satisfies the following constraint. The vectors $\bfu+(1,1,1,1)$ and $\bfu + (0,0,1,1)$ can be satisfied by one bucket from the buckets $1-3$.
	
\end{enumerate}

These four types are disjoint and their union covers all the nonzero linear combinations of $\cS_1$. From the symmetry of the first four information bits and the last four bits, the linear combinations of $\cS_2$ are divided in the same way.
It is possible to see that every two buckets from buckets $1-3$ can satisfy each possible linear combination of the first four bits. In the same way, every two buckets from buckets $4-6$ can satisfy each possible linear combination of the last four bits. Also, the last bucket can satisfy each vector $(\bfu,\bfu)$, where $\bfu \in \{(1,1,0,0),(0,0,1,1),(1,1,1,1)\}$.

If one of the vectors $\{\bfv_1^1,\bfv_2^1\}$ is included in $\cT_1$ (assume it is $\bfv_1^1$) and one of the vectors $\{\bfv_1^2,\bfv_2^2\}$ is included in $\cT_1$ (assume it is $\bfv_1^2$), then these two vectors can be satisfied by one bucket from $1-3$ and one bucket from $4-6$. Then the remaining two buckets of $1-3$ can satisfy $\bfv_2^1$ and the remaining two buckets of $4-6$ can satisfy $\bfv_2^2$. Therefore, in this case the request $R$ is satisfied by disjoint sets.

If there exist $2\leq q_1,q_2 \leq 4$ where $\bfv_1^1\in \cT_{q_1}$ and $\bfv_1^2\in \cT_{q_2}$. Then, there exists a vector $\bfu'$ where $\bfv_1^1 + \bfu'$ can be satisfied by one bucket from  buckets $1-3$ and $\bfv_1^2 + \bfu'$ can be satisfied by one bucket from  buckets $4-6$. Thus, the code can satisfy $\bfv_1^1$ and $\bfv_1^2$, that consist the request $\bfv_1$, by one bucket from $1-3$, one bucket from $4-6$, and the last bucket, which satisfies the request $(\bfu',\bfu')$ for each possible $\bfu'$. Then, the remaining two buckets of $1-3$ can satisfy $\bfv_2^1$ and the remaining two buckets of $4-6$ can satisfy $\bfv_2^2$. Similarly, if there exist $2\leq q_1,q_2 \leq 4$ where $\bfv_2^1\in \cT_{q_1}$ and $\bfv_2^2\in \cT_{q_2}$, the code can satisfy the requests $\bfv_1$ and $\bfv_2$ by disjoint sets.

The last case is when $\{\bfv_1^1,\bfv_2^1\}\subseteq \cT_1$ and $\{\bfv_1^2,\bfv_2^2\}\subseteq \cT_q$, where $2\leq q\leq 4$ (or $\{\bfv_1^1,\bfv_2^1\}\subseteq \cT_q$ and $\{\bfv_1^2,\bfv_2^2\}\subseteq \cT_q$).
In the beginning we satisfy $\bfv_1^1$ by one bucket from $1-3$. Then, take a vector $\bfu''$, such that $\bfv_2^2 + \bfu''$ can be satisfied by one bucket, denote it by $b_1$. The vector $\bfv_2^1 + \bfu''$ can be satisfied by the remaining two buckets from $1-3$, denote them by $b_2,b_3$. Then, the request $R_2 = \{\bfv_2^1,\bfv_2^2\}$ can be satisfied by $\{b_1,b_2,b_3,7\}$ (where $7$ is the last bucket). Lastly, the request $\bfv_1^2$ can be satisfied by the remaining two buckets from $4-6$. Thus, we can conclude that there exists $2$ recovering sets for each possible request, and hence, $FB_{2,2}(8,2) \leq 7$.
\end{IEEEproof}

The result in Theorem~\ref{theorem:FB2282} can be generalized to different values of $s$.
\begin{corollary}\label{cor:FB22s2}
$\log_{7}(2^{s-1}\cdot (2^s - 1)) \leq FB_{2,2}(s,2) \leq 7\cdot\left\lceil \frac{s}{8} \right\rceil.$
\end{corollary}
\begin{IEEEproof}
The upper bound is derived from \Tref{theorem:FB2282}, and \Tref{theorem:Basic}\eqref{theorem:partas}. The lower bound is obtained from \Tref{theorem:LBFB}, where $FB_{2,2}(s,2) \geq m$ where $m$ is the smallest positive integer such that $ \sum_{i=2}^{m} {m \choose i} \cdot {i \brace 2}  \cdot \left(\sum_{j=1}^{2} {2 \choose j} \right)^{i} \geq  {2^s \choose 2}$. It is known that ${i \brace 2} = 2^{i-1} - 1$. Thus, $\sum_{i=2}^{m} {m \choose i} \cdot (2^{i-1} - 1)  \cdot 3^{i} \geq  2^{s-1}\cdot (2^s - 1)$. For each $i\geq 2$, $(2^{i-1} -1)\cdot 3^i \leq 6^i$. Hence, it must hold that $\sum_{i=0}^{m} {m \choose i} \cdot 6^i \geq \sum_{i=2}^{m} {m \choose i} \cdot 6^i \geq 2^{s-1}\cdot (2^s - 1)$. From the equality $\sum_{i=0}^{m} {m \choose i} \cdot x^i = (x+1)^m$, we get that $\sum_{i=0}^{m} {m \choose i} \cdot 6^i = 7^m \geq 2^{s-1} \cdot (2^s - 1)$. Thus, $FB_{2,2}(s,2) \geq m \geq \log_{7}(2^{s-1} \cdot (2^s - 1))$.
\end{IEEEproof}

According to Corollary~\ref{cor:FB22s2}, we get that for $s$ large enough $\log_{7}(2^{s-1}\cdot (2^s - 1)) = \log_{7} (2^{s-1}) + \log_{7} (2^{s}-1) \approx (s-1)\cdot \log_7(2) + s \cdot \log_7(2) = (2s-1) \cdot \log_7(2) \approx 0.71s \lesssim FB_{2,2}(s,2) \leq \left\lceil \frac{7s}{8} \right\rceil.$

In addition, the result in Theorem~\ref{theorem:FB2282} can be modified to different value of $t$.
\begin{corollary}\label{cor:FB3182}
$6 \leq FB_{3,1}(8,2) \leq 7$.
\end{corollary}
\begin{IEEEproof}
The lower bound is obtained from Theorem~\ref{theorem:LBFB}.
The upper bound is verified by Theorem~\ref{cor:coveringgtell}\eqref{cor13:part2}, where $FB_{3,1}(8,2) \leq FB_{2,2}(8,2) \leq 7$.
\end{IEEEproof}

\section{Specific Constructions of Array Codes}\label{sec:cons_spec}
In this section we discuss three constructions of array codes.

\subsection{Construction $A$}


We start with a construction given in~\cite[Th.20]{FVY15}, where it was proved in~\cite[Th.10]{CKYZ19} that this construction gives a PIR array code for any integer $t\geq2$. We study how it can be used also as batch and functional PIR array codes for $t=2$. 
First, the construction for the general case is presented.

\begin{construction}\label{cons10}
Let $t \geq 2$ be a fixed integer. The number of information bits is $s = t(t + 1)$, the number of cells in each bucket (the number of the rows) is $t$. The number of buckets is $m=m'+m''$, where $m' = {t(t+1) \choose t}$, and $m'' = {t(t+1) \choose t+1}/t$. In the first $m'$ buckets all the tuples of $t$ bits out of the $t(t + 1)$ information bits are stored, which needs ${t(t+1) \choose t}$ buckets. In the last $m''$ buckets we store all possible summations of $t + 1$ bits, such that each one of the $t(t+1)$ bits appears in exactly one summation in every bucket (in each summation there are $t+1$ bits and there are $t$ rows). There are ${t(t+1) \choose t+1}$ such summations and since there are $t$ rows then $t$ summations can be stored in each bucket, so the number of buckets of this part is $m'' = {t(t+1) \choose t+1}/t$.
\end{construction}

For any integer $t \geq 2$ denote the code that is obtained from Construction~\ref{cons10} by $\cC^A_t$.
Construction~\ref{cons10} for the case of $t=2$ is demonstrated in Table~\ref{ex_cons10}. 

\begin{table*}
\begin{center}
\caption{Construction~\ref{cons10} for $t=2$}\label{ex_cons10}
\begin{tabular}{ |c|c|c|c|c|c|c|c|c|c|c|c|c|c|c| }
 \hline
 1&2&3&4&5&6&7&8&9&10&11&12&13&14&15 \\ 
 \hline
 \hline
 $x_1$ & $x_1$ & $x_1$ & $x_1$ & $x_1$ & $x_2$ & $x_2$ & $x_2$ & $x_2$ & $x_3$ & $x_3$ & $x_3$ & $x_4$ & $x_4$ & $x_5$ \\ 
 \hline
 $x_2$ & $x_3$ & $x_4$ & $x_5$ & $x_6$ & $x_3$ & $x_4$ & $x_5$ & $x_6$ & $x_4$ & $x_5$ & $x_6$ & $x_5$ & $x_6$ & $x_6$ \\ 
\hline
\end{tabular}
\end{center}

\begin{center}
\begin{tabular}{ |c|c|c|c|c|c|c|c|c|c| } 
 \hline
 16&17&18&19&20&21&22&23&24&25 \\ 
 \hline
 \hline
$x_1x_2x_3$ & $x_1x_2x_4$ & $x_1x_2x_5$ & $x_1x_2x_6$ & $x_1x_3x_4$ & $x_1x_3x_5$ & $x_1x_3x_6$ & $x_1x_4x_5$& $x_1x_4x_6$& $x_1x_5x_6$\\ 
\hline
$x_4x_5x_6$ & $x_3x_5x_6$ & $x_3x_4x_6$ & $x_3x_4x_5$ & $x_2x_5x_6$ & $x_2x_4x_6$ & $x_2x_4x_5$ & $x_2x_3x_6$& $x_2x_3x_5$& $x_2x_3x_4$\\ 
 
 \hline
\end{tabular}
\end{center}
\end{table*}

Now we want to show that the code $\cC^A_2$ is a $(6,15,25,2,2)$ batch array code, by using several properties which are proved in the following three lemmas.
For each $i\in[6]$, denote by $\cF_i\subseteq [15]$ the subset of buckets from the first $15$ buckets, that have a cell with the singleton $x_i$. It holds that for any $i\in[6]$, $|\cF_i| = 5$, and for any different $i,j\in[6]$, $|\cF_i \cap \cF_j| = 1$. 
Assume that every multiset request $R$ of size $k=15$ is represented by a vector $(k_1,\ldots,k_6)$, where $k_i$ indicates the number of times $x_i$ appears in the multiset request and $k_1 \geq \cdots \geq k_6$.

\begin{lemma}\label{lemma:BEx9Aux1}
For any multiset request $(k_1,\ldots,k_6)$ of size $k=15$, the code $\cC^A_2$ can satisfy all the requests of bits $x_3,x_4,x_5,x_6$ by using only the first $15$ buckets.
\end{lemma}

\begin{IEEEproof}
The proof is divided into the following cases according to number of different information bits that appear in the request.

\noindent
{\bf Case 1:} If $k_3 = 0$, then none of the bits $x_3,x_4,x_5,x_6$ is requested and the property clearly holds. 

\noindent
{\bf Case 2:} If $k_4 = 0$, then it necessarily holds that $k_3\leq 5$. Assume by contradiction that $k_3>5$. Then, it holds that $k_1\geq k_2 > 5$, and hence, $k = k_1+k_2+k_3 > 15$, which is a contradiction. Thus $k_3 \leq 5$ and the code can use $k_3$ buckets from $\cF_3$.

\noindent
{\bf Case 3:} If $k_5 = 0$, then it necessarily holds that $k_4\leq k_3 \leq 4$. Assume by contradiction that $k_4>4$. Then, it holds that $k_1\geq k_2 \geq k_3 > 4$, and hence, $k = k_1+k_2+k_3+k_4 > 15$, which is a contradiction. Assume by contradiction that $k_3>4$, when $k_4 \geq 1$. Then, it holds that $k_1\geq k_2 > 4$, and hence, $k = k_1+k_2+k_3+k_4 > 15$, which is a contradiction. Thus $k_3 \leq 4$ and the code $\cC^A_2$ can satisfy the bit requests of $x_3$ by taking $k_3$ buckets from $\cF_3$. Then the code $\cC^A_2$ can satisfy the bit requests of $x_4$ by taking $k_4\leq 4$ buckets from $\cF_4 \setminus (\cF_4 \cap \cF_3)$, where $|\cF_4 \setminus (\cF_4 \cap \cF_3)| = 4$.

\noindent
{\bf Case 4:} If $k_6 = 0$, then it necessarily holds that $k_5\leq k_4 \leq 3$ and $k_3 \leq 4$. Assume by contradiction that $k_5>3$. Then, it holds that $k_1\geq k_2 \geq k_3 \geq k_4 \geq k_5 > 3$, and hence, $k = k_1+k_2+k_3+k_4+k_5 > 15$, which is a contradiction. Assume by contradiction that $k_4>3$, when $k_5 \geq 1$. Then, it holds that $k_1\geq k_2 \geq k_3 \geq k_4 > 3$, and hence, $k = k_1+k_2+k_3+k_4+k_5 > 15$, which is a contradiction. Assume by contradiction that $k_3>4$, when $k_5 + k_4 \geq 2$. Then, it holds that $k_1\geq k_2 \geq k_3 > 4$, and hence, $k = k_1+k_2+k_3+k_4+k_5 > 15$, which is a contradiction. Thus, $k_3 \leq 4$ and the code $\cC^A_2$ can satisfy the bit requests of $x_3$ by taking $k_3$ buckets from $\cF_3$. Also, $k_4 \leq 3$, then the code $\cC^A_2$ can satisfy the bit requests of $x_4$ by taking $k_4$ buckets from $\cF_4 \setminus (\cF_4 \cap \cF_3)$. Lastly, the code $\cC^A_2$ can satisfy the bit requests of $x_5$ by taking $k_5\leq 3$ buckets from $\cF_5 \setminus ((\cF_5 \cap \cF_4) \cup (\cF_5 \cap \cF_3))$, where $|\cF_5 \setminus ((\cF_5 \cap \cF_4) \cup (\cF_5 \cap \cF_3))| = 3$.

\noindent
{\bf Case 5:} If $k_6 > 0$, then it necessarily holds that $k_6\leq k_5 \leq 2$, $k_4 \leq 3$ and $k_3 \leq 4$. Assume by contradiction that $k_6>2$. Then, it holds that $k_1\geq k_2 \geq k_3 \geq k_4 \geq k_5 > 2$, and hence, $k = \sum_{i=1}^{6} k_i > 15$, which is a contradiction. Assume by contradiction that $k_5>2$ when $k_6 \geq 1$. Then, it holds that $k_1\geq k_2 \geq k_3 \geq k_4 > 2$, and hence, $k = \sum_{i=1}^{6} k_i > 15$, which is a contradiction. Assume by contradiction that $k_4>3$ when $k_6 + k_5 \geq 2$. Then, it holds that $k_1\geq k_2 \geq k_3 > 3$, and hence, $k = \sum_{i=1}^{6} k_i> 15$, which is a contradiction. Assume by contradiction that $k_3>4$ when $k_6 + k_5 + k_4 \geq 3$. Then, it holds that $k_1\geq k_2 > 4$, and hence, $k = \sum_{i=1}^{6} k_i> 15$, which is a contradiction. Thus, $1 \leq k_3 \leq 4$ and the code $\cC^A_2$ can satisfy the bit requests of $x_3$ by taking $k_3$ buckets from $\cF_3$. Then the code $\cC^A_2$ can satisfy the bit requests of $x_4$ by taking $k_4\leq 3$ buckets from $\cF_4 \setminus (\cF_4 \cap \cF_3)$. Then the code $\cC^A_2$ can satisfy the bit requests of $x_5$ by taking $k_5\leq 2$ buckets from $\cF_5 \setminus ((\cF_5 \cap \cF_4) \cup (\cF_5 \cap \cF_3))$. Lastly, the code $\cC^A_2$ can satisfy the bit requests of $x_6$ by taking $k_6\leq 2$ buckets from $\cF_6 \setminus ((\cF_6 \cap \cF_5) \cup (\cF_6 \cap \cF_4) \cup (\cF_6 \cap \cF_3))$, where $\left|\cF_6 \setminus ((\cF_6 \cap \cF_5) \cup (\cF_6 \cap \cF_4) \cup (\cF_6 \cap \cF_3))\right| = 2$.
\end{IEEEproof}

\begin{lemma}\label{lemma:BEx9Aux3}
In the code $\cC^A_2$, for any information bit $x_i$ and for any bucket $b_1 \in [15] \setminus \cF_i$, there exists a bucket $b_2, 16\leq b_2\leq 25$ such that $\{b_1,b_2\}$ is a recovering set of $x_i$. In addition, the $\left| [15] \setminus \cF_i \right|$ recovering sets are mutually disjoint. 
\end{lemma}

\begin{IEEEproof}
For any information bit $x_i$, the buckets of $[15]\setminus \cF_i$, are the buckets from the first $m'=15$ buckets that does not include $x_i$. Each bucket $b_1 \in  [15]\setminus \cF_i$ has two singletons $x_{j_1},x_{j_2}$ which are different than $x_i$. From the construction of the code $\cC^A_2$ we know that there exists a bucket $b_2$ from the last $10$ buckets that has the summation $x_i + x_{j_1} + x_{j_2}$. Thus, the subset $\{b_1,b_2\}$ is a recovering set of $x_i$.


We want to show that for any two different buckets $b'_1,b''_1 \in [15]\setminus \cF_i$, the recovering sets $\{b'_1,b'_2\}$ and $\{b''_1,b''_2\}$ of $x_i$ are disjoint. It holds that $\{b'_1\} \cap \{b''_1,b''_2\} = \emptyset$ because it holds that $b'_1 \neq b''_1$ and $b'_1 \neq b''_2$ because $b'_1\in[15]$ but $b''_2 \notin[15]$. In addition, $\{b'_2\} \cap \{b''_1,b''_2\} = \emptyset$ because it holds that $b'_2 \notin [15]$ but $b''_1\in [15]$ and $b'_2 \neq b''_2$ because each bucket in the last $10$ buckets has exactly one summation with $x_i$.
\end{IEEEproof}

For any information bit $x_i,i\in[6]$ denote by $R_{b}^{i}$ the recovering set that uses bucket $b\in[15]$ and can satisfy $x_i$. For example, $R_1^1 = \{1\}$ and $R_{12}^1 = \{12,22\}$.

\begin{lemma}\label{lemma:BEx9Aux2}
For the two information bits $x_1,x_2$, the buckets $\{10,11,\ldots,15\}$ are divided into $3$ pairs, $\cP = \{(10,15)$, $(11,14)$,$(12,13)\}$, such that for any pair $(b_1,b_2)\in \cP$, it holds that $\left|R_{b_1}^{1} \cap R_{b_{2}}^{2}\right| > 0$ and $\left|R_{b_1}^{2} \cap R_{b_{2}}^{1}\right| > 0$.
\end{lemma}

\begin{IEEEproof}
For the first pair, $(10,15)$, it holds that $R_{10}^1 = \{10,20\}, R_{10}^2 = \{10,25\},R_{15}^1 = \{15,25\}$, and $R_{15}^2 = \{15,20\}$. Then, it holds that $\left|R_{10}^{1} \cap R_{15}^{2}\right| $ $=  | \{10,20\} \cap$ $\{15,20\}| > 0$ and $\left|R_{10}^{2} \cap R_{b_15}^{1}\right| = |\{10,25\} \cap$ $\{15,25\}| > 0$. Similarly, the claim holds also for the pairs $(11,14)$ and $(12,13)$.
\end{IEEEproof}

Now, we are ready to show that the code $\cC^A_2$ is a $(6,15,25,2,2)$ batch array code.

\begin{theorem}\label{theorem:BExample9}
The code $\cC^A_2$ is a $(6,15,25,2,2)$ batch array code. In particular, $B_{2,2}(6,15) = 25$.
\end{theorem}

\begin{IEEEproof}
The lower bound is derived from Theorem~\ref{theorem:PIRLB}\eqref{theorem:part4}, $B_{2,2}(6,15) \geq \frac{30\cdot 6 \cdot7}{(4)^2 + 36 - 4 + 4} > 24$.
The upper bound is derived from the code $\cC^A_2$. Let $(k_1,\ldots,k_6)$ be a multiset request of size $k = 15$. The first step is to satisfy all the requests of bits $x_3,x_4,x_5,x_6$ according to Lemma~\ref{lemma:BEx9Aux1} by using only the first $m' = 15$ buckets. Then, the remaining requests are of the bits $x_1,x_2$. Denote by $\alpha_1,\alpha_2$ the number of the remaining buckets from the first $m'=15$ buckets that include $x_1,x_2$ as singleton, but not both of them, respectively. Then, take $\min\{k_2,\alpha_2\}$ buckets as a recovering sets of $x_2$ and take $\min\{k_1,\alpha_1\}$ buckets as recovering sets of $x_1$. The first bucket which contains the singletons $x_1,x_2$ is not used yet. 
Denote by $r$ the number of bit requests from the multiset request that were satisfied so far. Furthermore, denote by $k'_1,k'_2$ the number of remaining bit requests of $x_1,x_2$, respectively, where $k'_1 = k_1 - \min\{k_1,\alpha_1\}$ and $k'_2 = k_2 - \min\{k_2,\alpha_2\}$. After this step we still have $15-r$ buckets in the first $m'=15$ buckets, including the first bucket and all the last $m''=10$ buckets. Therefore, for $x_1$ and $x_2$ there are $15-r$ possible recovering sets. 

The second step is to satisfy the remaining $15-r$ bit requests from the multiset request. If $k'_1 = 0$ or $k'_2=0$, then it is possible to satisfy them by using the remaining $k-r=15-r$ recovering sets of $x_1$ or $x_2$. 
Otherwise, $k'_1 > 0$ and $k'_2>0$. 
So far we used all the buckets from the set $(\cF_1 \cup \cF_2)\setminus \{1\}$ which is of size $8$ and another $p$ buckets from the subset $\{10,11,\ldots,15\}$. Thus,  $k'_1+k'_2 = 7 - p$. 
Let $\cG \subseteq \{10,11,\ldots,15\}$ be the subset of buckets from $\{10,11,\ldots,15\}$ that were not used in the first step and let $p = 6 - |\cG|$. According to Lemma~\ref{lemma:BEx9Aux3}, there are at least $7-p$ remaining recovering sets for each bit of $\{x_1,x_2\}$, which are the set $\{1\}$ and the sets of $R_b^i$ where $b\in\cG$ and $i \in [2]$. According to Lemma~\ref{lemma:BEx9Aux2}, the buckets $\{10,11,\ldots,15\}$ are divided into $3$ pairs, where the $b$-th bucket is paired with the $(25-b)$-th bucket, for $10\leq b\leq 15$. 
The subset $\cG$ is partitioned into two subsets, $\cU_1 = \{b\in\cG : (25-b) \in \cG\}$ and $\cU_2 = \{b\in\cG : (25-b) \notin \cG\}$. Let $\beta_1 = |\cU_1|$ and $\beta_2 = |\cU_2|$. The following cases are considered.



\noindent
{\bf Case 1:} If $p$ is even and $k'_1$ is even (or $k'_2$ is even). Since $p$ is even, it is deduced that $\beta_2$ is even as well. Assume that $k'_1$ is even, then also $(k'_1 - \beta_2)$ is even. In order to satisfy $x_1$ we can take $\min\{\beta_2,k'_1\}$ recovering sets that use $\min\{\beta_2,k'_1\}$ buckets from $\cU_2$. We can see that $\beta_1 + \beta_2 = 6-p$ and $k'_1 \leq 6-p = \beta_1 + \beta_2$ then $k'_1 - \beta_2 \leq \beta_1$.
If $k'_1>\beta_2$, then we can satisfy the remaining requests of $x_1$ with $(k'_1 - \beta_2)/2$ pairs of buckets from $\cU_1$, where for each bucket $b$ from the $(k'_1 - \beta_2)$ buckets we can take $R^1_b$ as a recovering set for $x_1$. It is possible to show that each recovering set for $x_1$ that uses a bucket from $\cU_2$ intersects with only one recovering set for $x_2$ that uses a bucket from $\cG$. Also, each pair of recovering sets for $x_1$ that uses a pair of bucket from $\cU_1$ intersects with only two recovering sets for $x_2$ that use buckets from $\cG$. Thus, from the $7-p$ recovering sets of $x_2$ it is not possible to use only $\max \{k'_1, \beta_2 + 2\cdot \frac{k'_1 - \beta_2}{2}\} = k'_1$ of them. Thus it is possible to use the remaining $7-p - k'_1 = k'_2$ to satisfy the $k'_2$ requests of $x_2$. The case when $k'_1$ is odd but $k'_2$ is even can be solved similarly while changing between $x_1$ and $x_2$.

\noindent
{\bf Case 2:} If $p$ is odd and $k'_1$ is odd (or $k'_2$ is odd). Then $\beta_2$ is odd. Assume that $k'_1$ is odd, then also $(k'_1 - \beta_2)$ is even and the rest is similar to Case 1.


\noindent
{\bf Case 3:} If $p$ is even and $k'_1,k'_2$ are odd. Then start with satisfying $x_1$ with a recovering set $\{1\}$. Then we still have an even number of remaining requests of $x_1$ that must be satisfied, and the rest is similar to Case 1.

\noindent
{\bf Case 4:} If $p$ is odd and $k'_1,k'_2$ are even. Then start with satisfying $x_1$ with a recovering set $\{1\}$. Then we still have an odd number of remaining requests of $x_1$ that must be satisfied, and the rest is similar to Case 2.

Thus, we can conclude that the code can satisfy each multiset of $15$ information bits, and hence, $B_{2,2}(6,15) = 25$. 
\end{IEEEproof}
In addition it is possible to show that the code $\cC^A_2$ is a $(6,11,25,2,2)$ functional PIR array code.
\begin{theorem}\label{theorem:FPExample9}
The code $\cC^A_2$ is a $(6,11,25,2,2)$ functional PIR array code. In particular, $21 \leq FP_{2,2}(6,11) \leq 25.$
\end{theorem}

\begin{IEEEproof}
The lower bound is obtained from \Tref{theorem:LBFP4}, where $FP_{2,2}(6,11)\geq \frac{2\cdot 11 \cdot 63}{3 + 63} = 21$.
The upper bound can be obtained from the code $\cC^A_2$. Given a request $R$, a linear combination of the information bits, that the code $\cC^A_2$ must satisfy $k=11$ times by disjoint recovering sets.
Because of the symmetry of $x_i,i\in[6]$, it is sufficient to check requests according to their length (number of information bits). Thus, the proof is divided into the following cases according to number of information bits that appear in the request. 

\noindent
{\bf Case 1:} If the request contains one information bit then it is the case of PIR.

\noindent
{\bf Case 2:} If the request contains two information bits, then assume that it is $x_1+x_2$. 
	Then the recovering sets are the following $\{\{1\}$, $\{2,6\}$, $\{3,7\}$, $\{4,8\}$, $\{5,9\}$, $\{16,11\}$, $\{17,10\}$, $\{18,13\}$, $\{19,12\}$, $\{20,25\}$, $\{21,24\}$, $\{22,23\}\}$.

\noindent
{\bf Case 3:} If the request contains three information bits, then assume that it is $x_1+x_2+x_3$. 
Then the recovering sets are the following $\{\{16\}$, $\{1,2\}$, $\{17,10\}$, $\{18,11\}$, $\{19,12\}$, $\{20,7\}$, $\{21,8\}$, $\{22,9\}$, $\{23,5\}$, $\{24,4\}$, $\{25,3\}\}$.

\noindent
{\bf Case 4:} If the request contains four information bits, then assume that it is $x_3+x_4+x_5+x_6$. 
Then the recovering sets are the following $\{\{16,2\}$, $\{17,3\}$, $\{18,4\}$, $\{19,5\}$, $\{20,25\}$, $\{21,24\}$, $\{22,23\}$, $\{10,15\}$, $\{11,14\}$, $\{12,13\}$, $\{6,7,8,9\}\}$.

\noindent
{\bf Case 5:} If the request contains five information bits, then assume that it is $x_2+x_3+x_4+x_5+x_6$. 
Then the recovering sets are the following $\{\{16,1\}$, $\{17,2\}$, $\{18,3\}$, $\{19,4\}$, $\{20,5\}$, $\{21,11\}$, $\{22,12\}$, $\{23,13\}$, $\{24,14\}$, $\{25,15\}$, $\{6,7,8,9\}\}$.

\noindent
{\bf Case 6:} If the request contains all the information bits, that it is $x_1+x_2+x_3+x_4+x_5+x_6$. 
Then the recovering sets are the following $\{\{16\}$, $\{17\}$, $\{18\}$, $\{19\}$, $\{20\}$, $\{21\}$, $\{22\}$, $\{23\}$, $\{24\}$, $\{25\}$, $\{1,10,15\}$, $\{2,8,14\}$, $\{3,9,11\}$, $\{4,7,12\}$, $\{5,6,13\}\}$.
\end{IEEEproof}

\subsection{Construction B}

Next we generalize an example given in~\cite{FVY15} of a PIR code for any integer $r\geq3$ and study how it can be used also as batch array codes. We first present the construction for the general case.

\begin{construction}\label{construction:cons8}
Let $r\geq 3$ be a fixed integer, the number of information bits is $s = r(r+1)$, the number of the buckets is $m = r+1$, and the number of the cells in each bucket is $t = (r-1)r + 1$. The information bits are partitioned into $r+1$ parts each of size $r$, denote by $\cS_i$ the part $i$ of the bits. For each $i\in[r+1]$, write the linear combination $\sum_{j\in\cS_i} x_j$ to bucket $i$. For each $i,i\in[r+1]$ write each one of the subsets of size $r-1$ of $\cS_i$ as singletons in a different bucket other than bucket $i$.
\end{construction}
For any integer $r \geq 3$ denote the code that is obtained from Construction~\ref{construction:cons8} by $\cC^B_r$.
Construction~\ref{construction:cons8} for the case of $r=3$ is demonstrated in Table~\ref{ex_cons8}. It is possible to show that for any $r\geq 3$ the code $\cC^B_r$ is an $(r^2+r,r,r+1,r^2-r+1,r-1)$ PIR array code.
 
\begin{table}
\begin{center}
\caption{Construction~\ref{construction:cons8} for $r=3$}\label{ex_cons8}
\begin{tabular}{ |c|c|c|c| } 
 \hline
 1 & 2 & 3 & 4 \\
 \hline
 \hline
 $x_1x_2x_3$ & $x_1$ & $x_2$ & $x_1$ \\ 
 \hline
 $x_4$ & $x_2$ & $x_3$ & $x_3$ \\
 \hline
 $x_6$ & $x_4x_5x_6$ & $x_4$ & $x_5$\\
 \hline
 $x_7$ & $x_7$ & $x_5$ & $x_6$ \\
 \hline
 $x_8$ & $x_9$ & $x_7x_8x_9$ & $x_8$ \\
 \hline
 $x_{10}$ & $x_{10}$ & $x_{11}$ & $x_{9}$ \\
 \hline
 $x_{11}$ & $x_{12}$ & $x_{12}$ & $x_{10}x_{11}x_{12}$\\
  
 \hline
\end{tabular}
\end{center}
\end{table}

\begin{theorem}\label{theorem:PExample8}
For any integer $r \geq 3$ the code $\cC^B_r$ from Construction~\ref{construction:cons8} is an $(r^2+r,r,r+1,r^2-r+1,r-1)$ PIR array code. In particular, $$\frac{r\cdot(4r^2 + 3r -1)}{4r^2 - r +1} \leq P_{r^2-r+1,r-1}(r^2+r,r) \leq r+1.$$
\end{theorem}

\begin{IEEEproof}
The lower bound can be obtained by using \Tref{theorem:PIRLB}\eqref{theorem:part2}, 
\begin{align*}
P_{r^2-r+1,r-1}&(r^2+r,r) \geq P_{r^2-r+1,r^2-r+1}(r^2+r,r) &\\
& \geq \frac{r\cdot (r^2+r)(4r-1)}{(4r-1)(r^2 - r +1) + (2r-1)^2} &\\
& = \frac{r(4r^3 - r^2 + 4r^2 - r)}{4r^3 - 4r^2 + 4r - r^2 + r -1 +4r^2-4r+1} &\\
& = \frac{r^2(4r^2 + 3r - 1)}{4r^3 - r^2 + r} = \frac{r\cdot(4r^2 + 3r -1)}{4r^2 - r +1}.
\end{align*}

The upper bound is verified by using the code $\cC^B_r$. There are $s = r(r+1)$ information bits, and the number of buckets is $m = r+1$. For each $i\in[m]$, there exists a cell with the linear combination $\sum_{q\in\cS_i}x_q$ and another $r(r-1)$ cells to store one $(r-1)$-subset from each $\cS_j,j\in[r+1]$, where $j\neq i$. Thus, the number of the rows is $r^2 -r +1$.

Let $x_j$ be a request that the code $\cC^B_r$ must satisfy by $r$ disjoint recovering sets. Assume that $x_j \in \cS_i, i\in[r+1]$.
There are $r-1$ buckets which include $x_j$ as a singleton, because $x_j$ appears in $r-1$ subsets of length $r-1$ of part $\cS_i$. Thus, each bucket of the $r-1$ buckets is taken as a recovering set, while reading only one cell from it. In addition, in the $i$-th bucket there exists a cell with $\sum_{q\in\cS_i} x_q$, which includes $x_j$. The $(r-1)$-subset, $\cS_i \setminus \{x_j\}$, is written in a bucket $p$, which is different from bucket $i$, and is different from the buckets that were taken so far (because $x_j \notin \cS_i \setminus \{x_j\}$). Thus, the set $\{i,p\}$ is a recovering set of $x_j$, and it is sufficient to read from bucket $i$ one cell, which is $\sum_{q\in\cS_i} x_q$ and to read $r-1$ cells with the $r-1$ bits of $\cS_i \setminus \{x_j\}$ from bucket $p$. Thus, there exist $r$ disjoint recovering sets for $x_j$, where at most $r-1$ cells are read from each bucket.
\end{IEEEproof}

Next we want to show that for any integer $r\geq 3$ the code $\cC^B_r$ is an $(r^2+r,r,r+1,r^2-r+1,r-1)$ batch array code, by using a property stated in the following lemma.
\begin{lemma}\label{lemma:BEx8Aux1}
For any integer $r\geq 3$ it holds that every two buckets of the code $\cC^B_r$ can form a recovering set of every bit $x_i$ by reading at most $r-1$ cells from each bucket.
\end{lemma}

\begin{IEEEproof}
Given a pair of buckets from $\cC^B_r$, for simplicity we assume that they are the first two buckets.
The first bucket has a cell with $\sum_{i\in\cS_{1}} x_i$, and has exactly $r-1$ bits as singletons from each $\cS_j, 2\leq j \leq r+1$. Hence, the first bucket does not include exactly one of the information bits from each $\cS_j,2\leq j \leq r+1$. Thus, the number of bits that do not appear as singletons in the first bucket is $2r$. Hence, the first bucket can satisfy each information bit except to these $2r$ bits, by reading exactly one cell.

The second bucket contains $r-1$ bits out of the $r$ bits of $\cS_{1}$ as singletons. Thus, each one of these $(r-1)$ bits from $\cS_{1}$ can be satisfied by reading each one of them as a singleton from the second bucket. Also, the remaining bit of $\cS_{1}$ can be satisfied by reading the $r-1$ singletons of $\cS_{1}$ from the second bucket with the cell $\sum_{i\in\cS_{1}} x_i$ in the first bucket.

The first two buckets include different $(r-1)$-subsets of each part other than $\cS_1,\cS_2$. Then, the information bit that does not appear as a singleton cell or as part of the cell $\sum_{i\in\cS_{1}} x_i$ in the first bucket, definitely appears as a singleton cell or in the cell $\sum_{i\in\cS_{2}} x_i$ in the second bucket. Then, each bit $x_q\in \cS_{j}$ where $3\leq j\leq r+1$ can be satisfied by reading it as a singleton from the second bucket. There are $r-1$ such bits, and thus, it remains to show that the code can satisfy the bit $x_{q_1}\in \cS_{2}$ that is not part of the $(r-1)$-subset of singletons which are stored in the first bucket. We can satisfy $x_{q_1}$ by reading the $r-1$ singletons of $\cS_{2}$ from the first bucket with the cell $\sum_{i\in\cS_{2}} x_i$ in the second bucket. Thus, the first two buckets of the code $\cC^B_r$ can form a recovering set of every bit $x_i$. Similarly, it holds for any two buckets of the code $\cC^B_r$.
\end{IEEEproof}

Now, we are ready to show that for any integer $r\geq 3$ the code $\cC^B_r$ is $(r^2+r,r,r+1,r^2-r+1,r-1)$ batch array code.

\begin{theorem}\label{theorem:BExample8}
For any integer $r \geq 3$ the code $\cC^B_r$ from Construction~\ref{construction:cons8} is an $(r^2+r,r,r+1,r^2-r+1,r-1)$ batch array code. In particular, $$\frac{r\cdot(4r^2 + 3r -1)}{4r^2 - r +1} \leq B_{r^2-r+1,r-1}(r^2+r,r) \leq r+1.$$
\end{theorem}
\begin{IEEEproof}
The lower bound is follows from the lower bound of $P_{r^2-r+1,r-1}(r^2+r,r)$.
The upper bound is achieved by using Contruction~\ref{construction:cons8}. Let $R = \{x_{i_1}, x_{i_2}, \ldots , x_{i_{r}}\}$ be a multiset request of $r$ information bits. First, we want to show that the code $\cC^B_r$ can satisfy the first $r-1$ bits of the request by using only $r-1$ buckets. From Construction~\ref{construction:cons8} it is known that each information bit $x_i$ appears as a singleton in $r-1$ buckets out of the $r+1$ buckets. 
Thus, in each subset of buckets of size at least $3$, there is at least one bucket that contains a cell with $x_i$. Therefore, the first $r-1$ bits of the request can be read by singletons from $r-1$ different buckets.
 


After the first step, we still have $2$ buckets and from Lemma~\ref{lemma:BEx8Aux1} it is known that these two buckets can satisfy each $x_i$, in particular $x_{i_{r}}$.
\end{IEEEproof}

According to Theorem~\ref{theorem:PExample8} and Theorem~\ref{theorem:BExample8} it can be verified that for any $r\geq 3$, 
$r < \frac{r\cdot(4r^2 + 3r -1)}{4r^2 - r +1} \leq P_{r^2-r+1,r-1}(r^2+r,r) \leq B_{r^2-r+1,r-1}(r^2+r,r) \leq r+1$. Thus, we conclude that Construction~\ref{construction:cons8} gives optimal PIR and batch array codes.

\subsection{Construction C}

We now present our third construction, and study how it can be used as PIR and functional PIR array codes for specific parameters.
\begin{construction}\label{consAdd}
Let $s \geq 2$ be a fixed integer. The number of information bits is $s$, the number of cells in each bucket (the number of the rows) is $2$. We write each two nonzero disjoint linear combinations of total size at most $s$, and hence, we need $m = \sum_{i=2}^{s} ({s \choose i} \cdot {i \brace 2})$ buckets. Then,
\begin{align*}
m = \sum_{i=2}^{s}\hspace{-0.3ex}\left({s \choose i}  {i \brace 2}\right) \hspace{-0.3ex}=\hspace{-0.3ex} \sum_{i=2}^{s}\hspace{-0.3ex} \hspace{-0.3ex}{s \choose i}  (2^{i-1}-1)\hspace{-0.3ex} = \frac{3^s+1}{2} - 2^s.
\end{align*}
\end{construction}

For any integer $s \geq 2$ denote the code that is obtained from Construction~\ref{consAdd} by $\cC^C_s$.
Construction~\ref{consAdd} for the case of $s=4$ is demonstrated in Table~\ref{ex_add_cons} and provides the following results. First, we show that the code $\cC^C_4$ is a $(4,16,25,2,1)$ PIR array code.

\begin{table*}
\begin{center}
\caption{Construction~\ref{consAdd} for $s=4$}\label{ex_add_cons}
\begin{tabular}{ |c|c|c|c|c|c|c|c|c|c|c|c|c|c| } 
 \hline
 1&2&3&4&5&6&7&8&9&10&11&12&13&14 \\ 
 \hline
 \hline
 $x_1$ & $x_1$ & $x_1$ & $x_2$ & $x_2$ & $x_3$ & $x_1$ & $x_1$ & $x_1$ & $x_2$ & $x_2$ & $x_2$ & $x_3$ & $x_3$  \\ 
 \hline
 $x_2$ & $x_3$ & $x_4$ & $x_3$ & $x_4$ & $x_4$ & $x_2x_3$ & $x_2x_4$ & $x_3x_4$ & $x_1x_3$ & $x_1x_4$ & $x_3x_4$ & $x_1x_2$ & $x_1x_4$  \\ 
 \hline
\end{tabular}
\end{center}

\begin{center}
\begin{tabular}{ |c|c|c|c|c|c|c|c|c|c|c| } 
 \hline
 15&16&17&18&19&20&21&22&23&24&25 \\ 
 \hline
 \hline
$x_3$ & $x_4$ & $x_4$ & $x_4$ & $x_1$ & $x_2$ & $x_3$ & $x_4$ & $x_1x_2$& $x_1x_3$& $x_1x_4$\\ 
\hline
$x_2x_4$ & $x_1x_2$ & $x_1x_3$ & $x_2x_3$ & $x_2x_3x_4$ & $x_1x_3x_4$ & $x_1x_2x_4$ & $x_1x_2x_3$ & $x_3x_4$& $x_2x_4$& $x_2x_3$\\ 
 \hline
\end{tabular}
\end{center}
\end{table*}

\begin{theorem}\label{theorem:P21416} 
The code $\cC^C_4$ from Construction~\ref{consAdd} is a $(4,16,25,2,1)$ PIR array code. In particular, $23 \leq P_{2,1}(4,16)$ $ \leq 25.$
\end{theorem}
\begin{IEEEproof}
The lower bound is obtained using Theorem~\ref{theorem:PIRLB}\eqref{theorem:part2}, $P_{2,1}(4,16) \geq P_{2,2}(4,16) \geq \frac{16\cdot 4 \cdot 5}{5\cdot 2 + 4} > 22$. The upper bound is verified using the code $\cC^C_4$. Let $x_i,i\in[4]$ be a request, that the code $\cC^C_4$ must satisfy $16$ times. From the symmetry of the code, assume that $x_i = x_1$. The following are the recovering sets of $x_1$, where from each bucket only one cell is read. $\{\{1\}$,$\{2\}$, $\{3\}$, $\{7\}$, $\{8\}$, $\{9\}$, $\{19\}$, $\{10,6\}$, $\{11,5\}$, $\{13,4\}$, $\{14,18\}$, $\{15,17\}$, $\{16,12\}$, $\{20,23\}$, $\{21,24\}$, $\{22,25\}\}$.
\end{IEEEproof}

Next, we show that the code $\cC^C_4$ is a $(4,14,25,2,2)$ functional PIR array code.

\begin{theorem}\label{theorem:FP22414}
The code $\cC^C_4$ from Construction~\ref{consAdd} is a $(4,14,25,2,2)$ functional PIR array code. In particular,
$24 \leq FP_{2,2}(4,14) \leq 25.$
\end{theorem}

\begin{IEEEproof}
The lower bound is obtained using Theorem~\ref{theorem:LBFP4}, $FP_{2,2}(4,14) \geq \frac{2\cdot 14 \cdot 15}{15 + 3} > 23$. The upper bound is verified using the code $\cC^C_4$. Let $R$ be a linear combination request, that the code $\cC^C_4$ must satisfy $14$ times. From the symmetry of the code, the proof is divided into the following cases according to the number of information bits that appear in $R$. If the number of information bits that appear in $R$ is $p$ then we assume that the request is $x_1+x_2+\cdots+x_p$.

\noindent
{\bf Case 1:} 
The recovering sets are the following $\{\{1\}$, $\{2\}$, $\{3\}$, $\{7\}$, $\{8\}$, $\{9\}$, $\{19\}$, $\{10,6\}$, $\{11,5\}$, $\{13,4\}$, $\{14,18\}$, $\{15,17\}$, $\{16,12\}$, $\{20,23\}$, $\{21,24\}$, $\{22,25\}\}$.

\noindent
{\bf Case 2:} 
The recovering sets are the following $\{\{1\}$, $\{13\}$, $\{16\}$, $\{23\}$, $\{2,4\}$, $\{3,5\}$, $\{7,10\}$, $\{8,11\}$, $\{9,12\}$, $\{14,15\}$, $\{17,18\}$, $\{19,20\}$, $\{21,22\}$, $\{24,25\}\}$.

\noindent
{\bf Case 3:} 
The recovering sets are the following. $\{\{7\}$, $\{10\}$, $\{13\}$, $\{22\}$, $\{1,24\}$, $\{2,23\}$, $\{3,25\}$, $\{4,17\}$, $\{5,14\}$, $\{6,16\}$, $\{8,20\}$, $\{9,21\}$, $\{11,12\}$, $\{18,19\}\}$.

\noindent
{\bf Case 4:} 
The recovering sets are the following. $\{\{19\}$, $\{20\}$, $\{21\}$, $\{22\}$, $\{23\}$, $\{24\}$, $\{25\}$, $\{1,6\}$, $\{2,5\}$, $\{3,4\}$, $\{7,11\}$, $\{8,10\}$, $\{9,13\}$, $\{12,16\}$, $\{14,18\}$, $\{15,17\}\}$.
\end{IEEEproof}

Construction~\ref{consAdd} for the case of $s=5$ is demonstrated in Table~\ref{ex_add_cons5} and provides the following result.

\begin{table*}
\begin{center}
\caption{Construction~\ref{consAdd} for $s=5$}\label{ex_add_cons5}
\begin{tabular}{ |c|c|c|c|c|c|c|c|c|c|c|c|c|c|c|c|c|c|c|c|c|c| } 
 \hline
 1&2&3&4&5&6&7&8&9&10&11&12&13&14&15&16&17&18&19&20&21&22 \\ 
 \hline
 \hline
 $x_1$ & $x_1$ & $x_1$ & $x_1$ & $x_2$ & $x_2$ & $x_2$ & $x_3$ & $x_3$ & $x_4$ & $x_1$ & $x_1$ & $x_1$ & $x_1$ & $x_1$ & $x_1$ & $x_2$ & $x_2$ & $x_2$ & $x_2$ & $x_2$ & $x_2$ \\ 
 \hline
 $x_2$ & $x_3$ & $x_4$ & $x_5$ & $x_3$ & $x_4$ & $x_5$ & $x_4$ & $x_5$ & $x_5$ & $x_2x_3$ & $x_2x_4$ & $x_2x_5$ & $x_3x_4$ & $x_3x_5$ & $x_4x_5$ & $x_1x_3$ & $x_1x_4$ & $x_1x_5$ & $x_3x_4$ & $x_3x_5$ & $x_4x_5$   \\ 
 \hline
\end{tabular}
\end{center}

\begin{center}
\begin{tabular}{ |c|c|c|c|c|c|c|c|c|c|c|c|c|c|c|c|c|c| } 
 \hline
 23&24&25&26&27&28&29&30&31&32&33&34&35&36&37&38&39&40 \\ 
 \hline
 \hline
$x_3$ & $x_3$ & $x_3$ & $x_3$ & $x_3$ & $x_3$ & $x_4$ & $x_4$ & $x_4$ & $x_4$ & $x_4$ & $x_4$ & $x_5$ & $x_5$ & $x_5$ & $x_5$ & $x_5$ & $x_5$ \\ 
\hline
$x_1x_2$ & $x_1x_4$ & $x_1x_5$ & $x_2x_4$ & $x_2x_5$ & $x_4x_5$ & $x_1x_2$ & $x_1x_3$ & $x_1x_5$ & $x_2x_3$ & $x_2x_5$ & $x_3x_5$ & $x_1x_2$ & $x_1x_3$ & $x_1x_4$ & $x_2x_3$ & $x_2x_4$ & $x_3x_4$\\ 
 \hline
\end{tabular}
\end{center}

\begin{center}
\begin{tabular}{ |c|c|c|c|c|c|c|c|c|c|c|c|c|c|c|c|c|c|c|c| } 
 \hline
 41&42&43&44&45&46&47&48&49&50&51&52&53&54 \\ 
 \hline
 \hline
$x_1$ & $x_1$ & $x_1$ & $x_1$ & $x_2$ & $x_2$ & $x_2$ & $x_2$ & $x_3$ & $x_3$ & $x_3$ & $x_3$ & $x_4$ & $x_4$  \\ 
\hline
$x_2x_3x_4$ & $x_2x_3x_5$ & $x_2x_4x_5$ & $x_3x_4x_5$ & $x_1x_3x_4$ & $x_1x_3x_5$ & $x_1x_4x_5$ & $x_3x_4x_5$ & $x_1x_2x_4$ & $x_1x_2x_5$ & $x_1x_4x_5$ & $x_2x_4x_5$ & $x_1x_2x_3$ & $x_1x_2x_5$  \\ 
 \hline
\end{tabular}
\end{center}

\begin{center}
\begin{tabular}{ |c|c|c|c|c|c|c|c|c|c|c| } 
 \hline
55&56&57&58&59&60&61&62&63&64&65 \\ 
 \hline
 \hline
$x_4$ & $x_4$ & $x_5$ & $x_5$ & $x_5$ & $x_5$ & $x_1$ & $x_2$ & $x_3$ & $x_4$ & $x_5$ \\ 
\hline
$x_1x_3x_5$ & $x_2x_3x_5$ & $x_1x_2x_3$ & $x_1x_2x_4$ & $x_1x_3x_4$ & $x_2x_3x_4$ & $x_2x_3x_4x_5$ & $x_1x_3x_4x_5$& $x_1x_2x_4x_5$ & $x_1x_2x_3x_5$ & $x_1x_2x_3x_4$ \\ 
 \hline
\end{tabular}
\end{center}

\begin{center}
\begin{tabular}{ |c|c|c|c|c|c|c|c|c|c|c|c|c|c|c| } 
 \hline
66&67&68&69&70&71&72&73&74&75&76&77&78&79&80 \\ 
 \hline
 \hline
$x_1x_2$ & $x_1x_2$ & $x_1x_2$ & $x_1x_3$ & $x_1x_3$ & $x_1x_3$ & $x_1x_4$ & $x_1x_4$ & $x_1x_4$ & $x_1x_5$ & $x_1x_5$ & $x_1x_5$ & $x_2x_3$ & $x_2x_4$ & $x_2x_5$ \\ 
\hline
$x_3x_4$ & $x_3x_5$ & $x_4x_5$ & $x_2x_4$ & $x_2x_5$ & $x_4x_5$ & $x_2x_3$ & $x_2x_5$ & $x_3x_5$ & $x_2x_3$ & $x_2x_4$ & $x_3x_4$ & $x_4x_5$ & $x_3x_5$ & $x_3x_4$ \\ 
 \hline
\end{tabular}
\end{center}

\begin{center}
\begin{tabular}{ |c|c|c|c|c|c|c|c|c|c| } 
 \hline
 81&82&83&84&85&86&87&88&89&90 \\ 
 \hline
 \hline
$x_1x_2$ & $x_1x_3$ & $x_1x_4$ & $x_1x_5$ & $x_2x_3$ & $x_2x_4$ & $x_2x_5$ & $x_3x_4$ & $x_3x_5$ & $x_4x_5$\\ 
\hline
$x_3x_4x_5$ & $x_2x_4x_5$ & $x_2x_3x_5$ & $x_2x_3x_4$ & $x_1x_4x_5$ & $x_1x_3x_5$ & $x_1x_3x_4$ & $x_1x_2x_5$ & $x_1x_2x_4$ & $x_1x_2x_3$ \\ 
 \hline
\end{tabular}
\end{center}

\end{table*}

\begin{theorem}\label{theorem:FP22548} 
The code $\cC^C_5$ from Construction~\ref{consAdd} is a $\allowbreak(5,48,90,2,2)$ functional PIR array code. In particular,
$88 \leq FP_{2,2}(5,48) \leq 90$.
\end{theorem}

\begin{IEEEproof}
The lower bound is obtained using Theorem~\ref{theorem:LBFP4}, $FP_{2,2}(5,48) \geq \frac{2\cdot 48 \cdot 31}{31 + 3} > 87$. The upper bound is verified using the code $\cC^C_5$. Let $R$ be a linear combination request that the code $\cC^C_5$ must satisfy $48$ times. From the symmetry of the code, the proof is divided into the following cases according to the number of information bits that appear in $R$. If the number of information bits that appear in $R$ is $p$ then we assume that the request is $x_1+x_2+\cdots+x_p$.

\noindent
{\bf Case 1:} The recovering sets are the following $\{\{1\}$, $\{2\}$, $\{3\}$, $\{4\}$, $\{11\}$, $\{12\}$, $\{13\}$, $\{14\}$, $\{15\}$, $\{16\}$, $\{41\}$, $\{42\}$, $\{43\}$, $\{44\}$, $\{61\}$, $\{17,26\}$, $\{18,32\}$, $\{19,38\}$, $\{20,23\}$, $\{21,29\}$, $\{22,35\}$, $\{24,33\}$, $\{25,39\}$, $\{27,30\}$, $\{28,36\}$, $\{31,40\}$, $\{34,37\}$, $\{45,8\}$, $\{46,9\}$, $\{47,10\}$, $\{49,6\}$, $\{50,7\}$, $\{48,51\}$, $\{53,5\}$, $\{52,54\}$, $\{55,67\}$, $\{57,72\}$, $\{58,69\}$, $\{59,66\}$, $\{64,56\}$, $\{65,60\}$, $\{62,78\}$, $\{63,79\}$, $\{71,81\}$, $\{73,82\}$, $\{83,80\}$, $\{68,85\}$, $\{70,88\}$, $\{74,86\}$, $\{75,90\}$, $\{76,89\}$, $\{77,87\}\}$.

\noindent
{\bf Case 2:} The recovering sets are the following $\{\{1\}$, $\{23\}$ ,$\{29\}$, $\{35\}$, $\{66\}$, $\{67\}$, $\{68\}$, $\{81\}$, $\{2,5\}$, $\{3,6\}$, $\{4,7\}$, $\{9,53\}$, $\{8,49\}$, $\{10,88\}$, $\{11,20\}$, $\{12,21\}$, $\{13,22\}$, $\{14,17\}$, $\{15,18\}$, $\{16,19\}$, $\{24,26\}$, $\{25,27\}$, $\{30,32\}$, $\{31,33\}$, $\{36,38\}$, $\{37,39\}$, $\{41,45\}$, $\{42,46\}$, $\{43,47\}$, $\{44,85\}$, $\{51,52\}$, $\{54,57\}$, $\{55,56\}$, $\{59,60\}$, $\{61,28\}$, $\{62,34\}$, $\{63,40\}$, $\{64,74\}$, $\{65,77\}$, $\{69,80\}$, $\{70,79\}$, $\{71,72\}$, $\{73,78\}$, $\{75,82\}$, $\{84,87\}$, $\{86,48\}$, $\{50,58\}$, $\{76,83\}\}$.

\noindent
{\bf Case 3:} The recovering sets are the following $\{\{11\}$, $\{17\}$, $\{23\}$, $\{53\}$, $\{57\}$, $\{90\}$, $\{1,8\}$, $\{2,6\}$, $\{3,32\}$, $\{4,38\}$, $\{5,16\}$, $\{7,36\}$, $\{9,29\}$, $\{10,89\}$, $\{30,66\}$, $\{12,40\}$, $\{13,34\}$, $\{14,39\}$, $\{15,33\}$, $\{18,80\}$, $\{19,79\}$, $\{20,37\}$, $\{21,31\}$, $\{22,88\}$, $\{24,76\}$, $\{73,86\}$, $\{26,74\}$, $\{27,68\}$, $\{28,87\}$, $\{35,63\}$, $\{41,64\}$, $\{42,65\}$, $\{43,81\}$, $\{44,47\}$, $\{45,69\}$, $\{46,60\}$, $\{48,82\}$, $\{49,56\}$, $\{50,59\}$, $\{51,78\}$, $\{52,85\}$, $\{54,67\}$, $\{55,70\}$, $\{58,77\}$, $\{61,75\}$, $\{62,71\}$, $\{72,84\}$, $\{25,83\}\}$.

\noindent
{\bf Case 4:} The recovering sets are the following $\{\{41\}$, $\{45\}$, $\{49\}$, $\{53\}$, $\{65\}$, $\{66\}$, $\{69\}$, $\{72\}$, $\{1,8\}$, $\{2,6\}$, $\{3,5\}$, $\{10,11\}$, $\{9,12\}$, $\{7,14\}$, $\{4,20\}$, $\{13,28\}$, $\{15,22\}$, $\{16,21\}$, $\{17,34\}$, $\{18,27\}$, $\{19,40\}$, $\{23,64\}$, $\{24,62\}$, $\{25,33\}$, $\{26,61\}$, $\{29,63\}$, $\{30,48\}$, $\{31,38\}$, $\{32,44\}$, $\{35,88\}$, $\{36,39\}$, $\{37,85\}$, $\{42,90\}$, $\{43,89\}$, $\{46,68\}$, $\{47,67\}$, $\{50,71\}$, $\{51,87\}$, $\{52,84\}$, $\{54,74\}$, $\{55,70\}$, $\{56,75\}$, $\{57,78\}$, $\{58,79\}$, $\{59,73\}$, $\{60,76\}$, $\{77,81\}$, $\{82,86\}\}$.

\noindent
{\bf Case 5:} The recovering sets are the following $\{\{61\}$, $\{62\}$, $\{63\}$, $\{64\}$, $\{65\}$, $\{81\}$, $\{82\}$, $\{83\}$, $\{84\}$, $\{85\}$, $\{86\}$, $\{87\}$, $\{88\}$, $\{89\}$, $\{90\}$, $\{66,4\}$, $\{67,3\}$, $\{68,2\}$, $\{69,7\}$, $\{70,6\}$, $\{71,1\}$, $\{72,9\}$, $\{73,5\}$, $\{74,17\}$, $\{75,10\}$, $\{76,8\}$, $\{77,18\}$, $\{78,11\}$, $\{79,12\}$, $\{80,13\}$, $\{41,40\}$, $\{42,34\}$, $\{43,28\}$, $\{44,19\}$, $\{45,39\}$, $\{46,33\}$, $\{47,27\}$, $\{48,14\}$, $\{49,38\}$, $\{50,32\}$, $\{51,20\}$, $\{52,15\}$, $\{53,37\}$, $\{54,26\}$, $\{55,21\}$, $\{56,16\}$, $\{57,31\}$, $\{58,25\}$, $\{59,22\}\}$.
\end{IEEEproof}

\section{Asymptotic Analysis of Array Codes}\label{sec:analysis}
The goal of this section is to provide a figure of merit in order to compare between the different constructions of array codes. For simplicity we consider the case where $\ell=t$, that is, it is possible to read all the bits in every bucket. Under this setup, it holds that $FP_{t,t}(s,k)\leq sk/t$ for all $s,k,$ and $t$. This motivates us to define the following values
$$\cR_{X}(t,k) = \limsup_{s \to \infty} \frac{X_{t,t}(s,k)}{sk/t},$$
where $X\in\{P,B,FP,FB\}$. 
The case where $t=1$ has been studied in several previous works. For example, for functional PIR array codes we have $\cR_{FP}(1,k) \geq \frac{1}{k\cdot H(1/k)}$ for any even integer $k\geq 4$~\cite[Th. 13]{ZYE19}. Also, for functional batch array codes it holds from~\cite[Th. 21]{ZYE19} that $\cR_{FB}(1,k) \leq \frac{1}{k\cdot H(c_{k})}$, where $c_1=\frac{1}{2}$ and $c_{k+1}$ is the root of the polynomial $H(z)=H(c_k)-zH(c_k)$.
For the case $k=1$ we have $\cR_{FB}(t,1) = \cR_{FP}(t,1) = 1$ from Theorem~\ref{theorem:ArrayCodek1}\eqref{theorem:ArrayCodek1tt}. According to the bounds and constructions studied in the paper, we can already summarize several results in the following theorems for $t=2$ and general values.
\begin{theorem}
\begin{enumerate}
\item $\cR_{FP}(2,2) \leq \cR_{FB}(2,2) \leq \frac{7}{8} = 0.875$, and $\cR_{FB}(2,2) \geq 0.71$.

\item $\cR_{FP}(2,11) \leq \frac{25}{33} = 0.758$.

\item $\cR_{FP}(2,14) \leq \frac{25}{28} = 0.893$.

\item $\cR_{FP}(2,48) \leq \frac{3}{4} = 0.75$.

\item $\cR_{P}(2,16) \leq \frac{25}{32} = 0.78125$.

\item $\cR_{B}(2,15) \leq \frac{5}{9} = 0.556$.
\end{enumerate}
\end{theorem}
\begin{IEEEproof}
\begin{enumerate}




\item From Theorem~\ref{theorem:FB2282} we have $FB_{2,2}(s,2) \leq 7\cdot\left\lceil \frac{s}{8} \right\rceil$. Thus, $\cR_{FB}(2,2) = \limsup_{s \to \infty} \frac{FB_{2,2}(s,2)}{2s/2} \leq \limsup_{s \to \infty} \frac{7\lceil s/8\rceil}{s}$ $ \leq \limsup_{s \to \infty} \frac{(7s/8) + 7}{s}$ $ = \frac{7}{8}$.

From Corollary~\ref{cor:FB22s2} we have $FB_{2,2}(s,2) \geq 0.71s$. Thus, $\cR_{FB}(2,2) = \limsup_{s \to \infty} \frac{FB_{2,2}(s,2)}{2s/2} \geq \limsup_{s \to \infty} \allowbreak\frac{0.71s}{s} = 0.71$.

\item From \Tref{theorem:FPExample9} we have $FP_{2,2}(6,11) \leq 25$. Then, it is possible to use \Tref{theorem:Basic}\eqref{theorem:partas} to get that $FP_{2,2}(s,11) \leq 25\cdot \left\lceil \frac{s}{6} \right\rceil$. Thus, $\cR_{FP}(2,11) = \limsup_{s \to \infty} \frac{FP_{2,2}(s,11)}{11s/2} \leq \limsup_{s \to \infty} \frac{25\lceil s/6\rceil}{11s/2}  \leq \limsup_{s \to \infty} \frac{(25s/6) + 25}{11s/2} = \frac{50}{66} = 0.758$.


\item From \Tref{theorem:FP22414} we have $FP_{2,2}(4,14) \leq 25$. Then, it is possible to use \Tref{theorem:Basic}\eqref{theorem:partas} to get that $FP_{2,2}(s,14) \leq 25\cdot \left\lceil \frac{s}{4} \right\rceil$. Thus, $\cR_{FP}(2,14) = \limsup_{s \to \infty} \frac{FP_{2,2}(s,14)}{14s/2} \leq \limsup_{s \to \infty} \frac{25\lceil s/4\rceil}{7s}  \leq \limsup_{s \to \infty} \frac{(25s/4) + 25}{7s} = \frac{25}{28} = 0.893$.

\item From \Tref{theorem:FP22548} we have $FP_{2,2}(5,48) \leq 90$. Then, it is possible to use \Tref{theorem:Basic}\eqref{theorem:partas} to get that $FP_{2,2}(s,48) \leq 90\cdot \left\lceil \frac{s}{5} \right\rceil$. Thus, $\cR_{FP}(2,48) = \limsup_{s \to \infty} \frac{FP_{2,2}(s,48)}{48s/2} \leq \limsup_{s \to \infty} \frac{90\lceil s/5\rceil}{24s} \leq \limsup_{s \to \infty} \frac{(90s/5) + 90}{24s} = \frac{90}{120} = \frac{3}{4} = 0.75$.

\item From~\Tref{theorem:P21416} we have $P_{2,1}(4,16) \leq 25$. Then, it is possible to use \Tref{theorem:Basic}\eqref{theorem:partas} and get that $P_{2,1}(s,16) \leq 25\cdot \left\lceil \frac{s}{4} \right\rceil$. Thus, $\cR_{P}(2,16) = \limsup_{s \to \infty} \frac{P_{2,2}(s,16)}{16s/2} \leq \limsup_{s \to \infty} \frac{25\lceil s/4\rceil}{8s}  \leq \limsup_{s \to \infty} \frac{(25s/4) + 25}{8s} = \frac{25}{32} = 0.78125$.

\item From~\Tref{theorem:BExample9} we have $B_{2,2}(6,15) = 25$. Then, it is possible to use \Tref{theorem:Basic}\eqref{theorem:partas} and get that $B_{2,2}(s,15) \leq 25\cdot \left\lceil \frac{s}{6} \right\rceil$. Thus, $\cR_{B}(2,15) = \limsup_{s \to \infty} \frac{B_{2,2}(s,15)}{15s/2} \leq \limsup_{s \to \infty} \frac{25\lceil s/6\rceil}{15s/2}  \leq \limsup_{s \to \infty} \frac{(25s/6) + 25}{15s/2} = \frac{25}{45} = 0.556$.
\end{enumerate}
\end{IEEEproof}

\begin{theorem}
\begin{enumerate}
\item For any $r\geq 3$, $\cR_{P}(r^2-r+1,r) \leq \frac{(r+1)(r^2 -r + 1)}{r(r^2+r)}$ (also for B).

\item  For any $t\geq 2$, $\cR_{P}(t,k) \leq\frac{m}{k(t+1)}$, where $k={t(t+1) \choose t}$ and $m = k+ \frac{{t(t+1) \choose t+1}}{t}$.

\item For any two integers $t$ and $k$, $\cR_{FB}(t,k) \leq \frac{1}{k\cdot H(c_{tk})}$, where $c_1=\frac{1}{2}$ and $c_{k+1}$ is the root of the polynomial $H(z)=H(c_k)-zH(c_k)$.


\item For any positive integers $t,k$ and $a$, $\cR_X(t,a\cdot k) \leq \cR_X(t,k)$, where $X\in\{P,B,FP,FP\}$.

\item For any positive integers $t,k$ and $a$, $\cR_X(t,k) \leq \cR_X(a\cdot t,k)$, where $X\in\{P,B,FP,FP\}$.
\end{enumerate}
\end{theorem}
\begin{IEEEproof}
\begin{enumerate}
\item  From \Tref{theorem:PExample8} we have for any $r \geq 3$, $P_{r^2-r+1,r-1}(r^2+r,r) \leq r+1$. Then, it is possible to use \Tref{theorem:Basic}\eqref{theorem:partas} to get that $P_{r^2-r+1,r-1}(s,r) \leq (r+1)\cdot \left\lceil \frac{s}{r^2+r} \right\rceil$. Thus, for a given $r$, it holds that 
\begin{align*}
\cR_{P}&(r^2-r+1,r) = \limsup_{s \to \infty} \frac{P_{r^2-r+1,r^2-r+1}(s,r)}{rs/(r^2-r+1)} &\\
 &\leq \limsup_{s \to \infty} \frac{P_{r^2-r+1,r-1}(s,r)}{rs/(r^2-r+1)} &\\
 &\leq \limsup_{s \to \infty} \frac{(r+1)\cdot \left\lceil \frac{s}{r^2+r} \right\rceil}{rs/(r^2-r+1)} &\\
  &\leq \limsup_{s \to \infty} \frac{\frac{(r+1)s}{r^2+r} + (r+1)}{rs/(r^2-r+1)} = \frac{(r+1)(r^2-r+1)}{r(r^2+r)}.
\end{align*}

\item From Theorem~\ref{theorem:PIRLB}\eq{theorem:part5} we have for any $t \geq 2$ and $p=t+1$, $P_{t,t}(t(t+1),k) \leq m$, where $k={t(t+1) \choose t}$ and $m = k+ \frac{{t(t+1) \choose t+1}}{t}$. Then, it is possible to use \Tref{theorem:Basic}\eqref{theorem:partas} to get that $P_{t,t}(s,k) \leq m\cdot \left\lceil \frac{s}{t(t+1)} \right\rceil$. Thus, for a given $t$, it holds that $\cR_{P}(t,k) = \limsup_{s \to \infty} \frac{P_{t,t}(s,k)}{sk/t} \leq \limsup_{s \to \infty} \frac{m\cdot \left\lceil \frac{s}{t(t+1)} \right\rceil}{sk/t} \leq \limsup_{s \to \infty} \frac{\frac{m\cdot s}{t(t+1)} + m}{sk/t} = \frac{m}{k(t+1)}$.

\item From Lemma~\ref{lemma:FBGadget}, we have $FB_{t,t}(s,k)\leq FB_{t,1}(s,k) \leq FB(\lceil s/t \rceil,t\cdot k)$. $\cR_{FB}(t,k) = \limsup_{s \to \infty} \frac{FB_{t,t}(s,k)}{sk/t} \leq \limsup_{s \to \infty}\frac{FB(\lceil s/t \rceil,t\cdot k)}{sk/t} = \limsup_{s \to \infty}\frac{FB(\lceil s/t \rceil,t\cdot k)}{s/t}\cdot \frac{1}{k}$. Thus, according to~\cite[Th. 21]{ZYE19}, $\cR_{FB}(t,k) \leq \frac{1}{k\cdot H(c_{tk})}$, where $c_1=\frac{1}{2}$ and $c_{k+1}$ is the root of the polynomial $H(z)=H(c_k)-zH(c_k)$.

\item From Theorem~\ref{theorem:Basic}\eqref{theorem:partak} we have that for any positive integer $a$ and any $X\in\{P,B,FP,FP\}$, $X_{t,t}(s,a\cdot k) \leq a \cdot X_{t,t}(s,k)$. Thus, $R_{X}(t,a\cdot k) = \limsup_{s \to \infty} \frac{X_{t,t}(s,a\cdot k)}{ska/t} \leq \limsup_{s \to \infty} \frac{a\cdot X_{t,t}(s,k)}{ska/t} = \limsup_{s \to \infty} \frac{X_{t,t}(s,k)}{sk/t} = \cR_{X}(t,k)$.

\item From Theorem~\ref{theorem:Basic}\eqref{theorem:partat} we have that for any positive integer $a$ and any $X\in\{P,B,FP,FP\}$, $a \cdot X_{a\cdot t,a\cdot t}(s,k) \geq X_{t,a \cdot t}(s,k) = X_{t,t}(s,k)$. Thus, $R_{X}(t,k) = \limsup_{s \to \infty} \allowbreak\frac{X_{t,t}(s,k)}{sk/t} \leq \limsup_{s \to \infty}\allowbreak\frac{a\cdot X_{a\cdot t,a\cdot t}(s,k)}{sk/t} = \limsup_{s \to \infty} \allowbreak\frac{X_{a\cdot t,a\cdot t}(s,k)}{sk/(at)} = \cR_{X}(a\cdot t,k)$.
\end{enumerate}
\end{IEEEproof}

\section{Locality Codes}\label{sec:Locality}
In this section we study a new family of array codes which is a special case of functional PIR array codes in the sense that each recovering set is of size at most $r$ and all the cells of each bucket can be read, i.e., $\ell=t$. This new family of array codes will be called \emph{locality functional array codes}. In order to find lower bounds and constructions for locality functional array codes we will use codes and designs in subspaces and covering codes.
\subsection{Definitions and Basic Constructions}
This section is studying the following family of codes.
\begin{definition}
An $(s,k,m,t,r)$ \textbf{locality functional array code} over $\Sigma$ is defined by an encoding map $\cE:\Sigma^s \rightarrow (\Sigma^t)^m$ that encodes $s$ information bits $x_1,\dots,x_s$ into a $t\times m$ array and a decoding function $\cD$ that satisfies the following property. For any request of a linear combination $\bfv$ of the information bits, there is a partition of the columns into $k$ recovering sets $S_1,\ldots,S_k \subseteq [m]$ where $|S_j| \leq r$ for any $j\in[k]$.


\end{definition}

We denote by $D(s,k,t,r)$ the smallest number of buckets $m$ such that an $(s,k,m,t,r)$ locality functional array code exists.
For the rest of the section, assume that the parameters $s,k,t$ and $r$ are positive integers such that $t\leq s$.
The following theorem summarizes several results on $D(s,k,t,r)$ based upon basic bound and constructions.

\begin{theorem}\label{basicLocality}
\begin{enumerate}
\item $D(s,k,t,r) \geq m^*$, where $m^*$ is the smallest positive integer such that $\sum_{i=1}^{\min\{r,m^*-k+1\}} {m^*\choose i}(2^t - 1)^i \geq k(2^s - 1)$.\label{section1}


\item For any integer $a$ where $1\leq a < t$, $D(s,k,t,r) \leq D(s-a,k, t - a ,r)$.

\item For every positive integers $s_1,s_2,r_1,r_2$ and $p$, $D(s_1+s_2,k,t,r_1+r_2) \leq D(s_1,k,t,r_1) + D(s_2,k,t,r_2)$. In particular, $D(ps,k,t,pr) \leq p \cdot D(s,k,t,r)$. \label{BaLocSec3}

\end{enumerate}
\end{theorem}

\begin{IEEEproof}
\begin{enumerate}
\item Similar to the proof of Theorem~\ref{theorem:LBFP1} but with minor changes. Here, all cells from each bucket can be read. Hence, for any positive integer $n$, there are $(2^t-1)^n$ nonzero linear combinations that can be obtained from $n$ buckets while using all the $n$ buckets. Also, each recovering set must be of size at most $\min\{r,m^*-k+1\}$. Thus, we get that $\sum_{i=1}^{\min\{r,m^*-k+1\}} {m^*\choose i}(2^t - 1)^i \geq k(2^s - 1)$.

\item Let $\cC$ be an $(s-1,k,m,t-1,r)$ locality functional array code with $m$ buckets such that each bucket has $t-1$ cells. For the $s$ information bits $x_1,\ldots,x_s$, we encode the first $s-1$ bits using the encoder of $\cC$ to get $m$ buckets where each bucket has $t-1$ cells. For each bucket, a new cell that stores $x_s$ is added. Assume that $R$ is the request which is a linear combination of the $s$ information bits. Let $R_1$ be the part of the request which is a linear combination of the first $s-1$ information bits. From the properties of $\cC$, for the request $R_1$, there exist $k$ disjoint recovering sets $\{\cS_1,\cS_2,\ldots,\cS_k\}$ such that $|\cS_j|\leq r$ for any $j\in[k]$. If $R=R_1$, then the same $\{\cS_1,\cS_2,\ldots,\cS_k\}$ are recovering sets for $R$. If $R = x_s$, we can take the first $k$ buckets as $k$ recovering sets each of size 1. If $R$ includes $x_s$, then the same $\{\cS_1,\cS_2,\ldots,\cS_k\}$ are recovering sets for $R$, where we can read $x_s$ from one of the buckets in each $\cS_j$. Thus, $D(s,k,t,r)\leq D(s-1,k,t-1,r)$ and we can get that $D(s,k,t,r) \leq D(s-a,k, t - a ,r)$ by induction on $a$.

\item Let $\cC_1$ be an $(s_1,k,m_1,t,r_1)$ locality functional array code and $\cC_2$ be an $(s_2,k,m_2,t,r_2)$ locality functional array code. The codes $\cC_1$ and $\cC_2$ are used to construct an $(s_1+s_2,k,m_1+m_2,t,r_1+r_2)$ locality functional array code by encoding the first $s_1$ bits using the encoder of $\cC_1$ and the last $s_2$ bits using the encoder of $\cC_2$. Assume that $R$ is the request which is a linear combination of the $s_1+s_2$ information bits. Let $R_1,R_2$ be the part of $R$ which is a linear combination of the first $s_1$, last $s_2$ information bits, respectively. According to $\cC_1,\cC_2$, there exist $k$ recovering sets $\{\cS^1_1,\cS^1_2,\ldots,\cS^1_k\},\{\cS^2_1,\cS^2_2,\ldots,\cS^2_k\}$ for $R_1,R_2$ such that each recovering set has size at most $r_1,r_2$, respectively. Then, the set $\cS^1_j \cup \cS^2_j$ for any $j\in[k]$ is a recovering set for $R$ with size at most $r_1+r_2$. Therefore, the sets $\{\cS^1_1\cup \cS^2_1,\cS^1_2\cup \cS^2_2,\ldots,\cS^1_k\cup \cS^2_k\}$ are $k$ recovering sets for $R$ such that the size of each recovering set is at most $r_1+r_2$. Thus, $D(s_1+s_2,k,t,r_1+r_2) \leq D(s_1,k,t,r_1) + D(s_2,k,t,r_2)$ and we can get that $D(ps,k,t,pr) \leq p \cdot D(s,k,t,r)$ by induction on $p$.
\end{enumerate}
\end{IEEEproof}



\subsection{Constructions Based on Subspaces}

In this section we show connections between the problem of finding the minimal number of buckets for locality functional array codes and several problems in subspaces. Subspaces were used in~\cite{SES19} to construct \emph{array codes} and to examine their locality and availability. The family of array codes that was defined in~\cite{SES19} is a linear subspace of $b\times n$ matrices over $\mathbb{F}_q$ such that each codeword is a $b\times n$ matrix where each entry is called a symbol. The \emph{weight} of each codeword was defined to be the number of nonzero columns in the codeword and the distance of the code is the minimal weight of a nonzero codeword. 

The problem that was presented in~\cite{SES19} was to examine locality and availability of array codes where two types of locality were defined. The first one is \emph{node locality}. A codeword column $j\in[n]$ has node locality $r_{nd}$ if it can be recovered by a linear combination of the symbols of the columns in a recovering set of size $r_{nd}$. If all codeword columns have node locality $r_{nd}$, then $r_{nd}$ is also called the node locality of the array code. The second type is \emph{symbol locality} $r_{sb}$ which is similar to node locality but instead of recovering the whole column, here only one symbol (entry of the codewords matrices) is needed to be recovered. Similarly, there are two types of availability. The node, symbol availability, denoted by $t_{nd},t_{sb}$ is the number of pairwise disjoint recovering sets of size at most $r_{nd},r_{sb}$ for any codeword column, symbol, respectively.

To simplify the problem, they flattened each $b\times n$ codeword into a vector of length $bn$ by reading the symbols of the codeword column by column from first to last entry. The $M \times bn$ generator matrix $G$, where each row is a flattened codeword, can represent the array code $C$, where the columns $(j-1)b+1,\ldots,jb$ of $G$ correspond to the symbols of the $j$-th codeword column of $C$ and these columns are called the $j$-th \emph{thick column} of $G$. By this way, the $j$-th thick column of $G$ which corresponds to the $j$-th codeword column of $C$, can be represented by $V_j$ which is a $b$-subspace of $\mathbb{F}_q^M$. Thus, an equivalent constraints of node and symbol locality can be formed using subspaces as stated in~\cite[Lemma 3]{SES19}, where a subset $\cS = \{j_1,\ldots,j_p\}\subseteq [n]\setminus\{j\}$ is a recovering set for the codeword column $j\in [n]$, if and only if $V_j \subseteq V_{j_1}+\cdots + V_{j_p}$. Similarly, $\cS$ is a recovering set for the symbol $(i,j), i\in[b],j\in[n]$ if and only if $\bfg_{(j-1)b+i} \in V_{j_1}+\cdots + V_{j_p}$, where $\bfg_{(j-1)b+i}$ is the $i$-th column in the $j$-th thick column of $G$ that corresponds to the $i$-th entry in the $j$-th codeword column of $C$.

In our work we are interested in the problem of recovering the requests which are all possible linear combinations of the information bits, which is different from the problem in~\cite{SES19} where the nodes or symbols that are part of the code are needed to be recovered. We can apply some of the results and constructions from~\cite{SES19} in our case. Recall that we defined $\Sigma = \mathbb{F}_2$. Let $\Sigma^s$ be a vector space of dimension $s$ over $\Sigma$. We can consider each bucket which has $t$ cells, as a subspace of $\Sigma^s$ with dimension $t$ and denote a subspace of dimension $t$ as a \emph{$t$-subspace}. The following claim is motivated by~\cite[Lemma 3]{SES19}.


\begin{claim}
The value of $D(s,k,t,r)$ is the smallest number $m$ of $t$-subspaces of $\Sigma^s$ such that there exists a partition of the subspaces into $k$ subsets, $\cS_1,\ldots,\cS_k$, that satisfies the following property. The size of each subset $\cS_i$ is at most $r$ and for every request $R$, which can be represented by a $1$-subspace $W$, it holds that for each $\cS_i$, $W \subseteq \Sigma_{j=0}^{r'} \cS_{i_j}$ where $\cS_{i_j}$ is the $j$-th subspace in $\cS_i$ and $|\cS_i| = r' \leq r$.

\end{claim}

Let $\bfx = (x_1,x_2,\ldots,x_s)$ be the vector of dimension $1\times s$ with the $s$ information bits and let $V$ be a $t$-subspace of $\Sigma^s$. It is said that a bucket with $t$ cells \emph{stores} a $t$-subspace $V$ if for a given basis $\cB=\{\bfv_1,\bfv_2,\ldots,\bfv_t\}$, the $i$-th cell $i\in[t]$ of the bucket stores the linear combination $\langle \bfv_i, \bfx\rangle$. Note that the choice of the basis $\cB$ does not matter and we can choose any basis of $V$. Each request $R$ which is a linear combination of the $s$ information bits can be represented by a $1$-subspace $W$ of $\Sigma^s$. It is said that a request is \emph{contained} in a bucket $b$ if the set $\{b\}$ is a recovering set for the request. Note that if $W$ is contained in a $t$-subspace $V$ then the request $R$ is contained in the bucket that stores $V$.

Let $\cG_q(s, t)$ denote the set of all $t$-dimensional subspaces of the vector space $\mathbb{F}_q^s$. The set $\cG_q(s, t)$ is often called the \emph{Grassmannian}~\cite{EZ19}. It is well known that
\begin{align*}
|\cG_q(s,t)| = \qbin{s}{t}{q} := \frac{(q^s-1)(q^{s-1}-1)\cdots(q^{s-t+1}-1)}{(q^{t}-1)(q^{t-1}-1)\cdots(q-1)},
\end{align*} 
where $\qbin{s}{t}{q}$ is the $q$-ary Gaussian coefficient~\cite{VW92}.
The following is a definition of \emph{spreads} from~\cite{GPSV17} which are partitions of vector spaces.

\begin{definition}\label{def:spread}
Let $s = at$. Then a set $\cS \subseteq \cG_q(s, t)$ is called a $t$-\textbf{spread} if all elements of $\cS$ intersect only trivially and they cover the whole space $\mathbb{F}^s_q$.
\end{definition}

It is known that the size of a $t$-spread of $\mathbb{F}^s_q$ is $\frac{q^s-1}{q^t-1}$ when $s$ is a multiple of $t$~\cite{GPSV17}. It is also follows that spreads do not exist when $t$ does not divide $s$. In case $s$ is not a multiple of $t$ there is a notion of \emph{partial spreads}, where a \emph{partial $t$-spread} of $\mathbb{F}_q^s$ is a collection of mutually disjoint $t$-subspaces. For the problem we are studying in this section, partial spreads cannot be used due to the fact that they do not necessarily cover the whole space. Thus, in order to deal with the cases when $t$ does not divide $s$ we use \emph{covering designs} which are defined as follows~\cite{EV11}.

\begin{definition}\label{def:CoveringDesign}
A \textbf{covering design} $\mathbb{C}_q(s, t, a)$ is a subset $\cS\subseteq \cG_q(s, t)$ such that each element of $\cG_q(s, a)$ is contained in at least one subspace from $\cS$. 
\end{definition}

The covering number $C_q(s, t, a)$ is the minimum size of a covering design $\mathbb{C}_q(s, t, a)$.
From~\cite[Th. 4.6]{EV11} we get that for any $1\leq t \leq s$,
\begin{equation}\label{eq:CovDesign}
C_q(s,t,1) = \left\lceil\frac{q^s-1}{q^t-1}\right\rceil.
\end{equation}

Note that when $t|s$, an optimal covering design $\mathbb{C}_q(s, t, 1)$ is exactly a $t$-spread of $\mathbb{F}^s_q$.
Now, we will define another family of partitions and another family of codes that can be used to construct locality functional array codes.
The following is a definition of \emph{$\lambda$-fold partitions} from~\cite{ESSSV11}.
\begin{definition}
Let $\lambda$ be a positive integer. A \textbf{$\lambda$-fold partition} of the vector space $V = \mathbb{F}^s_q$ is a multiset $\cS$ of subspaces of $V$ such that every nonzero vector in $V$ is contained in exactly $\lambda$ subspaces in $\cS$.
\end{definition}

Note that a $1$-fold partition of $\mathbb{F}^s_q$ that does not contain a subspace with dimension larger than $t$ is also a covering design $\mathbb{C}_q(s, t, 1)$.
Denote by $A_q(s,t,\lambda)$ the minimum size of a $\lambda$-fold partition of $\mathbb{F}^s_q$ that does not contain a subspace with dimension larger than $t$.
In~\cite{ESSSV11}, it is also possible to find results on $\lambda$-fold partitions. For example, there exists a construction of a $\left(\frac{2^t-1}{2^p-1}\right)$-fold partition of $\Sigma^s$ with $\frac{2^s-1}{2^p-1}$ $t$-subspaces where $p=$gcd$(s,t)$. Therefore, $A_2(s,t,\frac{2^t-1}{2^p-1}) \leq \frac{2^s-1}{2^p-1}$.
Lastly, the following is a definition of \emph{covering Grassmannian codes} from~\cite{EZ19}.
\begin{definition}
For every positive integers $\alpha$ and $\delta$ where $\delta+t \leq s$, an $\alpha$-$(s, t, \delta)_q^c$ \textbf{covering Grassmannian code} $\mathbb{C}$ is a subset of $\cG_q(s, t)$ such that each subset of $\alpha$ codewords of $\mathbb{C}$ spans a subspace whose dimension is at least $\delta+t$ in $\mathbb{F}_q^s$.
\end{definition}

The value $B_q(s,t,\delta;\alpha)$ will denote the maximum size of an $\alpha$-$(s, t, \delta)_q^c$ covering Grassmannian code.
The following theorem summarizes some bounds on $D(s,k,t,r)$ using spreads, covering designs, $\lambda$-fold partitions, and covering Grassmannian codes.

\begin{theorem}\label{SubspaceLocality}
For each $s,t,k$ and $r$ positive integers
\begin{enumerate}
\item $D(s,1,t,1) = C_2(s,t,1) = \left\lceil\frac{2^s-1}{2^t-1}\right\rceil$. \label{SubLocSec1}

\item $D(s,1,t,r) \leq r \cdot \left\lceil\frac{2^{s/r}-1}{2^t - 1}\right\rceil$, where $r|s$. \label{SubLocSec2}

\item $D(s,k,t,1) \leq A_2(s,t,k)$. \label{SubLocSec3}

\item $D(s,\lfloor B_2(s,t,s-t;r)/r \rfloor,t,r) \leq B_2(s,t,s-t;r)$.

\item $D(s,\qbin{s-1}{t-1}{2},t,1) \leq \qbin{s}{t}{2}$, where $t>1$.

\item $D(s,\left\lfloor \frac{2^{s}-2^t}{r\cdot 2^{t}-r}\right\rfloor +1,t,r) \leq \frac{2^{s}-1}{2^{t}-1}$, where $s = rt$. \label{SubLocSec5}

\end{enumerate}
\end{theorem}
\begin{IEEEproof}
\begin{enumerate}
\item To prove this part we use a construction motivated by~\cite[Construction 2]{SES19}. Let $\mathbb{C}$ be a $\mathbb{C}_2(s,t,1)$ covering design with $C_2(s,t,1)$ $t$-subspaces. To construct an $(s,C_2(s,t,1),t,1)$ locality functional array code, we take $C_2(s,t,1)$ buckets where each bucket stores one of the $t$-subspace from $\mathbb{C}$.
From Definition~\ref{def:CoveringDesign}, every $1$-subspace of $\Sigma^s$ is contained in at least one $t$-subspace from $\mathbb{C}$. Thus, each request $R$ which can be represented by a $1$-subspace of $\Sigma^s$, is contained in at least one bucket. Therefore, by using Equation~\eqref{eq:CovDesign} we get that $D(s,1,t,1) \leq C_2(s,t,1) = \left\lceil\frac{2^s-1}{2^t-1}\right\rceil$.

For the other direction, assume that $\cC$ is an $(s,1,m,t,1)$ locality functional array code with $m$ buckets. We construct a $\mathbb{C}_2(s,t,1)$ covering design with $m$ $t$-subspaces of $\Sigma^s$ that are stored in the $m$ buckets of $\cC$. Let $W$ be a $1$-subspace of $\Sigma^s$ that represents a request $R$ for the code $\cC$. From the property of the code $\cC$, there exists one bucket that contains $R$. Therefore, there exists one $t$-subspace in $\mathbb{C}$ that contains $W$. Thus, $C_2(s,t,1) \leq D(s,1,t,1)$.


\item This result holds from part \eqref{SubLocSec1} in this theorem and Theorem~\ref{basicLocality}\eqref{BaLocSec3}.

\item Let $\cS$ be a $k$-fold partition of $\Sigma^s$ that does not contain a subspace with dimension larger than $t$. Assume that $|\cS|=m$. To construct a locality functional array code, we take $m$ buckets where each bucket stores one of the subspaces from $\cS$. Assume that $R$ is the request which can be represented by a vector $\bfu$ of $\Sigma^s$. Then, from the property of the multiset $\cS$, the vector $\bfu$ is contained in exactly $k$ subspaces in $\cS$. Therefore, $R$ is contained in exactly $k$ buckets.
Thus, the $m$ buckets form an $(s,k,m,t,1)$ locality functional array code, and hence, $D(s,k,t,1) \leq A_2(s,t,k)$.


\item Let $\mathbb{C}$ be an $r$-$(s,t,s-t)_2$ covering Grassmannian code with $m$ $t$-subspaces of $\Sigma^s$. We take $m$ buckets where each bucket stores one of the $t$-subspaces from $\mathbb{C}$. Let $R$ be the request. From the property of the code $\mathbb{C}$, every subset of $r$ $t$-subspaces of $\mathbb{C}$ spans the whole space $\Sigma^s$. Hence, every subset of $r$ buckets contains $R$. Therefore, we can partition the $m$ buckets into $\lfloor m/r \rfloor$ parts, where each part contains $R$, and hence, there exist $\lfloor m/r \rfloor$ recovering sets for $R$. Thus, the construction with the $m$ buckets forms an $(s,\lfloor m/r \rfloor,m,t,r)$ locality functional array code.

\item To prove this part we use a construction motivated by~\cite[Construction 1]{SES19}. We construct an $(s,\qbin{s-1}{t-1}{2},\qbin{s}{t}{2},t,1)$ locality functional array code by taking $\qbin{s}{t}{2}$ buckets where each bucket has $t$ cells and stores one of the $t$-subspaces of $\Sigma^s$. Every $1$-subspace of $\Sigma^s$ is contained in exactly $\qbin{s-1}{t-1}{2}$ $t$-subspaces. Therefore, every request $R$ which can be represented by a $1$-subspace is contained in exactly $\qbin{s-1}{t-1}{2}$ buckets. Thus, we get that $D(s,\qbin{s-1}{t-1}{2},t,1)$ $\leq \qbin{s}{t}{2}$.

\item Let $s = rt$ and $\cS$ be a $t$-spread of $\Sigma^s$ such that $|\cS|$ = $\frac{2^s-1}{2^{t}-1}$. To construct a locality functional array code we store each $t$-subspace in $\cS$ in a bucket with $t$ cells. Assume that $R$ is the request which can be represented by a $1$-subspace $W$ of $\Sigma^s$. From the property of spreads, there exists a subspace in $\cS$ that includes $W$. Therefore, there exists a bucket that contains $R$ which can form a recovering set of size $1$. Then, partition the remaining $\frac{2^s-1}{2^{t}-1} - 1 = \frac{2^s - 2^t}{2^t-1}$ buckets into $\left\lfloor \frac{2^{s}-2^t}{r\cdot 2^{t}-r}\right\rfloor$ parts where each part has size $r$. Each part $\cP_i$ has $r$ mutually disjoint $t$-subspaces $U_{i_1},U_{i_2},\ldots,U_{i_r}$. Hence, $\sum_{j=1}^{r} U_{i_j} = \Sigma^s$. Thus, each part $\cP_i$ is a recovering set of $R$ of size $r$. Then, there exist $1+ \left\lfloor \frac{2^{s}-2^t}{r\cdot 2^{t}-r}\right\rfloor$ recovering sets each of size at most $r$ and the code is an $(s,\left\lfloor \frac{2^{s}-2^t}{r\cdot 2^{t}-r}\right\rfloor+1,\frac{2^s-1}{2^{t}-1},t,r)$ locality functional array code.


\end{enumerate}
\end{IEEEproof}

The following is an example of Theorem~\ref{SubspaceLocality}\eqref{SubLocSec3}.

\begin{example}\label{ex:foldLoc}
In this example we will use an example of a $2$-fold partition from~\cite{ESSSV11} in order to construct a locality functional array code. Let $s = 3$. The following multiset $\cS$ of subspaces of $\Sigma^3$ is a $2$-fold partition that does not contain a subspace with dimension larger than $t=2$.

$\cS = \{\{100,011,111\},\allowbreak\{010,001,011\},\allowbreak\{001,110,111\},\allowbreak\{110,010,100\},\allowbreak\{101\},\allowbreak\{101\}\}$.

We represent each element in $\Sigma^3$ as a binary vector of length $3$ and every subspace in $\cS$ by its elements except the zero vector. It holds that any binary vector of length $3$ is contained in exactly two subspaces in $\cS$, and hence, $A_2(3,2,2)\leq 6$. 
We construct a $(3,2,6,2,1)$ locality functional array code with the following buckets that are obtained from $\cS$.
\begin{center}
\begin{tabular}{ |c|c|c|c|c|c| } 
 \hline
 1 & 2 & 3 & 4 & 5 & 6\\
 \hline
 \hline
 $x_3$ & $x_2$ & $x_1$ & $x_2x_3$ & $x_1x_3$ & $x_1x_3$\\ 
 \hline
 $x_1x_2$ & $x_1$ & $x_2x_3$ & $x_2$ & & \\
 \hline
 \end{tabular}
\end{center}

For example, if the request is $x_1+x_2$, then the recovering sets are $\{\{1\},\{2\}\}$.
\end{example}

The following is an example of Theorem~\ref{SubspaceLocality}\eqref{SubLocSec5}.

\begin{example}
For $s=4,t=2$ and $r=2$, the following set $\cS$ is a $2$-spread of $\Sigma^4$ of size $\frac{2^4-1}{2^2-1} = 5$.

$\cS = \{\{0001,0010\},\allowbreak\{0100,1000\},\allowbreak\{0101,1010\},\allowbreak\{1001\allowbreak,0111\},\allowbreak\{0110\allowbreak,1011\}\}$.

We represent each element in $\Sigma^4$ as a binary vector of length $4$ and every $2$-subspace as a basis with $2$ vectors.
We construct a $(4,3,5,2,2)$ locality functional array code with the following buckets that are obtained from $\cS$.
\begin{center}
\begin{tabular}{ |c|c|c|c|c| } 
 \hline
 1 & 2 & 3 & 4 & 5\\
 \hline
 \hline
 $x_1$ & $x_3$ & $x_1x_3$ & $x_1x_4$ & $x_2x_3$\\ 
 \hline
 $x_2$ & $x_4$ & $x_2x_4$ & $x_1x_2x_3$ & $x_1x_2x_4$\\
 \hline
 \end{tabular}
\end{center}

For example, if the request is $x_1+x_2$, then the recovering sets are $\{\{1\},\{2,3\},\{4,5\}\}$.
\end{example}



\subsection{Bounds and Constructions based upon Covering Codes}
In this section we show how covering codes are used to construct locality functional array codes and to get lower bounds for $D(s,k,t,r)$. For the rest of the section we assume that $\bfx = (x_1,x_2,\ldots,x_s)$ is the vector of dimension $1\times s$ with the $s$ information bits. For the case of $t=1$ the following result can be obtained. Remember that $h[s,r]_q$ is the smallest length of a linear covering code over $\mathbb{F}_q$ with covering radius $r$ and redundancy $s$.

\begin{theorem}\label{th:LocCovk=1}
$D(s,1,1,r) = h[s,r]$.
\end{theorem}

\begin{IEEEproof}
There exists an $[h[s,r], h[s,r] - s,r]$ linear covering code with some parity check matrix $H$. To construct a locality functional array code we store in each bucket the linear combination $\langle \bfh_i, \bfx\rangle$ where $\bfh_i$ is the $i$-th column of $H$. Assume that $R$ is the request which can be represented by a binary vector $\bfu \in \Sigma^s$. From Property~\ref{property:covering}, we know that the vector $\bfu $ can be represented as the sum of at most $r$ columns of $H$. Therefore, there exists a recovering set of size at most $r$ for the request $R$. The number of buckets is the number of columns of $H$ which is $h[s,r]$. Thus, $D(s,1,1,r) \leq h[s,r]$.
The lower bound can be obtained from Corollary~\ref{cor:LocToCov} which will appear later.
\end{IEEEproof}

We can generalize the connection of covering codes and locality functional array codes with general $t$. We start by defining a partition of matrices.
\begin{definition}
A \textbf{$t$-partition} of a matrix $H$ is a collection $\cP$ of subspaces of dimension $t$ with the property that every column vector of $H$ is contained in at least one member of $\cP$. A $t$-partition is called \textbf{strict} if every column vector of $H$ is contained in exactly one member of $\cP$.

\end{definition}

The next theorem shows the connection between covering codes and locality functional array codes with $k=1$.
\begin{theorem}\label{th:CoveringToLocality}
Let $H$ be a parity check matrix for an $[n,n-s,r]$ covering code, and let $p$ be the smallest size of a $t$-partition of $H$. Then, $D(s,1,t,r) \leq p$.
\end{theorem}

\begin{IEEEproof}
Let $H$ be a parity check matrix of a given $[n,n-s,r]$ covering code. Let $\cP$ be a $t$-partition of $H$, that contains $p$ subspaces of dimension $t$. We construct an $(s,1,p,t,r)$ locality functional array code $\cC$ by storing each $t$-subspace from $\cP$ in one bucket with $t$ cells. 
Let $\bfu\in \Sigma^s$ be a request which represents the linear combination $\langle \bfu, \bfx\rangle$ of the $s$ information bits. From Property~\ref{property:covering}, we know that there exists a vector $\bfy\in\Sigma^n$ such that $H\cdot \bfy = \bfu$, where $w = w_H(\bfy) \leq r$. If $w_H(\bfy) = r' \leq r$, then the request $\bfu$ is equal to the sum of $r'$ columns of $H$ and denote them by $\bfh_{i_1},\bfh_{i_2},\ldots,\bfh_{i_r'}$. We know that each $1$-subspace with a basis $\{\bfh_{i_j}\},j\in[r']$ is contained in one subspace from the partition $\cP$, and hence, the vector $\bfh_i$ is contained in one bucket of $\cC$. Thus, we can get all the $r'$ columns from at most $r'\leq r$ buckets.
\end{IEEEproof}


Now, the method to get locality functional array codes from covering codes over $\mathbb{F}_q$ is established. We follow an example from~\cite{BPW89} and for that we use the following definition in the rest of this section.
\begin{definition}\label{def:BasisHT}
Let $\cB = \{1, \epsilon, \epsilon^2,\ldots, \epsilon^{w-1}\}$ be a basis for $\mathbb{F}_{2^w}$ over $\Sigma$ where $\epsilon$ is a primitive element of $\mathbb{F}_{2^w}$. For each $i\in[0,2^w-2]$ let $(\epsilon^i)_w$ be the binary column vector of length $w$ that represents the element $\epsilon^i$ of $\mathbb{F}_{2^w}$ with respect to the basis $\cB$. Let $\cU_0$ be the binary matrix of size $(w\times (2^{w}-1))$ that has in column number $i,i\in[0,2^{w}-2]$ the vector $(\epsilon^i)_w$. For each $i\in[0,2^{w}-2]$, let $\cU_i$ be the matrix which is obtained from $\cU_0$ by cyclically rotating its columns $i$ places to the left. Note that for each $i\in[0,2^{w}-2]$ the first column in matrix $\cU_i$ is the vector $(\epsilon^i)_w$.

For an element $\epsilon^i$ over $\mathbb{F}_{2^w}$ let $\cT(\epsilon^i) = \cU_i$ be a matrix over $\Sigma$ of size $(w \times (2^w-1))$ and let $\cT(0)$ be the $(w\times (2^{w}-1))$ zeros matrix. We define the same transformation for vectors and matrices, where for a matrix $M_1$ of size $(a\times b)$ over $\mathbb{F}_{2^w}$ let $\cT(M_1) = M_2$ be the matrix over $\Sigma$ of size $(aw \times b(2^w-1))$ that is obtained from $M_1$ by replacing each element $\alpha$ of $\mathbb{F}_{2^w}$ in the matrix $M_1$ by its appropriate $(w\times (2^{w}-1))$ matrix $\cT(\alpha)$.
\end{definition}

The following is an example to demonstrate Definition~\ref{def:BasisHT}.

\begin{example}
Let $\cB = \{1,\epsilon^1,\epsilon^2\}$ be a basis for $\mathbb{F}_{2^3}$ over $\Sigma$, where $\epsilon$ is a primitive element of $\mathbb{F}_{2^3}$ chosen to satisfy the primitive polynomial $x^3+x+1$, and hence, $\epsilon^3 = \epsilon + 1$. Then, the coordinates of the successive powers of $\epsilon$ with respect to $\cB$ are the columns of the matrix $\cU_0$
\begin{equation*}
\cU_0=
\begin{bmatrix}
1 & 0 & 0 & 1 & 0 & 1 & 1 \\
0&1&0&1&1&1&0\\
0&0&1&0&1&1&1
\end{bmatrix}
.
\end{equation*}

For example, the following matrix is $\cT(\epsilon^1)$
\begin{equation*}
\cT(\epsilon^1)=
\begin{bmatrix}
0&0&1&0&1&1&1\\
1&0&1&1&1&0&0\\
0&1&0&1&1&1&0
\end{bmatrix}
.
\end{equation*}
\end{example}

We show that the transformation defined in Definition~\ref{def:BasisHT} is a linear transformation.

\begin{lemma}\label{lemma:linearT}
The transformation $\cT:\mathbb{F}_{2^w} \rightarrow \mathbb{F}_2^{w\times (2^w-1)}$ is a linear transformation.
\end{lemma}
\begin{IEEEproof}
We want to show that for any $\epsilon^{i_1},\epsilon^{i_2} \in \mathbb{F}_{2^w}$, $\cT(\epsilon^{i_1})+\cT(\epsilon^{i_2}) = \cT(\epsilon^{i_1}+\epsilon^{i_2})$. Assume that $\epsilon^{i_1}+\epsilon^{i_2} = \epsilon^{i_3}$. From Definition~\ref{def:BasisHT}, we know that $\cT(\epsilon^{i_1})+\cT(\epsilon^{i_2}) = \cU_{i_1}+\cU_{i_2}$. From Definition~\ref{def:BasisHT}, for every $j\in[2^w-1]$, the $j$-th column of $\cU_{i_1},\cU_{i_2},\cU_{i_3},\cU_{i_1}+\cU_{i_2}$ is $(\epsilon^{i_1+j})_w,(\epsilon^{i_2+j})_w,(\epsilon^{i_3+j})_w,(\epsilon^{i_1+j} + \epsilon^{i_2+j})_w$, respectively.
Also, $\epsilon^{i_1+j} + \epsilon^{i_2+j}= \epsilon^{j}(\epsilon^{i_1} + \epsilon^{i_2}) = \epsilon^{i_3+j}$. Thus, the $j$-th column of $\cU_{i_3}$ is equal to the $j$-th column of $\cU_{i_1}+\cU_{i_2}$ for all $j\in[2^w-1]$. Thus, $\cT(\epsilon^{i_1})+\cT(\epsilon^{i_2}) = \cU_{i_1}+\cU_{i_2} = \cU_{i_3} = \cT(\epsilon^{i_1}+\epsilon^{i_2})$.
\end{IEEEproof}

The same transformation $\cT$ that was defined for vectors and matrices in Definition~\ref{def:BasisHT} is also a linear transformation following similar proof as for Lemma~\ref{lemma:linearT}.
The following result can be found in~\cite[Lemma 3.1]{BPW89}, but we want to prove it in a different way, by constructing a specific parity check matrix in order to use it in other claims.
\begin{lemma}\label{lemma:2^wTo2}
Let $H$ be a parity check matrix of an $[n,n-s,r]_{2^w}$ covering code. Then, the matrix $\cT(H)$ is a parity check matrix of a binary $[(2^w - 1)n, (2^w - 1)n - ws,r]$ covering code. In particular, $h[ws,r] \leq (2^w - 1)\cdot h[s,r]_{2^w}$.
\end{lemma}

\begin{IEEEproof}
Let $\cC$ be an $[n,n-s,r]_{2^w}$ covering code and let $H$ be a parity check matrix of the code $\cC$ of size $(s\times n)$. 
We want to show that the matrix $H' = \cT(H)$ is a parity check matrix of a binary $[(2^w - 1)n, (2^w - 1)n - ws,r]$ covering code. The size of $H'$ is $(ws\times (2^w-1)n)$. Given a binary column vector $\bfu$ of length $ws$, we show that there are at most $r$ columns of $H'$ that their sum is $\bfu$. 

The vector $\bfu$ can be partitioned into $s$ vectors of length $w$ where $\bfu = (\bfu_1,\bfu_2,\ldots,\bfu_s)^\intercal$. Each vector $\bfu_i$ of length $w$ can represent an element of $\mathbb{F}_{2^w}$ according to the basis $\cB$ from Definition~\ref{def:BasisHT}. Hence, $\bfu = ((\epsilon^{i_1})_w^\intercal,(\epsilon^{i_2})_w^\intercal,\ldots,(\epsilon^{i_s})_w^\intercal)^\intercal$ 
and from the $s$ elements we can get a column vector $\bfv = (\epsilon^{i_1},\epsilon^{i_2},\ldots,\epsilon^{i_s})^\intercal$ of dimension $s\times 1$ over $\mathbb{F}_{2^w}$. The first column in each $\cU_i,i\in[0,2^w-2]$ is the vector $(\epsilon^i)_w$. Then, from the construction of $\cT(\bfv)$, the first column of the matrix $\cT(\bfv)$ is the vector $\bfu$.

From the property of the code $\cC$, it is known that there exists a vector $\bfy\in\mathbb{F}_{2^w}^n$ such that $H\cdot \bfy = \bfv$, where $w_H(\bfy) \leq r$. Let $\cA = \{i: i\in [n], y_i \neq 0\}$ and note that $|\cA|\leq r$. Let $\bfh_i$ be the $i$-th column of $H$. Then, $\sum_{i\in\cA} y_i\bfh_i = \bfv$. For each $i\in\cA$ we define $\bfh'_i = y_i \bfh_i$ and from the linearity of the transformation $\cT$ we have $\cT(\bfv) = \cT(\sum_{i\in\cA} \bfh'_i) = \sum_{i\in\cA} \cT(\bfh'_i)$. Thus, the vector $(\sum_{i\in\cA} \cT(\bfh'_i))_1 = \sum_{i\in\cA} \cT(\bfh'_i)_1 = \bfu$, where $\cT(\bfh'_i)_1$ is the first column of the matrix $\cT(\bfh'_i)$. 

For each $i\in \cA$, assume that $y_i = \epsilon^{j_i}$. Then, the first column of the matrix $\cT(\bfh'_i)$ is the $j_i$-th column of the matrix $\cT(\bfh_i)$. Thus, $\sum_{i\in\cA} \cT(\bfh_i)_{j_i} = \bfu$, where $\cT(\bfh_i)_{j_i}$ is the $j_i$-th column of the matrix $\cT(\bfh_i)$. For each $i\in\cA$, the matrix $\cT(\bfh_i)$ has size $(ws\times (2^w-1))$ and it is a sub matrix of $H'$ that starts in the column number $(2^w-1)(i-1)+1$ of $H'$. Hence, the $j_i$-th column of the matrix $\cT(\bfh_i)$ is the column number $(2^w-1)(i-1)+j_i$ of the matrix $H'$. Therefore, $\sum_{i\in\cA} \bfh'_{(2^w-1)(i-1)+j_i} = \bfu$, where $\bfh'_i$ is the $i$-th column of $H'$. Thus, the vector $\bfu$ is a sum of $|\cA| \leq r$ columns of $H'$ and the matrix $H'$ is a parity check matrix of a binary $[(2^w - 1)n, (2^w - 1)n - ws,r]$ covering code.
\end{IEEEproof}

An upper bound on the value of $D(s,1,t,r)$ can be obtained in the next theorem using non-binary covering codes.
\begin{theorem}\label{th:covToLoc}
For any positive integer $w$ such that $t|w$, $D(ws,1,t,r) \leq \dfrac{(2^w - 1)h[s,r]_{2^w}}{2^t - 1}$.
\end{theorem}
\begin{IEEEproof}
Let $\cC$ be an $[n, n- s,r]_{2^w}$ covering code over $\mathbb{F}_{2^w}$, where $n=h[s,r]_{2^w}$. Let the matrix $H$ be a parity check matrix of $\cC$ of size $(s \times n)$. From Lemma~\ref{lemma:2^wTo2}, we get that there exists a binary $[(2^w - 1)n, (2^w - 1)n - ws,r]$ covering code with parity check matrix $H' = \cT(H)$. We want to find the smallest size of a $t$-partition of $H'$.



Let $j_1,j_2,j_3\in[0,2^w-2]$ be such that $\epsilon^{j_1} + \epsilon^{j_2} = \epsilon^{j_3}$. Then, in the matrix $\cU_0$ from Definition~\ref{def:BasisHT}, it holds that the sum of the $j_1$-th and $j_2$-th columns is the $j_3$-th column. In the matrix $\cU_i,i\in[0,2^w-2]$ the $j_1$-th, $j_2$-th, $j_3$-th column is ($\epsilon^i \cdot \epsilon^{j_1})_w, (\epsilon^i \cdot \epsilon^{j_2})_w, (\epsilon^i \cdot \epsilon^{j_3})_w$, respectively. It holds that $\epsilon^i \cdot \epsilon^{j_1} + \epsilon^i \cdot \epsilon^{j_2} = \epsilon^i \cdot (\epsilon^{j_1} + \epsilon^{j_2}) = \epsilon^i \cdot \epsilon^{j_3}$. Thus, we can conclude that in the matrix $\cU_i,i\in[0,2^w-2]$ it also holds that the sum of the $j_1$-th and $j_2$-th columns is the $j_3$-th column. Let $(\cU_i)_{j}$ be the $j$-th column of $\cU_i$.
Assume that a basis that includes the columns $\{(\cU_0)_{j_1},(\cU_0)_{j_2},\ldots,(\cU_0)_{j_t}\}$ spans the columns $\{(\cU_0)_{j_1},(\cU_0)_{j_2},\ldots,(\cU_0)_{j_{2^t-1}}\}$ of the matrix $\cU_0$. Then, the basis that includes the columns $\{(\cU_i)_{j_1},(\cU_i)_{j_2},\ldots,(\cU_i)_{j_t}\}$ spans the columns $\{(\cU_i)_{j_1},(\cU_i)_{j_2},\ldots,(\cU_i)_{j_{2^t-1}}\}$ of the matrix $\cU_i,i\in[0,2^w-2]$.

The matrix $\cU_0$ includes all the nonzero column vectors of length $w$, which means that it includes the space $\mathbb{F}^w_2\setminus \{0\}$. It is given that $t|w$. Hence, there exists a $t$-spread of $\mathbb{F}^w_2$. Thus, there exists a strict $t$-partition $\cP$ of $\cU_0$ with $p = \frac{2^w-1}{2^t-1}$ $t$-subspaces.
Each subspace of $\cP$ is represented by a basis of $t$ column vectors of $\cU_0$ and denote them by $\{\{(\cU_0)_{j^1_1},\allowbreak(\cU_0)_{j^1_2},\allowbreak\ldots,(\cU_0)_{j^1_t}\},\allowbreak\{(\cU_0)_{j^2_1},\allowbreak(\cU_0)_{j^2_2},\allowbreak\ldots,\allowbreak(\cU_0)_{j^2_t}\},\allowbreak\ldots,\allowbreak \{(\cU_0)_{j^p_1},\allowbreak(\cU_0)_{j^p_2},\allowbreak\ldots,\allowbreak(\cU_0)_{j^p_t}\}\}$. The $p$ $t$-subspaces $\{\{(\cU_i)_{j^1_1},\allowbreak(\cU_i)_{j^1_2},\allowbreak\ldots,\allowbreak(\cU_i)_{j^1_t}\},\allowbreak\{(\cU_i)_{j^2_1},\allowbreak(\cU_i)_{j^2_2},\allowbreak\ldots,\allowbreak(\cU_i)_{j^2_t}\},\allowbreak\ldots,\allowbreak \{(\cU_i)_{j^p_1},\allowbreak(\cU_i)_{j^p_2},\allowbreak\ldots,\allowbreak(\cU_i)_{j^p_t}\}\}$ form a strict $t$-partition of $\cU_i$.
For each $i\in[n]$ let $\bfh_i$ be the $i$-th column of the matrix $H$. The matrix $\cT(\bfh_i)$ includes $s$ matrices of size $(w \times (2^w-1))$ that all have the same partition regarding the column numbers. Hence, the partition $\{\{((\cU_{i_1})^\intercal_{j^1_1},\allowbreak(\cU_{i_2})^\intercal_{j^1_1},\allowbreak\ldots,\allowbreak(\cU_{i_s})^\intercal_{j^1_1})^\intercal,\allowbreak\ldots,\allowbreak((\cU_{i_1})^\intercal_{j^1_t},\allowbreak(\cU_{i_2})^\intercal_{j^1_t},\allowbreak\ldots,\allowbreak(\cU_{i_s})^\intercal_{j^1_t})^\intercal\},\allowbreak\ldots,\allowbreak \{((\cU_{i_1})^\intercal_{j^p_1},\allowbreak(\cU_{i_2})^\intercal_{j^p_1},\allowbreak\ldots,\allowbreak(\cU_{i_s})^\intercal_{j^p_1})^\intercal,\allowbreak\ldots,\allowbreak((\cU_{i_1})^\intercal_{j^p_t},\allowbreak(\cU_{i_2})^\intercal_{j^p_t},\allowbreak\ldots,\allowbreak(\cU_{i_s})^\intercal_{j^p_t})^\intercal\}\}$
is a strict $t$-partition of $\cT(\bfh_i)$ with $p=\frac{2^w-1}{2^t-1}$ $t$-subspaces.
Therefore, there exits a strict $t$-partition of the matrix $H'$ with $\frac{(2^w-1)n}{2^t-1}$ $t$-subspaces. Thus, By using Theorem~\ref{th:CoveringToLocality} we get that $D(ws,1,t,r) \leq \dfrac{(2^w - 1)h[s,r]_{2^w}}{2^t - 1}$.
\end{IEEEproof}

We can use Theorem~\ref{th:covToLoc} to find upper bounds on the value of $D(s,1,t,r)$ by using previous bounds on the size of non-binary covering codes.
\begin{example}\label{ex:covToLocEx}
\begin{enumerate}
\item In~\cite{DO01} a $[1097,1097-8,2]_{2^3}$ covering code is provided. Thus, $h[8,2]_{2^3} \leq 1097$. Then, from Theorem~\ref{th:covToLoc}$, D(3\cdot 8,1,3,2) = D(24,1,3,2) \leq \frac{2^3 - 1}{2^3-1}h[8,2]_{2^3} = 1097$. For a lower bound, we can use Theorem~\ref{basicLocality}\eqref{section1} to get $D(24,1,3,2) \geq 828$. 

\item For $r=3$, the following result can be obtained from~\cite[Theorem 4.3]{DGMP08}. For $q=4$ and $p = 3$, $h[s=3p+2,3]_{q} \leq (9\cdot q^{2} + 2\frac{q^2-1}{q-1}) = 154$. Hence, $h[11,3]_{2^2} \leq 154$. From Theorem~\ref{th:covToLoc}, $D(22,1,2,3)\leq 154$. For a lower bound, we can use Theorem~\ref{basicLocality}\eqref{section1} to get $D(22,1,2,3) \geq 99$. 
\end{enumerate}
\end{example}

The following is another use of Theorem~\ref{th:covToLoc} to find bounds on the value of $D(s,1,t,r)$ using another general family of non-binary covering codes.
\begin{corollary}\label{th:th58}
For any positive integers $w$ and $t$, where $t|w$, $D(4w,1,t,2) \leq \dfrac{(2^w - 1)(2^{w+1}+1)}{2^t - 1}$.
\end{corollary}

\begin{IEEEproof}
In~\cite[Theorem 3.2]{BPW89} there exists a construction of a $(4\times (2^{w+1}+1))$ parity check matrix $H$ of a $[2^{w+1}+1,2^{w+1}+1 - 4,2]_{2^w}$ covering code over $\mathbb{F}_{2^w}$. Therefore, $h[4,2]_{2^w} \leq 2^{w+1}+1$. From Theorem~\ref{th:covToLoc} we get $D(4w,1,t,2) \leq \dfrac{(2^w - 1)(2^{w+1}+1)}{2^t - 1}$.
\end{IEEEproof}

For any positive integers $w$ and $t$, where $t|w$ we have $D(4w,1,t,2)\leq 2\cdot \frac{2^{2w}-1}{2^t-1}$ from Theorem~\ref{SubspaceLocality}\eqref{SubLocSec2}, and from Corollary~\ref{th:th58} we get $D(4w,1,t,2) \leq \frac{(2^w - 1)(2^{w+1}+1)}{2^t - 1}$. Thus, we can save $2\cdot \frac{2^{2w}-1}{2^t-1} - \frac{(2^w - 1)(2^{w+1}+1)}{2^t - 1} = \frac{2^{w}-1}{2^t-1}$ buckets.

The following is an example of a locality functional array code that is obtained from Corollary~\ref{th:th58}.
\begin{example}
For the case of $w=4$ and $t=2$, we have $s=4w=16$. Let $V$ be $\mathbb{F}_{2^w}^4=\mathbb{F}_{16}^4$. To get a basis for $V$ as a vector space over $\Sigma$, we first choose a basis $\cB = \{1, \epsilon, \epsilon^2, \epsilon^3\}$ for $\mathbb{F}_{16}$ over $\Sigma$ where $\epsilon$ is a primitive element of $\mathbb{F}_{16}$ chosen to satisfy the primitive polynomial $x^4+x+1$. It holds that $\epsilon^4 = \epsilon + 1$. Then, the coordinates of the successive powers of $\epsilon$ with respect to the basis $\cB$ are the columns of the matrix
\setcounter{MaxMatrixCols}{20}
\begin{equation*}
\cU_0=
\begin{bmatrix}
1 & 0 & 0 & 0 & 1 & 0 & 0 & 1 & 1 & 0 & 1&0&1&1&1\\
0&1&0&0&1&1&0&1&0&1&1&1&1&0&0\\
0&0&1&0&0&1&1&0&1&0&1&1&1&1&0\\
0 & 0 & 0 & 1 & 0 & 0 &1 &1&0&1&0&1&1&1&1
\end{bmatrix}
.
\end{equation*}

In~\cite[Theorem 3.2]{BPW89}, there exists a construction of a parity check matrix of a $[33,33-4,2]_{2^w}$ covering code.

\setcounter{MaxMatrixCols}{20}
\begin{equation*}
H = 
\begin{bmatrix}
1 & 1 & 1 & \cdots & 1 & 1 & 0 & 0 & 0 & 0 & \cdots & 0\\
1 & \epsilon^1 & \epsilon^2 & \cdots & \epsilon^{14} & 0 &1&0&0&0&\cdots&0\\
1 & \epsilon^2 & \epsilon^4 & \cdots & \epsilon^{13} & 0&0&0&1&1&\cdots&1\\
1 & 0 & 0 & \cdots & 0 & 0 & 0 &1 &1 & \epsilon^1 & \cdots & \epsilon^{14}

\end{bmatrix}
.
\end{equation*}




Let $(\cU_i)_{j}$ be the $j$-th column of $\cU_i$. The following is a strict $t$-partition of $\cU_i$, $\cP_i=\{\{(\cU_i)_1,\allowbreak(\cU_i)_6,\allowbreak(\cU_i)_{11}\},\allowbreak\{(\cU_i)_2,\allowbreak(\cU_i)_7,\allowbreak(\cU_i)_{12}\},\allowbreak\{(\cU_i)_3,\allowbreak(\cU_i)_8,\allowbreak(\cU_i)_{13}\},\allowbreak\{(\cU_i)_4,\allowbreak(\cU_i)_9,\allowbreak(\cU_i)_{14}\},\allowbreak\{(\cU_i)_5,\allowbreak(\cU_i)_{10},\allowbreak(\cU_i)_{15}\}\}$, where we represent every subspace in $\cP$ by its elements except the zero vector. In addition, each subspace can be represented by a basis of two vectors.

From Lemma~\ref{lemma:2^wTo2} we get that $H' = \cT(H)$ is a parity check matrix of a binary $[495,495-16,2]$ covering code. Recall that in the transformation $\cT$, each element of $\mathbb{F}_{16}$ is replaced with an appropriate matrix $\cU_i$ of size $(4\times 15)$. Each column in $H$ has $4$ elements of $\mathbb{F}_{16}$ and is replaced with $4$ matrices such that each matrix $\cU_i$ of size $(4\times 15)$ that has a strict $t$-partition $\cP_i$ with $5$ subspaces. Each column in $H$ is a $(16\times 15)$ matrix in $H'$, which can be stored in $5$ buckets such that each bucket stores one subspace from the partition, and hence, the $33$ columns of $H$ can be stored in $33*5 = 165$ buckets. Thus, we get that $D(16,1,2,2) \leq 165$.
\end{example}

Next, another possible way to obtain locality functional array codes from covering codes is presented. First, we define a possible modification for matrices that we will use in order to construct new parity check matrices for covering codes from given parity check matrices.
\begin{definition}
Given a matrix $H$ of size $(n\times s)$, its $i$-th \textbf{modified matrix} denoted by $H^{(i)}$ of size $(n+1\times s)$ is the matrix that has the same rows of $H$ except of row $i$, where it has the complement of row $i$ of $H$, with an additional column with only $1$ in row $i$.
\end{definition}
 
The next theorem shows that for a given parity check matrix of a covering code, the modified matrix is also a parity check matrix of another covering code. Even though the following seems to be a basic property, we could not find its proof, and hence, we add the following proof for completeness. 

\begin{theorem}\label{th:CovMat}
For a parity check matrix $H$ for a binary $[n,n-s,2]$ covering code and an integer $i$, the $i$-th modified matrix $H^{(i)}$ is also a parity check matrix of a binary $[n+1,n+1-s,2]$ covering code.

\end{theorem}
\begin{IEEEproof}
Let $H$ be a parity check matrix of an $[n,n-s,2]$ covering code. For a given $i\in[s]$, let $H^{(i)}$ be the $i$-th modified matrix of $H$. The size of $H^{(i)}$ is $(s\times(n+1))$. From Property~\ref{property:covering}, for each vector $\bfv\in\Sigma^s$ there exists a vector $\bfy\in \Sigma^n$ such that $H\cdot \bfy = \bfv$ where $w_H(\bfy) \leq 2$. Let $\bfh_i,\bfh'_i$ be the $i$-th column of $H,H^{(i)}$, respectively. If $w_H(\bfy)=2$, assume that $\bfv = \bfh_{j_1}+\bfh_{j_2}$. The column vector $\bfh'_j$ is different from the column vector $\bfh_j$ only in row $i$, where $\bfh'_j$ has the complement of the element in row $i$ in $\bfh_j$. Thus, it holds that $\bfv = \bfh'_{j_1}+\bfh'_{j_2}$.

If $w_H(\bfy)=1$, assume that $\bfv = \bfh_j$. From the construction of $H^{(i)}$, it holds that $\bfh_j = \bfh'_j + \bfh'_{n+1}$. Therefore, we can get $\bfv$ as a sum of two columns of $H^{(i)}$. Thus, $H^{(i)}$ is a parity check matrix of a binary $[n+1,n+1-s,2]$ covering code.
\end{IEEEproof}

One possible way to use Theorem~\ref{th:CovMat} to get locality functional array codes is shown next.

\begin{theorem}\label{th:LOCs=7}
$D(7,1,2,2) = 7$.
\end{theorem}
\begin{IEEEproof}
From~\cite[Theorem 1]{GDT91} and the example after it, we can get a construction of a parity check matrix for a binary $[19,19-7,2]$ covering code. The following is a parity check matrix $H$ of the code.
\begin{equation*}
\begin{scriptsize}
\begin{bmatrix}
0&0&0&1&1&1&1&1&1&1&1&1&1&1&1&1&1&1&1 \\
0&1&1&0&0&1&1&0&0&1&1&0&0&1&1&0&0&0&0 \\
1&0&1&0&1&0&1&0&1&0&1&0&1&0&1&0&0&0&0 \\
0&0&0&0&0&1&1&0&1&0&1&0&1&1&0&0&0&1&1 \\
0&0&0&0&1&1&0&0&0&1&1&0&1&0&1&0&1&0&1 \\
0&0&0&0&0&0&0&0&0&0&0&1&1&1&1&1&1&1&1 \\
0&0&0&0&0&0&0&1&1&1&1&0&0&0&0&1&1&1&1 \\
\end{bmatrix}
.
\end{scriptsize}
\end{equation*}

The following is the matrix $H^{(1)}$, the first modified matrix of $H$ where the first row is the complement of the first row of $H$ and a new column with only $1$ in the first entry is added.

\noindent
\scalebox{0.84}{%
$\begin{bmatrix}
1&1&1&0&0&0&0&0&0&0&0&0&0&0&0&0&0&0&0&1 \\
0&1&1&0&0&1&1&0&0&1&1&0&0&1&1&0&0&0&0&0 \\
1&0&1&0&1&0&1&0&1&0&1&0&1&0&1&0&0&0&0&0 \\
0&0&0&0&0&1&1&0&1&0&1&0&1&1&0&0&0&1&1&0 \\
0&0&0&0&1&1&0&0&0&1&1&0&1&0&1&0&1&0&1&0 \\
0&0&0&0&0&0&0&0&0&0&0&1&1&1&1&1&1&1&1&0 \\
0&0&0&0&0&0&0&1&1&1&1&0&0&0&0&1&1&1&1&0 \\
\end{bmatrix}.$}

From Theorem~\ref{th:CovMat}, the matrix $H^{(1)}$ is a parity check matrix of a binary $[20,20-7,2]$ covering code. Note that the fourth column is all zero column which we can remove to get the following matrix $H^{(1)'}$ which is a parity check matrix of a binary $[19,19-7,2]$ covering code.
\begin{equation*}
\begin{scriptsize}
\begin{bmatrix}
1&1&1&0&0&0&0&0&0&0&0&0&0&0&0&0&0&0&1 \\
0&1&1&0&1&1&0&0&1&1&0&0&1&1&0&0&0&0&0 \\
1&0&1&1&0&1&0&1&0&1&0&1&0&1&0&0&0&0&0 \\
0&0&0&0&1&1&0&1&0&1&0&1&1&0&0&0&1&1&0 \\
0&0&0&1&1&0&0&0&1&1&0&1&0&1&0&1&0&1&0 \\
0&0&0&0&0&0&0&0&0&0&1&1&1&1&1&1&1&1&0 \\
0&0&0&0&0&0&1&1&1&1&0&0&0&0&1&1&1&1&0 \\
\end{bmatrix}
.
\end{scriptsize}
\end{equation*}

Let $\bfh'_j$ be the $j$-th column of the matrix $H^{(i)'}$. We can find a $2$-partition of the matrix $H^{(1)'}$. We will present the partition as a set of $7$ $2$-subspaces such that each subspace is presented by a basis with two columns of $H^{(i)'}$.  
The following is a possible $2$-partition of $H^{(i)'}$ $\cP = \{\{\bfh'_7,\bfh'_{11}\},\allowbreak\{\bfh'_8,\bfh'_{12}\},\allowbreak\{\bfh'_9,\bfh'_{13}\},\allowbreak\{\bfh'_{10},\bfh'_{14}\},\allowbreak\{\bfh'_4,\bfh'_5\},\allowbreak\{\bfh'_1,\bfh'_2\},\allowbreak\{\bfh'_3,\bfh'_{19}\}\}$. We can see that $14$ out of $19$ columns form the bases. It can be verified that $\bfh'_4 + \bfh'_5 = \bfh'_{6}$, $\bfh'_7+\bfh'_{11} = \bfh'_{15}$, $\bfh'_8+\bfh'_{12} = \bfh'_{16}$, $\bfh'_{10} + \bfh'_{14} = \bfh'_{17}$ and $\bfh'_9 + \bfh'_{13} = \bfh'_{18}$. Therefore, $\cP$ is a $2$-partition of $H^{(i)'}$ with size $7$. Thus, from Theorem~\ref{th:CoveringToLocality} we get that $D(7,1,2,2)\leq 7$.

For the lower bound, assume by contradiction that there exists a $(7,1,6,2,2)$ locality functional array code. Then, from Theorem~\ref{th:LocToCov} we get that $h[7,2] \leq 18$. But from~\cite{CHLL97} we have that $h[7,2] = 19$, which is a contradiction. Thus, $D(7,1,2,2)\geq 7$.
\end{IEEEproof}

Next, we show how to construct covering codes using locality functional array codes.

\begin{theorem}\label{th:LocToCov}
Let $\cC$ be an $(s,1,m,t,r)$ locality functional array code. Then, $h[s,r] \leq m\cdot(2^t - 1)$.
\end{theorem}
\begin{IEEEproof}
Assume that $\cC$ is an $(s,1,m,t,r)$ locality functional array code which has $m$ buckets such that in each bucket stored at most $t$ linear combinations of the $s$ information bits. From the $t$ cells in each bucket we can get at most $(2^t-1)$ different linear combinations. We can represent each linear combination as a binary vector of length $s$. Then, we construct an $(s\times m\cdot(2^t-1))$ parity check matrix $H$ where we have all the vectors that we get from the linear combinations of all the $m$ buckets as columns of the matrix. Let $\bfu\in \Sigma^s$ be a column vector of length $s$ which can represent a request for the code $\cC$. From the property of $\cC$, there exists a recovering set $S\subseteq [m]$ where $|S|\leq r$ that satisfies the request. Assume that $S = \{b_1,b_2,\ldots,b_{r'}\}$ where $r'\leq r$. From each bucket $b_i \in S$ we read a linear combination $\bfv_i$ of the $t$ cells which is a linear combination of the $s$ information bits. From the construction of $H$, the column vector $\bfv_i$ is a column in $H$.
Then, $\bfu = \sum_{i=1}^{r'} \bfv_i$, and hence, the vector $\bfu$ is a sum of at most $r$ columns of $H$. Thus, the matrix $H$ is a parity check matrix of a binary $[m\cdot(2^t-1),m\cdot(2^t-1) - s,r]$ covering code, and hence, $h[s,r] \leq m\cdot(2^t-1)$.
\end{IEEEproof}

Now we will use Theorem~\ref{th:LocToCov} to get a lower bound on the value of $D(s,1,t,r)$.
\begin{corollary}\label{cor:LocToCov}
$D(s,1,t,r) \geq \left\lceil\dfrac{h[s,r]}{2^t-1}\right\rceil$.
\end{corollary}
\begin{IEEEproof}
Assume by contradiction that $D(s,1,t,r) = m < \left\lceil\dfrac{h[s,r]}{2^t-1}\right\rceil$. The number of buckets $m$ is an integer. Then, $m < \dfrac{h[s,r]}{2^t-1}$. From Theorem~\ref{th:LocToCov} we have $h[s,r]\leq m\cdot (2^t-1) <  \dfrac{h[s,r]}{2^t-1} \cdot (2^t-1) = h[s,r]$ which is a contradiction.
\end{IEEEproof}

We can get upper bounds on the value $h[s,r]$ from~\cite{CHLL97}. For example, $h[2s-1,2]\geq 2^s-1$ for any $s\geq 3$ and we can conclude that $D(2s-1,1,t,2) \geq \left\lceil\dfrac{2^s-1}{2^t-1}\right\rceil$.


\section{Conclusion}\label{sec:conc}
In this work we studied constructions and bounds of several families of codes. We defined and presented functional PIR array codes, functional batch array codes, and locality functional array codes. Lower bounds on the smallest number of buckets of these codes were given. Several upper bounds on the smallest number of buckets were shown based on general constructions, specific constructions, subspaces, and covering codes. In Table~\ref{sumTable}, we provide a summary of most of the results that appear in the work.
The first column specifies the family of codes that the result refers to. Denote a PIR array code, batch array code, functional PIR array code, functional batch array code, locality functional array code by $P,B,FP,FB,L$, respectively. The next five columns specify the values of the parameters of the codes. The following two columns refer to lower and upper bounds on the codes and the last column includes notes such as constraints on the parameters and where the results appeared in the work.
Lastly, we note that there are plenty of problems which remain for future research, such as generalizing the specific constructions and finding new bounds for different parameters.

\begin{table*}
\begin{center}
\caption{Summary of the results}\label{sumTable}
\begin{tabular}{ |c|c|c|c|c|c|c|c|c| } 
 \hline
   Code & $s$ & $k$ & $t$ & $\ell$ & $r$ & Lower bound & Upper bound & notes \\ &&&&&&&&\\
   \hline
   \hline
   $FP/FB$ & $s$ & 1 & $t$ & $t$ & $-$ & $\left\lceil \frac{s}{t} \right\rceil$ & $\left\lceil \frac{s}{t} \right\rceil$ &\\&&&&&&&&\Tref{theorem:ArrayCodek1}\\
   \hline
   $FP/FB$ & $s$ & 1 & $t$ & $1$ & $-$ & $\left\lceil \frac{s}{\log_2(t+1)}\right\rceil$ &  $\left\lceil \frac{s}{\lfloor \log_2(t+1)\rfloor}\right\rceil$ & \\&&&&&&&&\Tref{theorem:ArrayCodek1}\\
   \hline
   $FP/FB$ & $s$ & 1 & $t$ & $t/2$ & $-$ & $\frac{s}{t} + 1$ & $\frac{s}{t} + 1$ & $t$ is even, $\frac{s}{t}$ is integer, and $\frac{s}{t} \leq t-1$\\&&&&&&&&\Tref{theorem:ArrayCodek1}\\
   \hline
   $FP/FB$ & $s_1+s_2$ & 1 & $t$ & $1$ & $-$ & $\left\lceil \frac{s_1+s_2}{\log_2(t+1)}\right\rceil$ & $\left\lceil\frac{s_1}{\left\lfloor \log_2(t+1) \right\rfloor}\right\rceil + 1$ & $2^{s_2} -1 \hspace{-0.5ex}\leq \left(\left\lceil\frac{s_1}{\left\lfloor \log_2(t+1) \right\rfloor}\right\rceil+1\right)\cdot $  \\&&&&&&&& $(t - (2^{\lfloor \log_2(t+1) \rfloor}-1))$ \Tref{theorem:FPIRell1}\\
   \hline
   $FP/FB$ & $s$ & 1 & $t$ & $\alpha t$ & $-$ & $$ & $\left\lceil \frac{s}{t-g[t, \alpha t]} \right\rceil $ & $0 < \alpha < 1$ \\ &&&&&&&&\Tref{theorem:ArrayCodek1}\\
   \hline
   $FB$ & $s$ & k & $t$ & $1$ & $-$ & & $FB(\frac{s}{t}, t\cdot k)$ & $\frac{s}{t}$ is integer \\ &&&&&&&& \Lref{lemma:FBGadget}\\
   \hline
   $FB$ & $8$ & 2 & $2$ & $2$ & $-$ & $6$ & $7$ & \\ &&&&&&&&\Tref{theorem:FB2282}\\
   \hline
   $FB$ & $s$ & 2 & $2$ & $2$ & $-$ & $\log_{7}(2^{s-1}\cdot (2^s - 1))$ & $7\cdot \left\lceil \frac{s}{8} \right\rceil$ & \\ &&&&&&&&\Cref{cor:FB22s2}\\
   \hline
   $P$ & $r^2 + r$ & $r$ & $r^2 - r + 1$ & $r-1$ & $-$ & $r+1$ & $r+1$ & $r \geq 3$ \\ &&&&&&&&\Tref{theorem:PExample8}\\
   \hline
   $B$ & $r^2 + r$ & $r$ & $r^2 - r + 1$ & $r-1$ & $-$ & $r+1$ & $r+1$ & $r \geq 3$ \\ &&&&&&&&\Tref{theorem:BExample8}\\
   \hline
   $B$ & $6$ & $15$ & $2$ & $2$ & $-$ & $25$ & $25$ & $$ \\ &&&&&&&&\Tref{theorem:BExample9}\\
   \hline
   $FP$ & $6$ & $11$ & $2$ & $2$ & $-$ & $21$ & $25$ & $$ \\ &&&&&&&&\Tref{theorem:FPExample9}\\
   \hline
   $P$ & $4$ & $16$ & $2$ & $1$ & $-$ & $23$ & $25$ & $$ \\ &&&&&&&&\Tref{theorem:P21416}\\
   \hline
   $FP$ & $4$ & $14$ & $2$ & $2$ & $-$ & $24$ & $25$ & $$ \\ &&&&&&&&\Tref{theorem:FP22414}\\
   \hline
   $FP$ & $5$ & $48$ & $2$ & $2$ & $-$ & $88$ & $90$ & $$ \\ &&&&&&&&\Tref{theorem:FP22548}\\
   \hline
   $L$ & $s$ & $1$ & $t$ & $t$ & $1$ & $\left\lceil\frac{2^s-1}{2^t-1}\right\rceil$ & $\left\lceil\frac{2^s-1}{2^t-1}\right\rceil$ & $$ \\ &&&&&&&&\Tref{SubspaceLocality}\\
    \hline
   $L$ & $s$ & $1$ & $t$ & $t$ & $r$ & $$ & $r\cdot \left\lceil\frac{2^{s/r}-1}{2^t-1}\right\rceil$ & $r|s$ \\ &&&&&&&&\Tref{SubspaceLocality}\\
   \hline
   $L$ & $s$ & $\qbin{s-1}{t-1}{2}$ & $t$ & $t$ & $1$ & $$ & $\qbin{s}{t}{2}$ & $$ \\ &&&&&&&&\Tref{SubspaceLocality}\\
   \hline
   $L$ & $s$ & $\left\lfloor \frac{2^{s}-2^t}{r\cdot 2^{t}-r}\right\rfloor +1$ & $t$ & $t$ & $r$ & $$ & $\frac{2^{s}-1}{2^{t}-1}$ & $s=rt$ \\ &&&&&&&&\Tref{SubspaceLocality}\\
   \hline
   $L$ & $3$ & $2$ & $2$ & $2$ & $1$ & $5$ & $6$ & $$ \\ &&&&&&&&Example~\ref{ex:foldLoc}\\
   \hline
   $L$ & $s$ & $1$ & $1$ & $1$ & $r$ & $h[s,r]$ & $h[s,r]$ & $$ \\ &&&&&&&&Theorem~\ref{th:LocCovk=1}\\
   \hline
   $L$ & $ws$ & $1$ & $t$ & $t$ & $r$ & $\left\lceil\dfrac{h[ws,r]}{2^t-1}\right\rceil$ & $\dfrac{(2^w - 1)h[s,r]_{2^w}}{2^t - 1}$ & $t|w$ \\ &&&&&&&&Corollary~\ref{cor:LocToCov}, Theorem~\ref{th:covToLoc}\\
   \hline
   $L$ & $4w$ & $1$ & $t$ & $t$ & $2$ & $$ & $\dfrac{(2^w - 1)(2^{w+1}+1)}{2^t - 1}$ & $t|w$ \\ &&&&&&&&Corollary~\ref{th:th58}\\
   \hline
   $L$ & $24$ & $1$ & $3$ & $3$ & $2$ & $828$ & $1097$ & $$ \\ &&&&&&&&Example~\ref{ex:covToLocEx}\\
   \hline
   $L$ & $22$ & $1$ & $2$ & $2$ & $3$ & $99$ & $154$ & $$ \\ &&&&&&&&Example~\ref{ex:covToLocEx}\\
   \hline
   $L$ & $7$ & $1$ & $2$ & $2$ & $2$ & $7$ & $7$ & $$ \\ &&&&&&&&Theorem~\ref{th:LOCs=7}\\

   \hline
\end{tabular}
\end{center}
\end{table*}






\end{document}